\newif\ifsubmodeapjs
\newif\ifsubmodeastroph

\submodeapjsfalse	
\submodeastrophtrue	

\ifsubmodeapjs
  \documentclass[12pt,preprint]{aastex}
  \usepackage{natbib,amsmath,verbatim}
  \received{}
  \revised{}
  \accepted{}
\fi
\ifsubmodeastroph
  \documentclass[12pt,preprint,iop]{emulateapj}
  \usepackage{natbib,amsmath,verbatim}

\fi

\newcommand{\etal}{{et al.~}}

\newcommand{\lta}{\lesssim}

\def\arcsec{\ensuremath{^{\prime\prime}}}

\def\Hubble{{\it Hubble}}
\def\Hubbles{{\it Hubble's}}

\newcommand{\note}[1]{}

\def\calacs{{\tt calacs}}
\def\calwf{{\tt calwf3}}
\def\drizzle{{\tt Drizzle}}
\def\pixfrac{{\tt pixfrac}}
\def\mosaicdrizzle{{\tt MosaicDrizzle}}
\def\multidrizzle{{\tt MultiDrizzle}}

\shorttitle{CANDELS: The Cosmic Assembly Near-infrared Deep Extragalactic Legacy Survey - HST Data Products}

\shortauthors{Koekemoer et al.}

\ifsubmodeastroph
  \submitted{Draft version April 30 2011}
  \submitted{Submitted to the Astrophysical Journal Supplement, 17 May 2011 --- Revised 18 Oct 2011}
\fi

\begin{document}

\title{CANDELS: The Cosmic Assembly Near-infrared Deep Extragalactic Legacy Survey ---
  The Hubble Space Telescope Observations, Imaging Data Products and Mosaics}

\author{
Anton M. Koekemoer\altaffilmark{1},
S. M. Faber\altaffilmark{2},
Henry C. Ferguson\altaffilmark{1},
Norman A. Grogin\altaffilmark{1},
Dale D. Kocevski\altaffilmark{2},
David C. Koo\altaffilmark{2},
Kamson Lai,\altaffilmark{2},
Jennifer M. Lotz\altaffilmark{1},
Ray A. Lucas\altaffilmark{1},
Elizabeth J. McGrath\altaffilmark{2},
Sara Ogaz,\altaffilmark{1},
Abhijith Rajan,\altaffilmark{1},
Adam G. Riess\altaffilmark{3},
Steve A. Rodney\altaffilmark{3},
Louis Strolger,\altaffilmark{4},
Stefano Casertano,\altaffilmark{1},
Marco Castellano,\altaffilmark{6},
Tomas Dahlen,\altaffilmark{1},
Mark Dickinson,\altaffilmark{5},
Timothy Dolch,\altaffilmark{3},
Adriano Fontana,\altaffilmark{6},
Mauro Giavalisco,\altaffilmark{7},
Andrea Grazian,\altaffilmark{6},
Yicheng Guo,\altaffilmark{7},
Nimish P. Hathi\altaffilmark{8},
Kuang-Han Huang,\altaffilmark{3,1},
Arjen van~der~Wel,\altaffilmark{9},
Hao-Jing Yan,\altaffilmark{10},
Viviana Acquaviva,\altaffilmark{11},
David M. Alexander\altaffilmark{12},
Omar Almaini,\altaffilmark{13},
Matthew L. N. Ashby\altaffilmark{14},
Marco Barden,\altaffilmark{15},
Eric F. Bell\altaffilmark{16},
Fr\'ed\'eric Bournaud,\altaffilmark{17},
Thomas M. Brown\altaffilmark{1},
Karina I. Caputi\altaffilmark{18},
Paolo Cassata,\altaffilmark{7},
Peter Challis,\altaffilmark{19},
Ranga-Ram Chary,\altaffilmark{20},
Edmond Cheung,\altaffilmark{2},
Michele Cirasuolo,\altaffilmark{18},
Christopher J. Conselice\altaffilmark{13},
Asantha Roshan Cooray\altaffilmark{21},
Darren J. Croton\altaffilmark{22},
Emanuele Daddi,\altaffilmark{17},
Romeel Dav\'e,\altaffilmark{23},
Duilia F. de~Mello\altaffilmark{24},
Loic de~Ravel,\altaffilmark{25},
Avishai Dekel,\altaffilmark{26},
Jennifer L. Donley\altaffilmark{1},
James S. Dunlop\altaffilmark{25},
Aaron A. Dutton\altaffilmark{27},
David Elbaz,\altaffilmark{28},
Giovanni G. Fazio\altaffilmark{14},
Alex V. Filippenko\altaffilmark{29},
Steven L. Finkelstein\altaffilmark{30},
Chris Frazer,\altaffilmark{21},
Jonathan P. Gardner\altaffilmark{24},
Peter M. Garnavich\altaffilmark{31},
Eric Gawiser,\altaffilmark{11},
Ruth Gruetzbauch,\altaffilmark{13},
Will G. Hartley\altaffilmark{13},
Boris H\"aussler,\altaffilmark{13},
Jessica Herrington,\altaffilmark{16},
Philip F. Hopkins\altaffilmark{29},
Jia-Sheng Huang,\altaffilmark{32},
Saurabh Jha,\altaffilmark{33},
Andrew Johnson,\altaffilmark{2},
Jeyhan S. Kartaltepe\altaffilmark{3},
Ali Ahmad Khostovan\altaffilmark{21},
Robert P. Kirshner\altaffilmark{14},
Caterina Lani,\altaffilmark{13},
Kyoung-Soo Lee,\altaffilmark{34},
Weidong Li,\altaffilmark{29},
Piero Madau,\altaffilmark{2},
Patrick J. McCarthy\altaffilmark{8},
Daniel H. McIntosh\altaffilmark{35},
Ross J. McLure\altaffilmark{36},
Conor McPartland,\altaffilmark{2},
Bahram Mobasher,\altaffilmark{37},
Heidi Moreira,\altaffilmark{38},
Alice Mortlock,\altaffilmark{13},
Leonidas A. Moustakas\altaffilmark{39},
Mark Mozena,\altaffilmark{2},
Kirpal Nandra,\altaffilmark{40},
Jeffrey A. Newman\altaffilmark{41},
Jennifer L. Nielsen\altaffilmark{35},
Sami Niemi,\altaffilmark{1},
Kai G. Noeske\altaffilmark{1},
Casey J. Papovich\altaffilmark{43},
Laura Pentericci,\altaffilmark{6},
Alexandra Pope,\altaffilmark{5},
Joel R. Primack\altaffilmark{2},
Swara Ravindranath,\altaffilmark{44},
Naveen A. Reddy\altaffilmark{5},
Alvio Renzini,\altaffilmark{45},
Hans-Walter Rix,\altaffilmark{9},
Aday R. Robaina\altaffilmark{46},
David J. Rosario\altaffilmark{2},
Piero Rosati,\altaffilmark{47},
Sara Salimbeni,\altaffilmark{48},
Claudia Scarlata,\altaffilmark{20},
Brian Siana,\altaffilmark{20},
Luc Simard,\altaffilmark{49},
Joseph Smidt,\altaffilmark{21},
Diana Snyder,\altaffilmark{2},
Rachel S. Somerville\altaffilmark{1},
Hyron Spinrad,\altaffilmark{29},
Amber N. Straughn\altaffilmark{24},
Olivia Telford,\altaffilmark{50},
Harry I. Teplitz\altaffilmark{20},
Jonathan R. Trump\altaffilmark{2},
Carlos Vargas,\altaffilmark{38},
Carolin Villforth,\altaffilmark{1},
Cory R. Wagner\altaffilmark{35},
Pat Wandro,\altaffilmark{2},
Risa H. Wechsler\altaffilmark{51},
Benjamin J. Weiner\altaffilmark{23},
Tommy Wiklind,\altaffilmark{1},
Vivienne Wild,\altaffilmark{36},
Grant Wilson,\altaffilmark{7},
Stijn Wuyts,\altaffilmark{14},
Min S. Yun\altaffilmark{7}
}

\altaffiltext{1}{Space Telescope Science Institute}

\altaffiltext{2}{UCO/Lick Observatory, University of California, Santa Cruz}

\altaffiltext{3}{The Johns Hopkins University}

\altaffiltext{4}{Western Kentucky University}

\altaffiltext{5}{National Optical Astronomy Observatories}

\altaffiltext{6}{INAF, Osservatorio Astronomico di Roma}

\altaffiltext{7}{University of Massachusetts, Amherst}

\altaffiltext{8}{Observatories of the Carnegie Institution of Washington}

\altaffiltext{9}{European Southern Observatory}

\altaffiltext{10}{The Ohio State University Research Foundation}

\altaffiltext{11}{Rutgers, The State University of New Jersey}

\altaffiltext{12}{Durham University}

\altaffiltext{13}{University of Nottingham}

\altaffiltext{14}{Max Planck Institute for Extraterrestrial Physics}

\altaffiltext{15}{Institute of Astro- and Particle Physics, University of Innsbruck}

\altaffiltext{16}{University of Michigan}

\altaffiltext{17}{Commissariat \`a l'\'Energie Atomique}

\altaffiltext{18}{SUPA, Institute for Astronomy, University of Edinburgh}

\altaffiltext{19}{Harvard College Observatory}

\altaffiltext{20}{California Institute of Technology}

\altaffiltext{21}{University of California, Irvine}

\altaffiltext{22}{Swinburne University of Technology}

\altaffiltext{23}{University of Arizona}

\altaffiltext{24}{NASA Goddard Space Flight Center}

\altaffiltext{25}{University of Edinburgh, Institute for Astronomy}

\altaffiltext{26}{Racah Institute of Physics, The Hebrew University}

\altaffiltext{27}{University of Victoria}

\altaffiltext{28}{CEA-Saclay/DSM/DAPNIA/Service d'Astrophysique}

\altaffiltext{29}{University of California, Berkeley}

\altaffiltext{30}{Texas A\&M Research Foundation}

\altaffiltext{31}{University of Notre Dame}

\altaffiltext{32}{Smithsonian Institution Astrophysical Observatory}

\altaffiltext{33}{Rutgers, the State University of New Jersey}

\altaffiltext{34}{Yale Center for Astronomy \& Astrophysics}

\altaffiltext{35}{University of Missouri - Kansas City}

\altaffiltext{36}{Royal Observatory Edinburgh}

\altaffiltext{37}{University of California, Riverside}

\altaffiltext{38}{Rutgers, the State University of New Jersey}

\altaffiltext{39}{Jet Propulsion Laboratory}

\altaffiltext{40}{Imperial College of Science}

\altaffiltext{41}{University of Pittsburgh}

\altaffiltext{42}{Space Telescope Science Insitute}

\altaffiltext{43}{Texas A\&M University}

\altaffiltext{44}{Inter-University Centre for Astronomy and Astrophysics}

\altaffiltext{45}{Osservatorio Astronomico di Padova}

\altaffiltext{46}{Institut de Ciencies del Cosmos}

\altaffiltext{47}{European Southern Observatory}

\altaffiltext{48}{University of Massachusetts, Amherst}

\altaffiltext{49}{Dominion Astrophysical Observatory}

\altaffiltext{50}{University of Pittsburgh, School of Engineering}

\altaffiltext{51}{Stanford University}

\begin{abstract}

This paper describes the Hubble Space Telescope imaging data products and data reduction procedures for the Cosmic Assembly Near-IR Deep Extragalactic Legacy Survey (CANDELS). This survey is designed to document the evolution of galaxies and black holes at $z\sim1.5-8$, and to study Type~Ia SNe beyond $z>1.5$. Five premier multi-wavelength sky regions are selected, each with extensive multiwavelength observations. The primary CANDELS data consist of imaging obtained in the Wide Field Camera 3 / infrared channel (WFC3/IR) and UVIS channel, along with the Advanced Camera for Surveys (ACS). The CANDELS/Deep survey covers $\sim\,$125$\;$square arcminutes within GOODS-N and GOODS-S, while the remainder consists of the CANDELS/Wide survey, achieving a total of $\sim\,$800$\;$square arcminutes across GOODS and three additional fields (EGS, COSMOS, and UDS). We summarize the observational aspects of the survey as motivated by the scientific goals and present a detailed description of the data reduction procedures and products from the survey. Our data reduction methods utilize the most up to date calibration files and image combination procedures. We have paid special attention to correcting a range of instrumental effects, including CTE degradation for ACS, removal of electronic bias-striping present in ACS data after SM4, and persistence effects and other artifacts in WFC3/IR. For each field, we release mosaics for individual epochs and eventual mosaics containing data from all epochs combined, to facilitate photometric variability studies and the deepest possible photometry. A more detailed overview of the science goals and observational design of the survey are presented in a companion paper.

\end{abstract}


%
%
%

%

\section {Introduction}\label{sec:intro}

In this paper we describe the \Hubble\ imaging and mosaic data products from the Cosmic Assembly Near-infrared Deep Extragalactic Legacy Survey (CANDELS), a 902-orbit Multi-Cycle Treasury (MCT) program aimed at documenting the evolution of galaxies and black holes from $z$$\,\sim$ 1.5 to 8, characterizing Type Ia supernovae (SNe~Ia) beyond $z > 1.5$ to better constrain the nature of dark energy, and probing galaxy evolution into the epoch of reionization. The CANDELS program uses the Wide Field Camera~3~/ Infra-Red channel (WFC3/IR) as its prime instrument, as well as the WFC3/UVIS channel, and obtains parallel observations with the Advanced Camera for Surveys (ACS). It is executing across Cycles 18, 19 and 20, and resulted from the combination of two approved proposals that were submitted in response to the special \Hubble\ MCT call for proposals 2009, which provided for large programs to address unique and broad science themes that could not be accommodated within the standard annual time-allocation process.

The structure of the survey includes essential elements of two MCT programs that were submitted separately: one, led by Ferguson, involved studying the full area of the GOODS-North and GOODS-South fields
	\citep{2004ApJ...600L..93G}\note{Giavalisco \etal 2004} 
to a uniform depth, including also UV imaging, and carrying out an extensive search for high-$z$ SNe~Ia; the other program, led by Faber, aimed at studying half the GOODS-North and GOODS-South areas to a greater depth, together with wider/shallower imaging of the Extended Groth Strip
	(EGS: \citealt{2007ApJ...660L...1D,2011.Newman}\note{Davis \etal 2007}, in prep),
	COSMOS \citep{2007ApJS..172....1S,2007ApJS..172..196K}\note{Scoville \etal 2007},
and the UKIDSS Ultra-Deep Survey
	(UDS: \citealt{2007MNRAS.379.1599L}\note{Lawrence \etal 2007};
	\citealt{2007MNRAS.380..585C}\note{Cirasuolo \etal.2007}),
while also permitting a search for SNe~Ia.

The combined CANDELS program obtains observations across all five fields, as well as including the SNe~Ia follow-up program, the UV imaging, and the multi-tier Deep+Wide observing strategy. Most of the observations use WFC3/IR as prime and ACS/WFC in parallel, and substituting UV imaging with WFC3/UVIS for parts of the orbit in \Hubbles\ continuous viewing zone (CVZ) that are too bright for WFC3/IR imaging. More detailed information is presented at our CANDELS website http://candels.ucolick.org, and in 
	\cite{2011.Grogin}\note{Grogin \etal 2011}
which provides a more detailed overview of the science goals and observing strategy. The outline of this paper is as follows: a brief outline of the major science goals is given in \S\ref{sec:science} to place the data products in context; a description of the fields is provided in \S\ref{sec:fields}; the observations are described in \S\ref{sec:observations}; the data products are presented in \S\ref{sec:data}; and we conclude in \S\ref{sec:summary}.

\section {Science Goals}\label{sec:science}

We summarize here the CANDELS science goals in the context of how they relate to the \Hubble\ data products. We refer to
	\cite{2011.Grogin}\note{Grogin \etal 2011}
for a more detailed description of the science goals, which include studies of galaxies in the reionization era (``cosmic dawn''), the growth and morphological transformation of galaxies during the era of peak star-formation and AGN activity (``cosmic high noon''), and measurements of high-$z$ SNe~Ia to constrain dark energy and measure supernova rate evolution.

\hfill\break
\noindent{\bf Cosmic Dawn.}
QSO observations 
	\citep{2006ARA&A..44..415F}\note{Fan \etal 2006} 
and WMAP 
	(\citealt{2007ApJS..170..335P, 2007ApJS..170..377S})\note{Page \etal 2007, Spergel \etal 2007}
indicate that the intergalactic medium was reionized between $t_0 = 0.5$ and 1$\,$Gyr. The IGM was seeded with metals to $Z  \sim 4 \times 10^{-4}$ within the first billion years, and the energy released by the stars that produced these metals appears sufficient to reionize the IGM 
	(\citealt{2001ApJ...561L.153S, 2006MNRAS.371L..78R})\note{Songaila 2001; Ryan-Weber \etal 2006}.
The bright end of the UV luminosity function (UVLF) of star-forming galaxies evolves rapidly at 4$\,<\,$$z$$\,<\,$7 (e.g.,
	\citealt{2004ApJ...600L..99D,
		2007ApJ...670..928B,
		2008ApJ...686..230B,
		2010ApJ...708L..69B}),
but the UV flux from these galaxies appears insufficient to explain reionization without extrapolations --- an important puzzle to solve. Current $z$$\,\sim\,$8 luminosity functions are based on only a handful of objects, most with $L_{UV} \lta L^\ast$. The CANDELS data in the $z_{850}Y_{105}J_{125}H_{160}$ filters can provide measurements of the bright end of the UVLF at $z$$\,\approx\,$7$\,-\,$8 and also permit robust LBG color selection at $\langle z \rangle =\,$5.8, 6.6, and 8.0, for $L^\ast$ for $z$$\,=\,$7  ($J_{125}$$\,\approx\,$27) and $1.5 L^\ast$ for $z$$\,=\,$8 ($H_{160}$$\,\approx\,$27), as well as fainter LBGs at higher $z$. This can constrain extinction via UV spectral slopes and improve measurements of their evolution.

Furthermore, the evolution of faint AGN at $z$$\,\gtrsim\,$6$\,-\,$7 can be directly probed by cross-correlating the drop-out samples with the deep X-ray data in these fields (e.g.,
	\citealt{2003AJ....125.1649F},	
	\citealt{2004ApJ...600L.123K},	
	\citealt{2008MNRAS.387..883A},	
	\citealt{2009A&A...507.1277B})	
for which the depth in the near-IR data is crucial. Moreover, photometric redshifts and non-LBG color criteria can help reveal whether there are non-star-forming galaxies lurking at these redshifts
	(\citealt{2005ApJ...635..832M},	
	\citealt{2008ApJ...676..781W},	
	\citealt{2007ApJ...665..257C},	
	\citealt{2007MNRAS.376.1054D}),	
thereby driving the photometric requirements of the CANDELS data products. The CANDELS data also allow fluctuations in the extragalactic background light (EBL) to be probed, potentially constraining the properties of the first generations of stars
	(\citealt{2004ApJ...606..611C},	
	\citealt{2010ApJ...710.1089F}),	
as well as enabling the use of clustering statistics to constrain the properties of dark matter halos
	(\citealt{2006ApJ...647..201C},	
	\citealt{2006ApJ...642...63L,2009ApJ...695..368L}),	
which drives the need to produce contiguous mosaics across each of the CANDELS fields.

\hfill\break
\noindent{\bf Cosmic High Noon.}
At $z$$\,\sim\,$2, star formation and nuclear activity within galaxies are at their peaks while the morphological differentiation of galaxies is well under way. A key question is what drives stellar mass buildup, bulge growth, and the emergence of passive ellipticals. To resolve this requires accurate mass function measurements well below $M^\ast$, achieved by robust SED fitting at rest-frame optical wavelengths, where the 4000$\,$\AA\ and Balmer breaks constrain accurate photo-$z$ measurements, stellar population ages, and $M/L$ ratios. The CANDELS data are designed to match the photometric depths of the existing {\it Spitzer} IRAC and {\it HST} ACS data, accurately determining the mass function of quiescent galaxies for $M > 10^9 M_\odot$ at $z$$\,\approx\,$2 and their contribution to the global mass density. Another key question in galaxy growth is the relative importance of mergers and structural instabilities. The discovery of large, rotating, clumpy, gas-rich disks at $z$$\,\sim\,$2 implies that disk instabilities may drive bulge formation more rapidly than previously thought
	(\citealt{2006ApJ...645.1062F,2009ApJ...706.1364F}\note{F\"orster Schreiber \etal 2006, 2009};
	\citealt{2006Natur.442..786G,2008ApJ...687...59G}\note{Genzel \etal 2006, 2008}).

The star formation rate also correlates strongly with stellar mass $M_{\star}$, and the zeropoint of the SFR$(M_{\star})$ relation declines steadily below $z$$\,\sim\,$2.5
	(\citealt{2007ApJ...670..156D}\note{Daddi \etal 2007};
	\citealt{2007A&A...468...33E}\note{Elbaz \etal 2007};
	\citealt{2007ApJ...660L..43N}\note{Noeske \etal 2007}).
The empirical evidence is now reminiscent of the ``clump-merging'' scenario for the growth of bulges in gas-rich disks from numerical simulations 		(e.g.,
	\citealt{1999ApJ...514...77N}\note{Noguchi \etal 1999};
	\citealt{2004ApJ...611...20I}\note{Immeli \etal 2004};
	\citealt{2007ApJ...670..237B}\note{Bournaud \etal 2007};
	\citealt{2008ApJ...688...67E}\note{Elmegreen \etal 2008}).
CANDELS can provide a census of clumps within galaxies along with their sizes and masses, and the resulting estimates of bulge formation rates can be compared to the timescale of clump migration driven by dynamical friction. Through comparison with the deep X-ray catalogs, CANDELS also provides detailed morphological information on AGN in this redshift range, tracking the connection between galaxy mergers and black hole growth.
Finally, the evolution of galaxies with very low specific SFRs (passive galaxies) is also of interest, with sources having been found out to at least $z$$\,=\,$2.5
	(e.g.,
	\citealt{2005ApJ...626..680D}\note{Daddi \etal 2005};
	\citealt{2006MNRAS.373L..36T}\note{Trujillo \etal 2006};
	\citealt{2008A&A...482...21C}\note{Cimatti\etal 2008};
	\citealt{2008ApJ...677L...5V}\note{van Dokkum \etal 2008}).
CANDELS enables the luminosity function of large numbers of these sources to be directly constrained, in addition to probing their morphogical structure to faint limits in the near-IR with better resolution than previous studies.

\hfill\break
\noindent{\bf Deep Ultraviolet Observations.}
An important feature of CANDELS is the fact that the GOODS-North field is in the {\it HST} CVZ, thus we use the bright day-side of the orbit to observe with WFC3/UVIS in the ultraviolet (F275W and F336W). This enables measurements of the Lyman-continuum (LyC) escape fraction from galaxies at $z$$\,\sim\,$ 2.5, identification of $\sim$350 Lyman-break galaxies at $z$$\,\sim\,$2, and measurements of the star-formation rate in low-luminosity dwarfs which may just be ``turning on'' at $z$$\,\sim\,$1
	(\citealt{1992MNRAS.255..346B}\note{Babul \& Rees 1992};
	\citealt{2000ApJ...539..517B}\note{Bullock \etal 2000}).
There are $\sim\,$40$\,-\,$50 UV-luminous LBGs ($L_{UV} > 0.25 L^*$) in this field at 2.38$\,<\,$$z$$\,<\,$2.55 (half with spec-$z$), which is the optimal redshift for constraints with the F275W filter, many of which may be bright enough to detect if $f_{\rm esc}$$\,>\,$0.5
	(e.g.,
	\citealt{2006ApJ...651..688S}\note{Shapley \etal 2006};
	\citealt{2009ApJ...692.1287I}\note{Iwata \etal. 2009}).
Importantly, these galaxies are at redshifts that allow H$\alpha$ measurements for an independent measure of the ionizing continuum. The resolved LyC distributions provide tests of different mechanisms for high $f_{esc}$\ including SNe winds
	(\citealt{2002MNRAS.337.1299C}\note{Clarke \& Oey 2001},
	\citealt{2002ApJ...577...11F}\note{Fujita \etal 2002}),
minor galaxy interactions
	\citep{2008ApJ...672..765G}\note{Gnedin \etal 2008},
and emission from globular cluster formation
	\citep{2002MNRAS.336L..33R}\note{Ricotti \etal 2002}.

\hfill\break
\noindent{\bf Supernova Cosmology.}
While SNe~Ia at $z$$\,\lesssim\,$1$\,-\,$1.5 have already provided startling evidence of dark energy
	(\citealt{1998AJ....116.1009R,
	1999ApJ...517..565P,
	2004ApJ...607..665R,
	2007ApJ...659...98R}),
CANDELS now provides a direct probe of 1.5$\,<\,$$z$$\,<\,$2.5 to test the nature of SNe~Ia progenitors and their possible evolution
	(\citealt{1998ApJ...497L..57R}\note{Ruiz-Lapuente \& Canal 1998};
	\citealt{2005A&A...433..807M}\note{Mannucci et al. 2005};
	\citealt{2009ApJ...707.1466K}\note{Kobayashi \& Nomoto 2009};
	\citealt{2008MNRAS.388..829G}\note{Greggio et al. 2008}),
which can be tested with CANDELS since the predicted rates diverge significantly at $z$$\,>\,$1.5. In addition, CANDELS yields SNe~Ia at $z$$\,\lesssim\,$1.5 which remain crucial tracers of the evolution in the dark energy equation of state $w$. Increasing the samples at 0.7$\,<\,$$z$$\,<\,$1 in the IR reduces the uncertainties in host extinction, thereby testing whether dust is a factor in the declining high-$z$ SNe~Ia rate. CANDELS also includes followup WFC3 or ACS grism observations to determine the supernovae type and redshift, and rest-frame $B$ and $V$  light curves for each SN (thus limiting the K-correction errors to below the random distance error of the SNe). We note also that related follow-up programs to CANDELS (including the SNe~Ia search) obtain grism data on these fields, but these are separate programs from the CANDELS imaging survey and are not discussed in further detail here.

\ifsubmodeastroph
\hfill\break
\begin{deluxetable*}{llllrlrlr}[h]
\tablecaption{\label{tab:filters}
	New {\it HST} WFC3 and ACS Filter Coverage in the CANDELS Fields}
\tablehead{%
Field		& Prog.		& Cycles	& \multicolumn{2}{c}{WFC3/IR}		& \multicolumn{2}{c}{WFC3/UVIS}		& \multicolumn{2}{c}{ACS/WFC}		\\
$\,$		& ID		&		& Filters		& \#Exp		& Filters		& \#Exp	& Filters		& \#Exp}
\startdata
UDS (Wide)	& 12064		& 18		& F125W F160W		& 352		& F350LP		& 24		& F606W F814W		& 376		\\
GOODS-S Deep	& 12060,1,2	& 18+19		& F125W F160W		& 696		& F350LP		& 174		& F606W F814W F850LP	& 870		\\
GOODS-S Wide	& 12060,1	& 18+19		& F125W F160W F105W	& 134		& F350LP		& 0		& F606W F814W F850LP	& 134		\\
EGS (Wide)	& 12063		& 18+20		& F125W F160W		& 360		& F350LP		& 18		& F606W F814W		& 378		\\
GOODS-N Deep	& 12442,3,4,5	& 19+20		& F125W F160W F105W	& 660		& F350LP		& 360		& F606W F814W F850LP	& 780		\\
GOODS-N Wide	& 12442,3,4	& 19+20		& F125W F160W F105W	& 240		& F350LP F275W F336W	& 40		& F606W F814W F850LP	& 240		\\
COSMOS (Wide)	& 12440		& 19		& F125W F160W		& 352		& F350LP		& 24		& F606W F814W		& 376		\hspace{-3pt}\vspace{2pt}\\
{\bf Total}	&		&		&			& {\bf 2794}	&			& {\bf 640}	&			& {\bf 3154}
\enddata
\end{deluxetable*}
\fi

\section {The CANDELS Fields}\label{sec:fields}

Here we summarize the general properties of the CANDELS fields and refer to
	\cite{2011.Grogin}\note{Grogin \etal 2011}
for a more detailed description. The CANDELS survey consists of a two-tier Deep+Wide survey designed to address the science goals discussed in \S\ref{sec:science} as well as providing a legacy dataset on these fields. The CANDELS Deep portion covers $\sim\,$125 square arcminutes to $\sim\,$10-orbit depth within the GOODS-North and GOODS-South Fields
	\citep{2004ApJ...600L..93G}\note{Giavalisco \etal 2004} 
including the E-CDFS
	\citep{2004ApJS..152..163R}	
as well as the WFC3 ERS2 field
	\citep{2011ApJS..193...27W}.	
The full area of the CANDELS survey covers a total of $\sim\,$800 square arcminutes, where the additional area includes the shallower Wide portion to $\sim\,$2-orbit depth around the Deep portions of GOODS, together with subsections of three additional fields, namely the Extended Groth Strip
	(EGS: \citealt{2007ApJ...660L...1D}\note{Davis \etal 2007}; Newman et al. 2011, in prep.),
	COSMOS \citep{2007ApJS..172....1S,2007ApJS..172..196K}\note{Scoville \etal 2007},
and the UKIDSS Ultra-Deep Survey
	(UDS: \citealt{2007MNRAS.379.1599L}\note{Lawrence \etal 2007};
	\citealt{2007MNRAS.380..585C}\note{Cirasuolo \etal.2007}).
When combined with the existing Ultra Deep Field
	(UDF: \citealt{2006AJ....132.1729B}\note{Beckwith \etal 2006})
within GOODS-South, CANDELS provides a unifying survey at three principal exposure time depths with approximately an order of magnitude difference between each. Another unifying aspect of the survey is that all five CANDELS fields are the targets of the {\it Spitzer} Extended Deep Survey (Fazio et al., in prep.) which is a 2108~hour program, together with a more recently approved 1200~hour follow-on program, covering each of these regions with {\it Spitzer} IRAC 3.6$\,\micron$ and 4.5$\,\micron$ imaging to a total depth of 12 hours per pointing.

Each of the five CANDELS fields has a wealth of additional imaging and spectroscopic ancillary data from X-rays to radio wavelengths, described in the aforementioned papers and their references. For the present work, we note in particular that all of them, except the UDS, have pre-existing {\it HST} data covering the field. In addition, all five fields have extensive pre-existing catalogs that can serve as astrometric and photometric reference standards as well as being combined with the catalogs from new {\it HST} data to obtain derived measurements of source properties including photometric redshifts, stellar masses and star formation histories.

\section {Observations}\label{sec:observations}

We summarize here the general layout of the observations, referring to
	\cite{2011.Grogin}\note{Grogin \etal 2011}
for a more detailed description. The {\it HST} observations in the CANDELS fields can be summarized as consisting of three complementary sets of imaging data: WFC3/IR, WFC3/UVIS and ACS/WFC imaging exposures.%
	\footnote{For current details on ACS see http://www.stsci.edu/hst/acs}
	\footnote{For\,current\,details\,on\,WFC3\,see\,http://www.stsci.edu/hst/wfc3}
In all cases, the WFC3 observations are taken as the prime observations, while the ACS data are obtained in parallel. The filter breakdown is shown in Table~\ref{tab:filters} for each of the different fields. We also discuss two generally different sets of observations, namely the GOODS fields, which contain the CANDELS-Deep pointings (together with a CANDELS-Wide ``flanking field'' in GOODS-South), and the other three fields (COSMOS, EGS and UDS) which only consist of the CANDELS-Wide component and thus have a somewhat different structure.

\subsection{Filters and Exposure Times}

We first describe the filter choice and exposure structure of the three Wide fields (COSMOS, EGS and UDS), each of which is covered using a mosaic grid of tiles and repeated over two epochs. During each epoch, Each tile is observed for one orbit ($\sim\,$2000\,s), divided into two exposures in F125W (at a depth of $\sim\,$1/3~orbit) and two exposures in F160W (at a depth of $\sim\,$2/3~orbit), together with parallel exposures using ACS/WFC in F606W and F814W. However, some WFC3/IR tiles near one end of the Wide mosaics are not covered by ACS parallels so a short 434$\,$s WFC3/UVIS F350LP exposure is inserted, creating a total of five WFC3 exposures per orbit for these tiles, and five ACS/WFC exposures in parallel.

The GOODS-North and GOODS-South fields contain the Deep portions of the CANDELS survey, with a total depth of at least 4~orbits in both WFC3/IR F125W and F160W and 3~orbits in F105W, spread across 10~epochs. Each single-orbit pointing, for each epoch, contains four WFC3/IR exposures (two F125W and two F160W), and one WFC3/UVIS (F350LP). In parallel, we also obtain five ACS/WFC exposures, where the primary requirement is to obtain at least 32,000~s depth in F814W. Once this requirement is met, the next ACS/WFC priorities are $\sim\,$2500~s of F850LP, followed by $\sim\,$5000~s in F606W, and finally any remaining depth is placed back into F814W. Since the GOODS-North field is in the CVZ, some portions of the orbit are too bright for observations using WFC3/IR so WFC3/UVIS exposures are substituted, using the F275W and F336W filters. In these cases the ACS parallels retain their structure as described for the remainder of the Deep observations.

Finally, the GOODS-South Deep portion has a wider ``flanking field'', similar to the Wide fields in its filter choice and exposure time requirements. This is divided into two epochs, achieving a depth of $\sim\,$1/3~orbit in F125W and $\sim\,$2/3~orbit in F160W per epoch, also using short WFC3/UVIS F350LP exposures where necessary, and with ACS/WFC F606W and F814W exposures in parallel. For the ACS/WFC exposures falling outside the Deep area, the exposure time requirements are firstly to obtain $\sim\,$2500~s in F814W, followed by $\sim\,$2500~s in F850LP, and distribute any remaining time according to the same priorities as for the Wide mosaics. The GOODS-North ``flanking field'' is similar in design, with a slightly different layout. Note that, since the ACS/WFC field of view is larger than WFC3/IR, the ACS parallel pointings overlap considerably and their effective exposure time on overlapping pointings can exceed the nominal 2$\,$orbit observing time, due to the dense packing of pointings for contiguous WFC3/IR coverage.

\begin{deluxetable}{lcc}
\tablecaption{\label{tab:zeropoints}
	CANDELS ACS and WFC3 Filter Zeropoints\tablenotemark{a}%
	}
\tablehead{%
Instrument/Camera	& Filter	& Zeropoint (ABmag)}
\startdata
ACS/WFC			& F606W		& 26.49		\\
ACS/WFC			& F814W		& 25.94		\\
ACS/WFC			& F850LP	& 24.84		\\
WFC3/UVIS		& F275W		& 24.14		\\
WFC3/UVIS		& F336W		& 24.64		\\
WFC3/UVIS		& F350LP	& 26.94		\\
WFC3/IR			& F105W		& 26.27		\\
WFC3/IR			& F125W		& 26.25		\\
WFC3/IR			& F160W		& 25.96
\enddata
\tablenotetext{a}{For current filter zeropoint information, please see the
following webpages:\hfill\break
http://www.stsci.edu/hst/acs/analysis/zeropoints\hfill\break
http://www.stsci.edu/hst/wfc3/phot\_zp\_lbn}
\end{deluxetable}

In Table~\ref{tab:zeropoints} we list the current zeropoints (in the AB magnitude system) for all the filters used in the CANDELS survey observations. We note that these are subject to change and we provide links to the instrument webpages where the most up-to-date zeropoints can be obtained in future. The primary uncertainties associated with these are related to the spectral characteristics of the standard stars that are used by staff at STScI to carry out the photometric zeropoint calibrations, as well as changes in the instrument and filters over time. Generally these are accurate to better than $\sim\,$1$\,-\,$2\%, and we present later in Section~\ref{sec:photom} a quantitative validation of this level of accuracy using the photometry from our CANDELS data, compared with photometric data from ground-based imaging.

\subsection{Mosaic Layout Design}

For each of the five CANDELS fields, the goal is to cover a contiguous area with WFC3/IR (thus, the larger ACS/WFC parallel exposures overlap somewhat to create deeper pointings), and to overlap as much as possible the  existing relevant ancillary datasets. Here we summarize the specific considerations for each of the fields and how they impact the overall design of the mosaic observations.

For the three Wide fields (COSMOS, EGS and UDS), the layout consists of a rectangular region, which for COSMOS and UDS comprises a grid of 4$\,\times\,$11 tiles ($\sim\,$8$\farcm$6$\,\times\,$23$\farcm$8), at spacing intervals designed to allow for maximal contiguous coverage in WFC3/IR without introducing gaps between tiles as a result of pointing errors. The exposures are all oriented so that the ACS/WFC parallels are offset along the long axis of the mosaic, thereby producing a similar-sized mosaic overlapping the bulk of the WFC3/IR mosaic, except at its ends where some tiles are covered only by WFC3/IR or by ACS/WFC, but not both. For the EGS field the mosaic is instead 3$\,\times\,$15 tiles ($\sim\,$6$\farcm$5$\,\times\,$32$\farcm$5), to optimize coverage with ancillary data.

For the GOODS-North and GOODS-South Deep regions, the layout consists of a smaller rectangular grid of 3$\,\times\,$5 tiles ($\sim\,$6$\farcm$5$\,\times\,$10$\farcm$8). In GOODS-South the WFC3/IR pointings are placed adjacent to the existing WFC3/IR ERS2 observations
	(Windhorst etal 2011).
Since the field is observed over 10 epochs, the orientation of the tiles rotates by $\sim\,$45$\,-\,$50$\arcdeg$ from one epoch to the next. This also changes the coverage of the parallel ACS/WFC observations, creating a net effect of a larger area covered by ACS/WFC to shallower depth, surrounding the deep central WFC3/IR data, which therefore has a slight deficit of ACS coverage.

Finally, the shallower ``flanking field'' region in GOODS-South covers $\sim\,$2$\,\times\,$4 tiles, divided into two epochs, and using a similar filter choice and exposure time strategy as the other Wide fields. In Goods-North the layout is similar, with differences due to the field geometry. The pointings are oriented such that the ACS/WFC parallels land mostly on the GOODS-South Deep region. The GOODS-South ``flanking field'' region is also covered with F105W to 1\,orbit depth which provides additional parallel ACS/WFC data for the central region.

\subsection{Sub-Pixel Dither Pattern}

Each mosaic tile is observed for one orbit during each epoch, where the prime WFC3/IR observations consist of four exposures. In most cases these four exposures consist of two exposures each in F125W and F160W, except for the Y-band visits where all four exposures are obtained in F105W. Due to the relatively large pixel scale of the WFC3/IR detector (0$\farcs$128/pixel at its central reference pixel), we offset the four WFC3/IR exposures in each orbit using a 4-point small-scale dither pattern to provide half-pixel subsampling of the point spread function (PSF) while also mitigating the impact of hot pixels and persistence. The strategy of dividing the 4-point dither pattern into two pointings with F125W and two with F160W was particularly motivated by the supernova science, where we carried out tests to ensure that good supernova subtraction could be achieved. Since the second epoch on these tiles also contain two exposures in F125W and F160W, we ultimately achieve a 4-point dither pattern in each filter, on all tiles, once the data from all the epochs are combined.

\begin{deluxetable}{cccc}
\tablecaption{\label{tab:dithers}
	CANDELS WFC3/IR Sub-pixel Dither Pattern}
\tablehead{%
Position & Filter	& X offset (pix)	&  Y offset (pix)}
\startdata
1	& F160W		& $-$2.5		& $-$2.5	\\
2	& F125W		& $-$4.5		& $+$2.0	\\
3	& F125W		& $+$2.0		& $+$2.5	\\
4	& F160W		& $+$4.0		& $-$2.0
\enddata
\end{deluxetable}

The dither pattern serves two complementary purposes: 1) provide non-integer shifts to subsample the PSF; 2) add integer components to these shifts in order to ensure that hot pixels and possible persistence from previous bright sources are moved around sufficiently. In particular, persistence is a concern, especially ``self-persistence'' from sources in previous CANDELS exposures executed as part of the dither pattern, since it typically tends to be more extended than a single pixel and is diffuse, thereby subtly impacting the photometry if it is not mitigated. For compact sources, the expected spatial extent has been quantified as a circle $\sim\,$2.5\,pixels in diameter (WFC3/IR), and therefore constrains the minimum size of the dither offset. On the other hand, due to the geometric distortion of the detector, an offset of a certain number of pixels at the center corresponds to a different number of pixels near the edge, and for sufficiently large shifts the subsampling can vary from half-pixel to integer several times between the center and the edge of the detector, introducing non-uniform sub-pixel sampling.

\begin{figure}[t]
\begin{center}
\ifsubmodeastroph	\includegraphics[width=8cm]{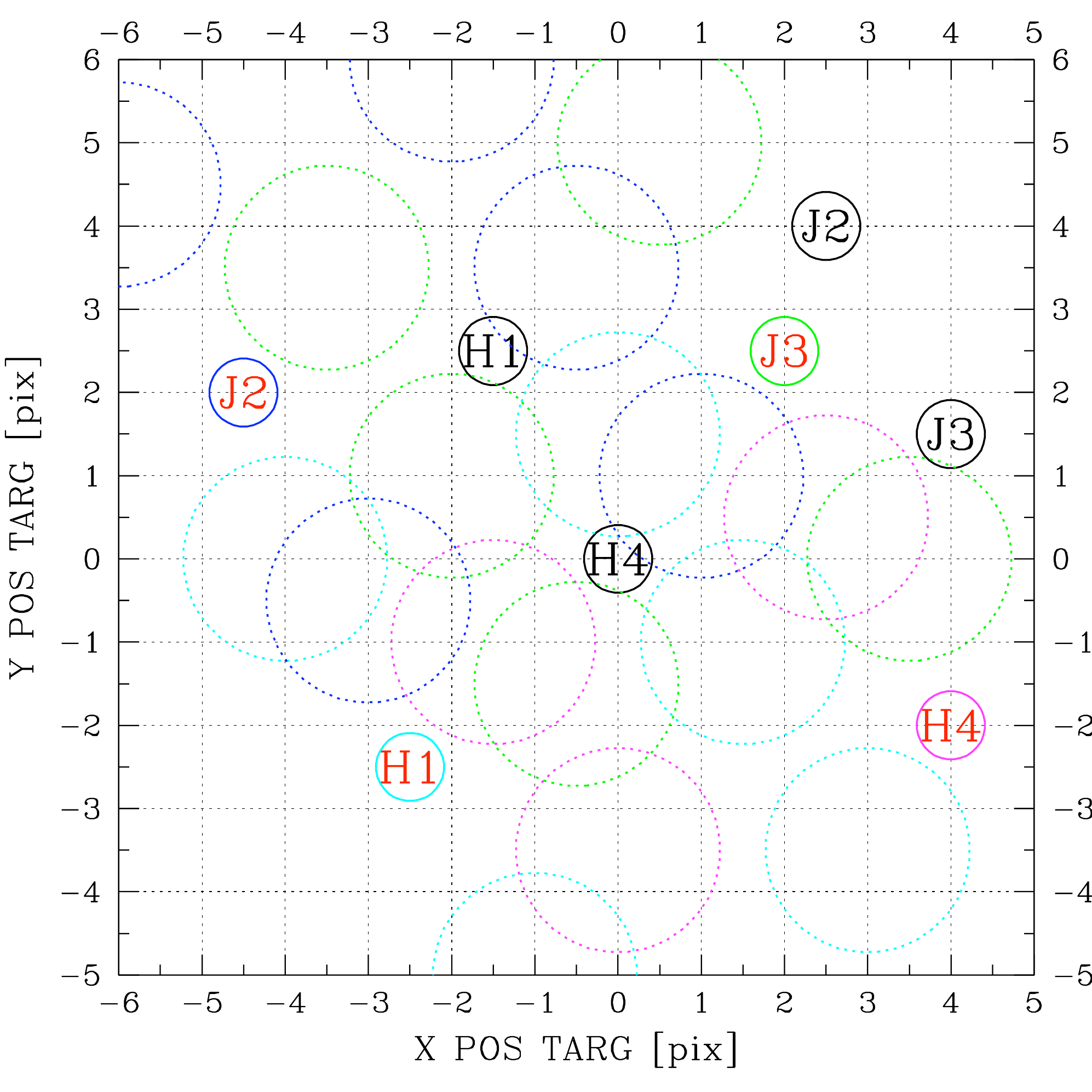}	\fi
\figcaption{\label{fig:dither}%
The CANDELS WFC3/IR sub-pixel dither pattern, indicated by the set of four red labels H1, J2, J3, H4, as well as the standard default dither pattern (indicated by the four black labels). The large circles indicate the 2.5-pixel diameter regions that are expected to be impacted by persistence from a point source, thus their overlap ends to be minimized. The new CANDELS 4-point dither pattern is chosen to be a few pixels wider, simultanteously still providing half-pixel subsampling for the WFC3/IR observations as well as introducing larger offsets to help mitigate the effects of persistence from a previous exposure, by ensuring that any persistent pixels from previously observed bright sources are moved around to different, non-repeated pixel locations and can therefore be excluded from the final image combination.}
\end{center}
\end{figure}

Therefore, the desire to retain uniform half-pixel sampling across the entire WFC3/IR detector during each epoch constrains the dither offsets to be as small as possible, which in this case is the minimum size needed to avoid issues from persistent regions $\sim\,$2.5\,pixels across. As a result, larger detector blemishes (such as the WFC3/IR ``deathstar'', a circular region of bad pixels $\sim\,$50$\,$pixels in diameter, or $\sim\,$6$\farcs$4) are not covered during a given epoch, and instead subsequent epochs at different orientations are used to provide coverage. The final pattern that satisfies both the persistence requirements and the PSF sampling requirements is a 4-point dither pattern with parameters shown in Table~\ref{tab:dithers}, and presented graphically in Figure~\ref{fig:dither}. This has the advantage of being large enough to avoid small-scale detector defects and persistence from compact sources, but small enough for the half-pixel subsampling to change by $\lesssim\,$0.1$\,-\,$0.2~pixels across the detector.

\subsection{Detector Characteristics and Read-out Modes}

All the near-infrared {\it HST} observations on the CANDELS fields are obtained using the WFC3/IR detector which consists of a 1024$\times$1024 Teledyne HgCdTe array, comprising a central 1014$\times$1014 pixel area for imaging (covering a region $\sim\,$130$\arcsec$ across, with a plate scale of 0$\farcs$128/pixel at its central reference pixel), surrounded by a 5-pixel wide strip of reference pixels along the edges which are unexposed to light and are used to track changes in bias level that may occur during an exposure. The exposures are all obtained using the WFC3 IR-FIX aperture, which samples the full imaging field of view of the WFC3/IR detector. All exposures are obtained in MULTIACCUM mode using either a SPARS50 or SPARS100 read-out sequence, enabling the pixels to be sampled non-destructively every 50 or 100 seconds respectively. These SPARS read samples are repeated for each sequence by specifying the NSAMP parameter, which range from 9$\,-\,$14 for the SPARS50 sequences and 6$\,-\,$10 for the SPARS100 sequences before reading out the array, depending on the scheduling constraints for each particular field and epoch, leading to exposure times in the range 450$\,-\,$1000$\,$seconds. In addition, each of these MULTIACCUM exposures is preceded by a short read 2.9 seconds in duration (the zeroth read), which serves as a measure of the bias structure across the array at the start of each exposure.

The optical and UV observations are all obtained using the WFC3/UVIS and ACS/WFC cameras,
which are rather similar in construction, each consisting of two thinned, back-illuminated CCDs with a usable area of of either 4096$\times$2048 (ACS) or 4096$\times$2051 (WFC3), located adjacent to one another to form an imaging area approximately square in size, with a small physical gap between the detectors. The WFC3/UVIS wavelength response is optimized to be UV sensitive, as compared with the ACS/WFC which we use for observations at longer wavelengths. Each detector has two amplifiers, one at each corner, with half the pixels being read out in parallel by each. The ACS/WFC CCDs were manufactured by SITe (Scientific Imaging Technologies) and are each physically 4144$\times$2048 pixels. The first and last 24 columns are read out as physical overscan pixels which are unexposed to light, while an additional 20 rows of virtual overscan are added while clocking out each column during read-out. This yields a usable image area of 4096$\times$2048 pixels for each of the two chips, provided by the WFC aperture which we use for all the ACS observations, with the physical gap between them corresponding to about a 50~pixel width ($\sim\,$2$\farcs$5), and covering a combined area $\sim\,$200$\arcsec$ in extent with a plate scale of 0$\farcs$05/pixel at the central reference pixel. The WFC3/UVIS CCDs were manufactured by e2v (formerly Marconi) and are each physically 4146$\times$2051 pixels, being read out using the first and last 25 columns as physical overscan pixels which are unexposed to light, while an additional 30 columns of serial virtual overscan are added at the boundary between the two amplifier regions on each CCD, together with an additional 19 rows of parallel virtual overscan added at the end of each column during read-out. Thus their usable image area, provided by the UVIS-IR-FIX and UVIS-CENTER apertures which we use for all the WFC3/UVIS observations, is 4096$\times$2051 pixels for each of the two chips, with the physical gap between them corresponding to about 30 pixels ($\sim\,$1$\farcs$2), and covering a combined area $\sim\,$160$\arcsec$ in extent with a plate scale of 0$\farcs$0396/pixel at the central reference pixel.

The WFC3/UVIS camera is used for the UV F275W and F336W observations in the CVZ for GOODS-North, as well as for the F350LP observations for all fields, which is a long-pass filter covering the detector response from $\sim\,$3500$\,$\AA\ to its red cutoff around 1$\,\micron$. The latter exposures are all 434\,s in length, with no CRSPLIT, consisting of only a single exposure per visit. The ACS exposures are all obtained in the F606W, F814W and F850LP filters, depending on which field is being observed, and are all obtained as parallel exposures to either the WFC3/IR or WFC3/UVIS exposures. As a result, they have a range of exposure times between 225\,s and 807\,s, driven by read-out times and overhead considerations associated with managing the buffer dumps of both ACS and WFC3 in parallel, depending on which field is being observed. In addition, the ACS/WFC coverage from one pointing to the next is also less homogeneous, both in terms of the filters used and also its depth, particularly for the three CANDELS Wide fields, due to the scheduling and buffer time management constraints.

\section{Data Processing and Products}\label{sec:data}

\begin{figure*}[t!]
\begin{center}
\ifsubmodeastroph	\includegraphics[height=18cm]{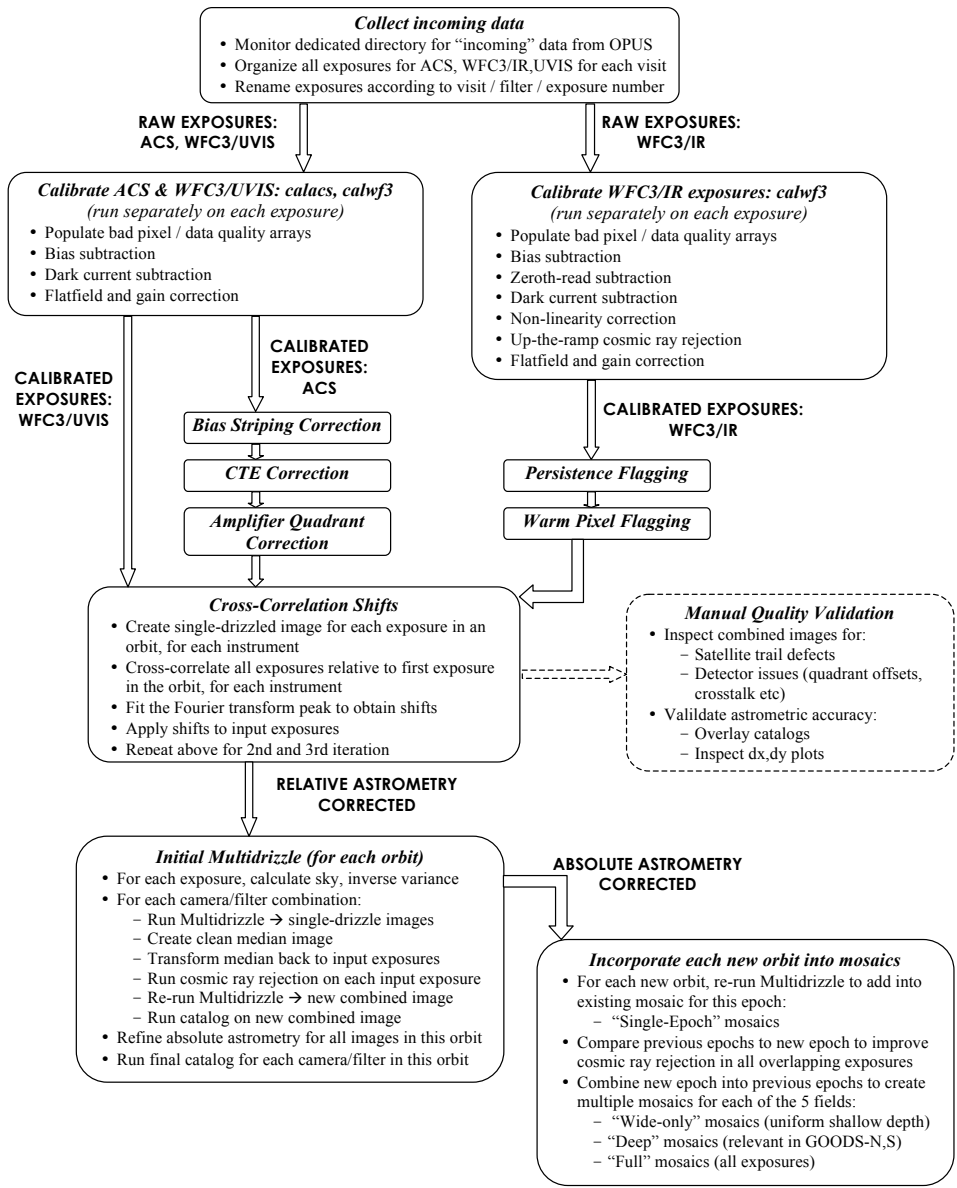}	\fi
\figcaption{\label{fig:pipeline}%
General overview of the ``\mosaicdrizzle'' pipeline used for the CANDELS data processing, showing each of the steps that are used to process each new dataset as it arrives, including automated calibration and \multidrizzle\ combination, as well as data quality validation, astrometric registration and mosaic combination.}
\end{center}
\end{figure*}

For each of the five CANDELS fields, the final data products consist of a set of mosaics for each camera/filter combination, processed using the ``\mosaicdrizzle'' pipeline (described here), which carries out astrometric registration and image combination using \multidrizzle\
	\citep{2002hstc.conf..337K}\note{Koekemoer et al. 2002},
and \drizzle\
	\citep{2002PASP..114..144F}\note{Fruchter \& Hook 2002}.
Here we describe the original data, together with each of the calibration and pipeline processing steps included in the ``\mosaicdrizzle'' pipeline (outlined in Figure~\ref{fig:pipeline}), as well as the properties of the final combined mosaic images. All the processing is done on a dedicated cluster of linux machines at STScI, using the Pyraf/STSDAS packages.

For all exposures in each visit, the raw files are processed first through a number of basic calibration steps that are needed to produce a final calibrated file for each exposure. Since these steps are somewhat different for the WFC3/IR detectors than for the CCD cameras, we discuss the WFC3/IR data calibration steps separately. The WFC3/UVIS and ACS/WFC calibration steps, however, are very similar to one another and are therefore discussed together, with differences highlighted where necessary.

\subsection{WFC3/IR Detector Calibration}

Here we describe the calibration steps carried out to convert the WFC3/IR data from raw counts to a set of final, flatfielded and flux-calibrated exposures, which are subsequently all combined to create the final mosaics.

\subsubsection{Standard Pipeline Calibration with calwf3}

The raw WFC3/IR exposures, still containing all the separate MULTIACCUM read samples, are calibrated using the Pyraf/STSDAS task \calwf%
	\footnote{Further documentation for all the PyRAF/STSDAS data reduction software is provided at http://stsdas.stsci.edu/}.
This task populates the bad pixel arrays using known bad pixel tables, followed by a bias subtraction for each read, based on the mean value of all the pixels in the 5-pixel wide reference pixel region. It then carries out a subtraction of the zeroth read in order to remove the bias structure across the detector, followed by a subtraction of the dark current reference files for either the SPARS50 or SPARS100 read-out sequences, as applicable. This is followed by the non-linearity correction and photometric keyword calculation, using the current filter throughput tables and detector quantum efficiency curves.

Once all the separate MULTIACCUM reads are calibrated as described above, they then move through the next steps in \calwf, namely up-the-ramp slope fitting and cosmic ray rejection. For each pixel, this step performs a linear fit to the accumulating counts that are sampled during each MULTIACCUM read, while rejecting outliers from the fit as being due to cosmic rays. Currently a threshold of 4 sigma is used for this rejection, and flags are populated in the data quality arrays corresponding to the read during which the cosmic ray occurred. A final count-rate value is then computed for each pixel using only the unflagged reads, and is stored as the count-rate in the final calibrated exposure, while the uncertainty in the slope of counts versus time is stored in the error extension of the image. The final steps of \calwf\ include multiplicative corrections for the detector gain and the flatfield structure across the detector, appropriate to each filter. Other factors affecting the fluxes include different pixel sizes due to detector distortion and are still present at this stage; these are corrected later when the geometric distortion is removed.

\subsubsection{Additional Corrections for Persistence, Warm Pixels and Flat Field Residuals}

In addition to the default calibrations, there are also a number of additional corrections carried out to further improve the WFC3/IR data, but which are not part of the standard \calwf\ pipeline. The CANDELS team has implemented these as additional steps in our automated image processing pipelines, and they are executed in conjunction with the standard \calwf\ calibration steps.

The first of these additional corrections concerns the presence of persistent flux in certain pixels due to bright sources having been observed in previous exposures, which can be a significant issue for the WFC3/IR detector.%
	\footnote{http://www.stsci.edu/hst/wfc3/ins\_performance/persistence/}
For the first few epochs of this program, darks from the WFC3 calibration program that execute just prior to CANDELS visits are used to aid in identifying and measuring problematic pixels. Pixels with persistent flux are then identified in these dark frames if they exceeded a count-rate threshold of five sigma above the mean, and are flagged in the following science exposures. For subsequent orbits during a visit, the CANDELS team can determine directly from the preceding CANDELS exposures which pixels may contain sufficient flux to cause persistence; the calibration darks are only used for the first orbit in a visit, when the previous data may be from another program and not necessarily accessible. During the initial calibrations for the CANDELS project, pixels flagged as having significant persistent flux are excluded when performing the mosaic combinations, but work is being done to see if the persistent flux can be modeled and subtracted instead, which would enable these pixels to be used.

Another correction that is implemented at this stage is the identification of additional ``warm'' pixels in the exposures, which might be fluctuating and therefore perhaps not present in the calibration reference dark files but only in the images. The calibration darks are used to identify these pixels if they exceed a threshold of five sigma above the mean, in which case they are flagged in the data quality arrays that are associated with each exposure, and are excluded from the final image combination.

Finally, the initial WFC3/IR data obtained for the CANDELS program were calibrated using an early generation of flatfield files in the STScI archive pipeline that did not fully correct for all the flatfield features present in the data. Specifically, these included the IR ``blobs'' that have appeared in the WFC3/IR channel since launch and were not present in the ground flats, as well as a $\sim\,$3$\,-\,$4\% residual large-scale variation in the overall structure of the flatfield files. We therefore created residual flatfield files which we applied in our pipelines to correct these effects after having run the data through the initial calibration. Both of these issues have subsequently been corrected by constructing an extensive set of sky flats for each of the different filters, and the resulting corrected flatfield reference files are now in the STScI archive pipeline and are used when the \calwf\ calibration software is run.

We have verified that the degree of flatness of the final sky backgrounds in the images is within $\sim\,$1$\,-\,$2\% of the mean sky level. In particular, after adoping the improved WFC3/IR flatfields that we have discussed here, we verified that the photometric repeatability across the detector was consistent to the same level of accuracy, with no significant deviations found beyond $\sim\,$1$\,-\,$2\%. We present a more detailed comparison in Section~\ref{sec:photom} between our WFC3/IR photometry and existing ground-based photometry, to illustrate the degree of consistency that is achieved.

The final, calibrated images for each WFC3/IR exposure are then used in the subsequent steps of astrometric alignment and \multidrizzle\ mosaic combination, as described in more detail in \S\ref{sec:multidrizzle}.

\begin{figure*}[t!]
\begin{center}
\ifsubmodeastroph
  \includegraphics[width=2in]{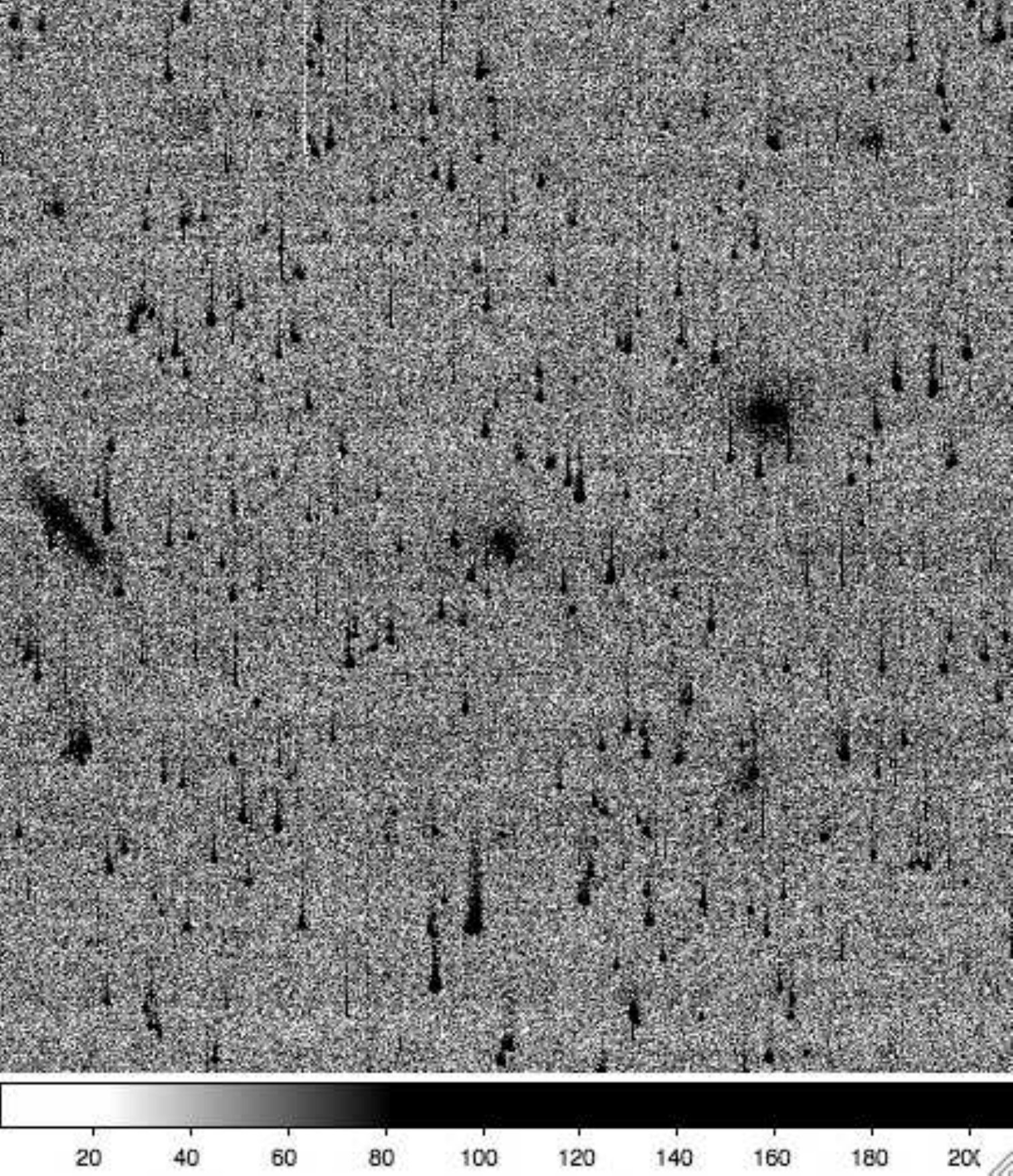}\hspace{0.2in}
  \includegraphics[width=2in]{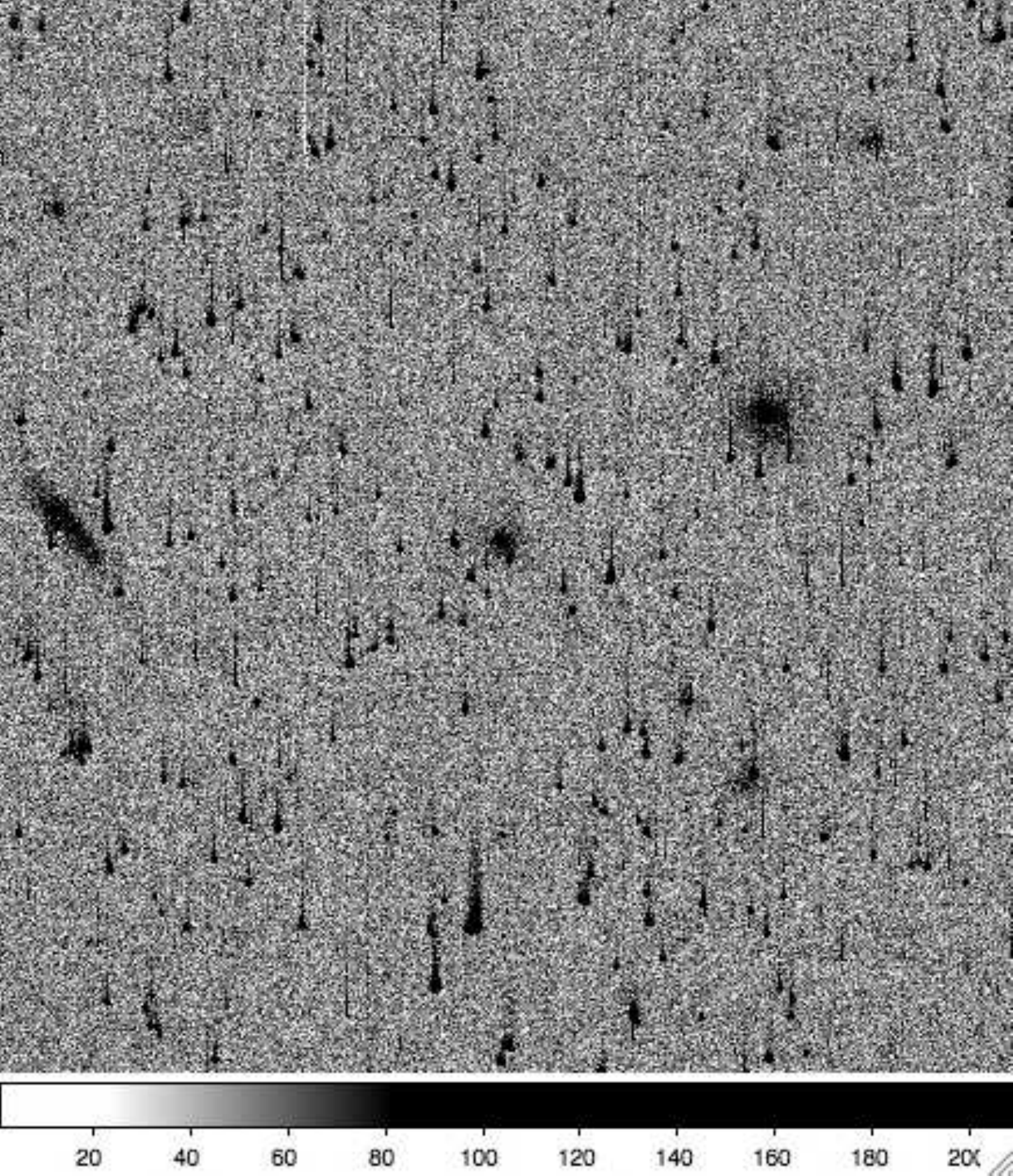}\hspace{0.2in}
  \includegraphics[width=2in]{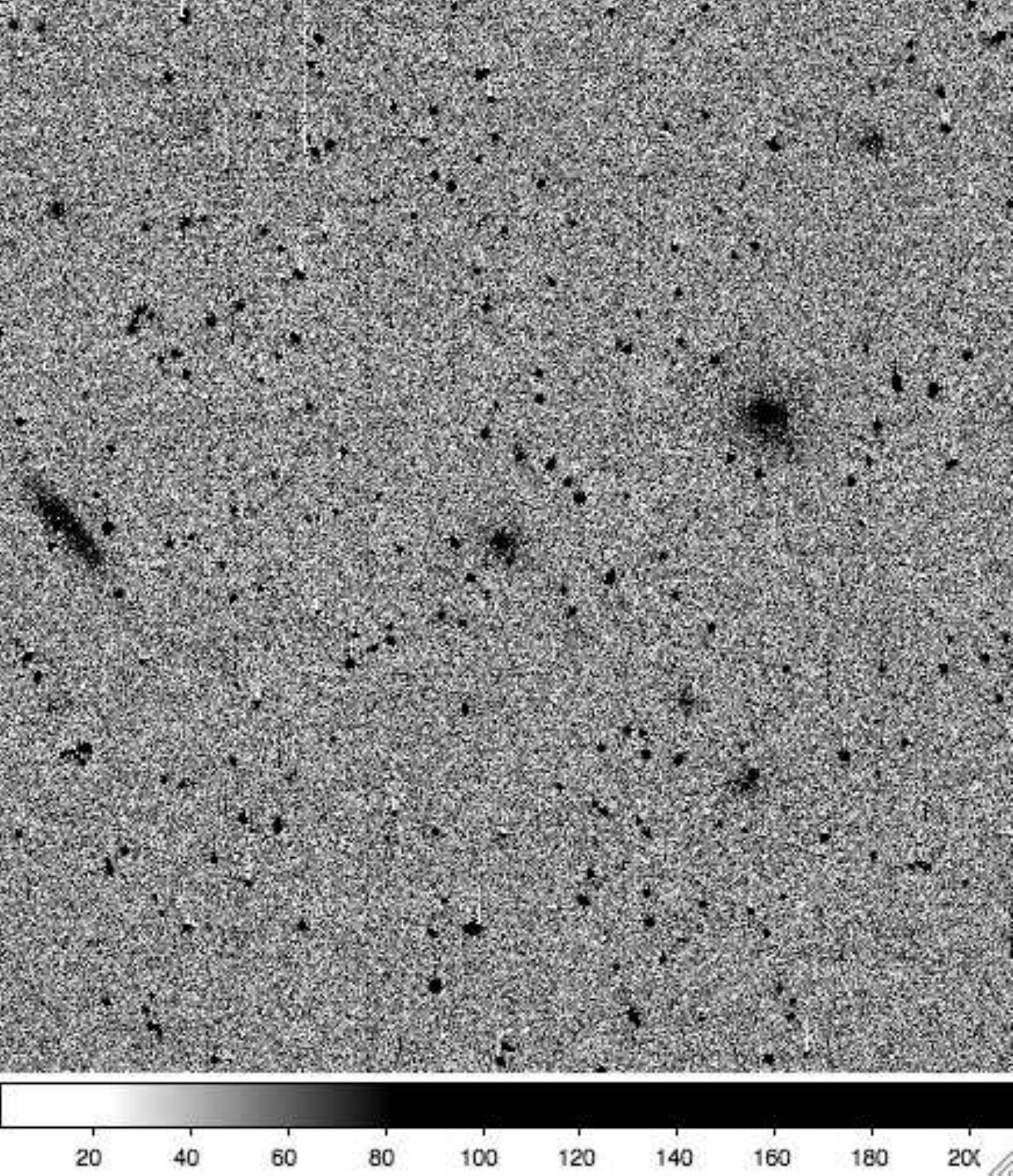}
\fi
\figcaption{\label{fig:bias_stripe_cte}%
Example showing the removal of bias stripe and CTE degradation in the ACS data: (left): original exposure after calibration with \calacs; (middle): same exposure after removal of the bias stripes
	\citep{2010.Grogin.bias};
(right): final dataset after subsequent correction for the CTE degradation
	\citep{2010PASP..122.1035A}.
}
\end{center}
\end{figure*}

\subsection{WFC3/UVIS and ACS/WFC Detector Calibration}

In this subsection the WFC3/UVIS and ACS/WFC observations are described together since both of these cameras have rather similar physical characteristics, with differences highlighted where necessary.

\subsubsection{Standard Pipeline Calibration with calwf3 and calacs}

Each of the raw ACS/WFC and WFC3/UVIS exposures are initially calibrated using the Pyraf/STSDAS tasks \calacs\ and \calwf\ respectively, for which the standard steps are quite similar, although additional steps are incorporated as described later in this section. For the standard calibration, in each case the CCD data first have an overall bias correction applied, which is measured from the physical and virtual overscan regions that are around the edges of the detector. This is done by carrying out a linear fit along the overscan rows and columns, to capture any potential slope across the rest of the detector that might be the result of a bias drift. Subsequently a bias reference file image is subtracted from each exposure to account for the remaining pixel-to-pixel bias structure in the detectors. In addition, the dark reference file is subtracted from each exposure to take into account the dark current structure across the detector and help mitigate warm pixels that may be present in the images. Finally, the multiplicative gain correction and flatfield structure reference files are applied, followed by photometric keyword calibration using the current filter throughput curves and detector sensitivity information, thereby resulting in a set of exposures for WFC3/UVIS and ACS/WFC that are calibrated according to the standard pipeline calibration.

\subsubsection{Additional Corrections for Bias Striping, CTE Degradation and Amplifier Quadrant Offsets}\label{sec:cte}

In addition to these default calibrations, a number of other corrections need to be carried out particularly on the ACS/WFC images, primarily related to the length of time that the detectors have been on orbit, and the changes in the detector read-out electronics as a result of the new CCD Electronics Box Replacement (CEB-R) that was installed during Servicing Mission 4 (SM4) to restore the instrument to operation. These additional steps, described below, are implemented in our automated CANDELS image processing pipeline and are executed in conjunction with the standard \calacs\ calibration steps, with some modifications.

The first of these additional effects is the removal of bias striping noise from the ACS/WFC exposures. This effect is introduced by the SIDECAR ASIC electronics (System Image, Digitizing, Enhancing, Controlling, and Retrieving / Application Specific Integrated Circuit, manufactured by Teledyne) which was part of the new CEB-R installed on ACS during SM4. This circuitry exhibits a low-frequency ($\sim\,$0.001$\,-\,$1~Hz) noise with a power spectrum similar to 1/f noise, which is introduced into one of the reference voltages for the ACS/WFC CCDs and subsequently manifests itself as a bias amplitude variation from one row to the next. However, this noise has the property of being relatively uniform across the detectors, with its amplitude distribution being Gaussian with $\sigma\,$=$\,$0.75$\,$e$^-$ (significantly less than the $\sim\,$3$\,-\,$4$\,$e$^-$ readnoise of these CCDs). The relative uniformity of the signal along each row enables this striping pattern to be characterized and largely removed using an algorithm
	\citep{2010.Grogin.bias}
which fits each CCD row independently, determining the background level using an iterative $\sigma\,$-clipping technique to reject non-sky pixels, followed by a hybrid mean and median estimator to compensate for lower-level signal from faint sources and cosmic ray hits. This algorithm has been implemented in our CANDELS image calibration pipeline in conjunction with the calibration stage and reduces this noise to $\lesssim\,$0.3$\,-\,$0.4$\,$e$^-$, thereby largely removing its impact on the data.

The second detector effect addressed in our pipelines is the impact of Charge Transfer Efficiency (CTE) degradation. This effect comes about because of the read-out scheme that is implemented for CCDs, whereby charge that has been detected by each pixel on the array is first transferred down all the remaining pixels in the same column, and subsequently across all the remaining columns to the amplifier where it is read out. As the charge packets are transferred from one pixel to the next, charge traps that are present in the pixels due to impurities and crystalline defects can capture some of the electrons, releasing them after a short period of time. This leads to a loss of flux in the original pixel and manifests itself as deferred-charge trails along the columns behind bright pixels in each exposure, while also producing a net astrometric shift up along in the column for bright sources. The effect becomes increasingly severe for pixels furthest from the amplifiers, which for these detectors are the pixels near the chip gaps. The charge trap population is created by cosmic ray bombardment and increases depending on the length of time for which a detector has been on orbit, which is now $\gtrsim\,$8$\,$years for the ACS/WFC. While WFC3/UVIS was installed more recently, indications are that its CTE is degrading more rapidly than ACS and it may also need to be corrected, though its level of CTE degradation is currently small and has not yet been characterized as fully as ACS, which we focus on here. Figure~\ref{fig:bias_stripe_cte} shows an example of the impact of the bias stripe and CTE degradation in the images.

We implement a correction for this CTE degradation in the CANDELS imaging pipelines, making use of a pixel-based algorithm that carries out a fit to the deferred-charge trail behind each bright pixel and redistributes the charge into the pixels in a way that is intended to represent the original flux as detected, before any CTE degradation
	(as implemented by
	\citealt{2010PASP..122.1035A}\note{Anderson \& Bedin 2010};
	see also
	\citealt{2010MNRAS.409L.109M}\note{Massey 2010}).
A key point about this algorithm is that it is effectively a deconvolution, by virtue of the fact that it restores the charge profiles of pixels along a particular column to their original shape, which is sharper and more concentrated than the observed profiles which have been smeared by the deferred charge trails. As such, the pixel-to-pixel noise in the final reconstructed image is also somewhat higher than in the original exposure. In addition, the algorithm includes the effect of read-out noise, which is introduced at the amplifier and does not participate in the deferred-charge smearing on the detector pixels, by treating this as a separate noise component from the noise on the detector. Tests to date have shown that this algorithm correctly reproduces the expected noise that would be present in the images if no CTE degradation had been present, and in addition restores both the photometry and the astrometric accuracy to levels that are comparable to images without CTE degradation. Therefore, this algorithm is included in our pipelines in conjunction with the standard detector imaging calibration steps. We will continue to review the performance of this correction as more data are collected.

Finally, the ACS/WFC detector amplifier quadrants exhibit additional bias-related offsets between one another that are not fully corrected during standard calibration. We implement a routine that fits for the differences between these, using an iterative clipping procedure to eliminate signal from astronomical sources and preserve only the background flux, which is then used to remove the residual amplifier quadrant differences and place all four quadrants on a uniform background level.

The resulting calibrated, flatfielded exposures for all the CCD observations are then used in the subsequent steps of alignment, cosmic ray rejection and image combination, as described in the next section.

\subsection{Relative Astrometry and Distortion}

Once all the individual WFC3/IR, WFC3/UVIS and ACS/WFC exposures have had their detector-level calibrations applied, they are next passed through the rest of the CANDELS calibration and mosaicing processing pipelines. The overall design of this pipeline is shown in Figure~\ref{fig:pipeline} and is described here in more detail. At the starting point in this process, the data from all three detectors,WFC3/IR, WFC3/UVIS and ACS/WFC have all been calibrated and are in a very uniform set of formats, effectively representing the incoming photons detected on each pixel. Therefore, from this stage onwards they are processed through very similar steps in the remainder of the pipeline that generate the higher-level products, and are all discussed together in this section, with occasional differences highlighted as appropriate. The first stage of this pipeline processes all the exposures in a given visit, for each of the different cameras, and addresses the relative shifts between exposures in each given single-orbit visit.

\begin{figure*}[t!]
\begin{center}
\ifsubmodeastroph
  \includegraphics[width=3in]{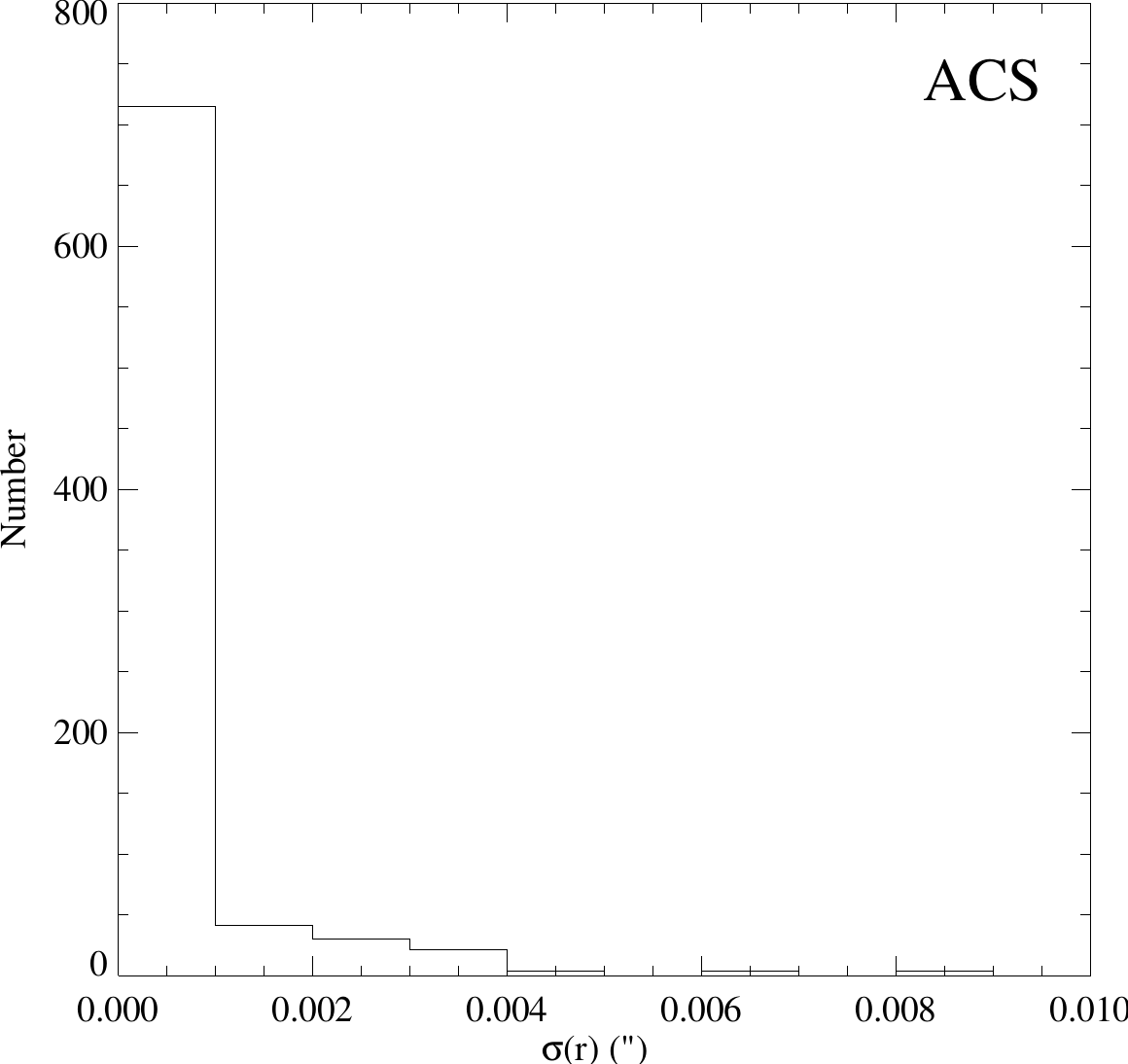} \hspace{0.5cm}
  \includegraphics[width=3in]{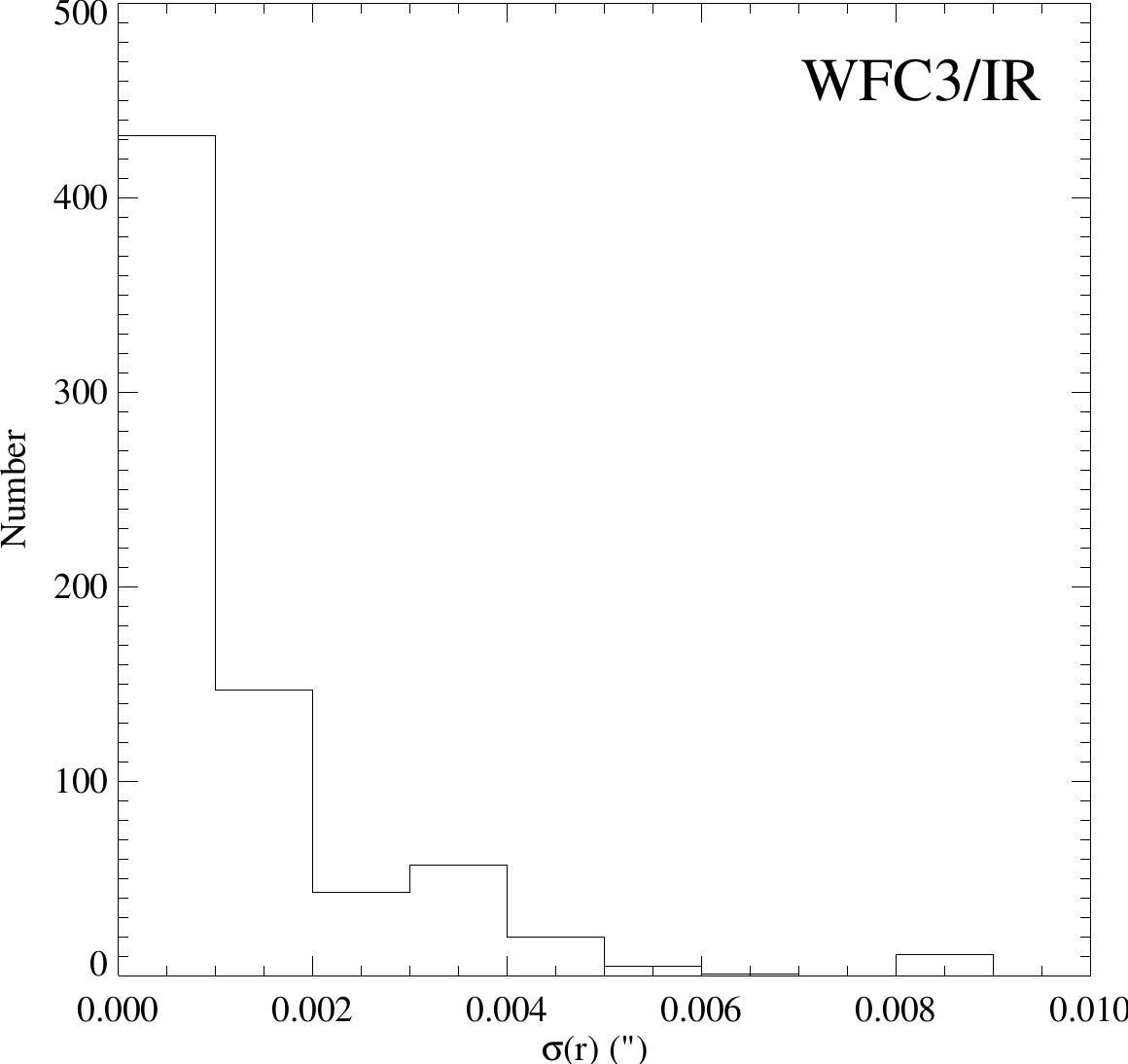}
\fi
\figcaption{\label{fig:crosscor}%
Distribution of uncertainties in the measured shifts obtained by cross-correlation, for all the exposures for a given instrument during a given orbit, for all the data obtained so far using ACS/WFC (left) and WFC3/IR (right). The measured uncertainties are typically $\sim\,$0.01$\,-\,$0.02~pixel, thus $\sim$0.5$\,-\,$1\,mas for ACS/WFC and $\sim$1$\,-\,$2\,mas for WFC3/IR due to its larger pixel size. This technique was used to correct for the shifts between exposures as a result of spacecraft pointing uncertainties in executing dither offsets (up to $\sim\,$5\,mas) as well as the more significant astrometric shifts introduced by filter changes ($\sim\,$10$\,-\,$25\,mas, depending on the instrument/filter) which are not accounted for by the distortion information and need to be solved for empirically.}
\end{center}
\end{figure*}

\subsubsection{Dithering and Pointing Uncertainties}

The four WFC3/IR exposures, together with their corresponding ACS parallels, are all obtained in a 4-point dither box pattern aimed at providing half-pixel subsampling for the WFC3/IR detectors along both axes of the pixel, together with a small integer pixel shift to ensure that bad pixels and other artifacts (e.g., persistence) are moved around. Due to the geometrical distortion of the detector, shifts that are too large correspond to a substantially different number of pixels along the edge than at the center where the shifts are defined, which would cause the pixel subsampling phase to change across the detector. Therefore the shifts are kept small enough to retain the intended half-pixel subsampling across much of the detector with no significant change in phase. This means, however, that large artifacts such as the WFC3/IR ``deathstar'', or the WFC3/UVIS and ACS/WFC detector chip gaps ($\sim\,$2$\,-\,$3$\arcsec$ across) are not covered by the dither pattern in a single orbit, therefore we rely on observations during subsequent epochs to cover these missing areas and provide coverage across the entire field. While the dithers are small, there is nonetheless a spacecraft positioning error associated with these small angle maneuvers, which is on the level $\sim\,$3$\,-\,$5$\,$mas (i.e., $\sim\,$0.05$\,-\,$0.1$\,$pixel for the WFC3/UVIS and ACS/WFC CCDs).

In addition to the slight positioning uncertainties introduced by the small angle maneuvers, an optical offset is introduced whenever a different filter is inserted into the optical path. This offset is currently not included in the geometric and astrometric information for the {\it HST} instruments, thereby leading to an apparent astrometric change which can be quite significant, on the order $\sim\,$0.2$\,-\,$0.3$\,$pixel depending on the filter and instrument. Moreover, during each orbit the spacecraft undergoes thermal expansion and contraction (``breathing'') due to changes in solar illumination, which lead to changes in the optical path length to the detectors, hence resulting in slight scale changes from one exposure to the next. This scale change is distinct from that which is produced by velocity aberration related to the orbital motion of the spacecraft around the earth (which is accurately known and can be corrected), and therefore needs to be treated separately. Finally, uncertainties in guide star reacquisition from one orbit to the next can lead to errors in position as well as small rotation uncertainties, while a full acquisition of a new guide star has astrometric uncertainties of $\sim\,$0$\farcs$3$\,-\,$0$\farcs$5 (reflecting the absolute astrometric uncertainties in the Guide Star Catalog~2, GSC-2:
	\citealt{2008AJ....136..735L}\note{Lasker \etal 2008}).

\subsubsection{Distortion Calibration Models}

In addition to the pointing and guidestar uncertainties related to the spacecraft, the astrometric accuracy also depends on the degree to which the detector geometric distortions are calibrated. For the ACS/WFC camera, the detector distortion is modeled using a 4th order polynomial which has been measured separately for each filter, including also a time-dependent evolution of the first-order linear skew terms
	\citep{2007.Anderson.ISR}	
which is independent of filter. In addition, the measured distortion on the detector deviates slightly from a polynomial description, and is captured in a ``distortion residual image'' which is combined with the polynomials when correcting for the distortion. For ACS/WFC, the global accuracy of the distortion solution is now within $\sim\,$0.02$\,-\,$0.03~pixel (Anderson 2007).

For the WFC3/IR and WFC3/UVIS detectors, the on-orbit distortion is less well characterized due to the relatively short time for which the instrument has been in operation. The latest distortion solutions have been delivered on 11 October 2010 for both cameras
	(see \citealt{2009.Kozhurina.ISR}	
for general reference) and contain empirical measurements of the distortion in all the filters used in the CANDELS program. In all cases the distortion is modeled as a 4th order polynomial, but there is not yet any evidence of possible skew term evolution, and there are also not yet any distortion residual images that are to be combined with the polynomials. The global accuracy of the present solutions is $\lesssim$0.1\,pixel in each camera, as verified also by our testing on the exposures. If these solutions improve in future, we will include them in subsequent reprocessing of the data.

\subsubsection{Cross-Correlation Shift Determination}

To solve for and remove the residual uncertainties in the spacecraft dither offsets between all the exposures in each orbit, a cross-correlation prcedure is applied to all the exposures for a given instrument in each orbit. The involves taking the first exposure as a reference, and cross-correlating all the subsequent exposures against that. Filters that are close in wavelength (e.g., F606W and F814W for ACS, or F125W and F160W for WFC3/IR) can be successfully cross-correlated because the strength of the cross-correlation signal from even a small sub-pixel shift is coherent across the image and is much stronger than any morphological differences in galaxies between adjacent filters. Thus, for example, if an orbit contains four WFC3/IR exposures (two F125W and F160W), then the first exposure would be used as a reference and the other three would be cross-correlated against it (since they are all obtained with the same instrument, even though the filters are different). Similarly, if the same orbit contains four parallel ACS exposures, then the first ACS exposure would be used as a reference and the other three would be cross-correlated against it; the ACS and WFC3 are treated completely independently since the detectors have quite different properties. Currently this technique is only applied within a single orbit since other astrometric errors such as orientation differences can be present between different orbits, or different visits, and these are solved for in a different part of the pipeline.

The cross-correlation procedure first passes all the exposures for each instrument (using all the filters) through a partial run of \multidrizzle, up to the point where single-drizzled images are produced for all the individual exposures. These images have had the instrument distortion removed, as well as having had the commanded spacecraft offsets applied, and are all aligned on the same pixel grid so that astronomical sources should be at the same pixel locations if no residual shifts were present. These images are then masked so that all empty or faint regions are set to zero, retaining only regions around objects that contain sufficient signal for the cross-correlation. Bright saturated sources, in particular stars with long charge-bleed columns, are also excluded since such columns degrade the quality of the cross-correlation solutions. The regions containing all the remaining objects are also tapered to avoid introducing artificial ringing in the cross-correlation step.

Each of these images is then cross-correlated relative to the first one, producing a Fourier transform image for each exposure. The Fourier transform image contains a strong peak, offset from its center by an amount that corresponds to the residual shift between the two exposures. The profile of this peak is fit using a two-dimensional fitting routine to determine its location and associated uncertainty, which is then directly translated into a shift between the two exposures. The uncertainties are typically less than a few hundredths of a pixel.

After a set of first-pass cross-correlation shifts has been obtained, these shifts are propagated back to the input exposures, in order to do a second-pass run of \multidrizzle\ which includes a cosmic-ray rejection step with the improved shifts. The pixels flagged as cosmic rays in each exposure then have their flux replaced with pixels from the clean drizzled image, then subsequently re-drizzled to a new set of single-drizzled images for each separate exposure, which are again masked and used as input for the second-pass cross-correlation step. This provides a much cleaner cross-correlation signal, since much of the noise in the first-pass cross-correlation is due to cosmic rays. Finally, the shifts from the second-pass cross-correlation are propagated into the image headers for a third-pass \multidrizzle, cosmic ray rejection and cross-correlation step, and are also stored as the final set of relative coordinates for all the exposures within a given orbit. At this stage the expected shifts are zero, hence any remaining shifts between the second and third cross-correlation iterations provide a good diagnostic of the residual uncertainties in the shift measurement.  

Figure~\ref{fig:crosscor} shows the distribution of the uncertainties in the measured shifts, for all CANDELS exposures obtained to date in WFC3/IR and ACS/WFC. These are derived from the uncertainties on the location of the cross-correlation peak, which are obtained during the fitting procedure. The uncertainties are typically $\sim\,$0.01$\,-\,$0.02~pixel, thus $\sim$0.5$\,-\,$1\,mas for ACS/WFC and $\sim$1$\,-\,$2\,mas for WFC3/IR due to its larger pixel size. Occasionally, the uncertainties are somewhat higher depending on the structure of the sources within a given exposure, or residual anomalies such as satellite trails or other defects that might be present within the sources used for the cross-correlation, but in all cases they are still less than 0\farcs01. This technique was used to correct for the shifts from one exposure to the next in each filter as a result of pointing uncertainties in executing dither offsets (up to $\sim\,$5\,mas) as well as the much larger astrometric shifts introduced by filter changes ($\sim\,$10$\,-\,$25\,mas, depending on the instrument/filter) which are not accounted for by the distortion information and need to be solved empirically. A comprehensive program of testing has been carried out to validate this routine, and its final implementation in the pipeline is able to correct the relative shift errors present between the exposures in each orbit to a level of accuracy better than a few milliarcseconds, thereby correcting the small errors introduced when the spacecraft executes small angle maneuvers for dither offsets, as well as correcting the offsets introduced by filter changes.

It should be noted that this level of accuracy refers specifically to the overall relative alignment from one exposure to the next during an orbit as determined by cross-correlation, and does not reflect the astrometric accuracy of individual sources across the images. These are still limited by the accuracy of the distortion models along with the absolute astrometric uncertainties, which are further discussed in \S\ref{sec:absolute_astrometry}.

\ifsubmodeapjs		\begin{deluxetable}{lllllll}\rotate	\fi
\ifsubmodeastroph	\begin{deluxetable*}{lllllll}[t!]	\fi
\tablecaption{\label{tab:astrom_catalogs}
	Astrometric Reference Catalogs Used for Each CANDELS field}
\tablehead{%
Field	& Area					& Telescope		& Filter	& Depth (5$\sigma$ AB)	& Resolution	& Reference}
\startdata
GOODS-N	& 34$\arcmin$$\,\times\,$27$\arcmin$	& Subaru/Suprime-Cam	& R		& 26.6		& 1$\farcs$1	& Capak et al. (2004)		\\
	&					& HST/ACS		& F850LP	& 27.4		& 0$\farcs$08	& Giavalisco et al.(2004)	\vspace{6pt}	\\
GOODS-S	& 34$\arcmin$$\,\times\,$33$\arcmin$	& ESO/2.2m WFI		& R		& 25.5		& 0$\farcs$8	& Arnouts et al.(2001)		\\
	&					& HST/ACS		& F850LP	& 27.4		& 0$\farcs$08	& Giavalisco et al.(2004)	\vspace{6pt}	\\
COSMOS	& 2$\arcdeg$$\,\times\,$2$\arcdeg$	& Subaru/Suprime-Cam	& $i^+$		& 26.2		& 0$\farcs$9	& Capak et al. (2007)		\\
	&					& HST/ACS		& F814W		& 27.2		& 0$\farcs$08	& Koekemoer et al. (2007)	\vspace{6pt}	\\
EGS	& 1$\arcdeg$$\,\times\,$1$\arcdeg$	& CFHT/12k		& R		& 24.7		& 0$\farcs$9	& Coil et al. (2004)		\\
	&					& HST/ACS		& F814W		& 27.2		& 0$\farcs$08	& Davis et al. (2007)		\vspace{6pt}	\\
UDS	& 1$\arcdeg$$\,\times\,$1$\arcdeg$	& UKIRT/WFCAM		& K		& 24		& 0$\farcs$8	& Lawrence et al. (2007)
\vspace{3pt}
\enddata
\ifsubmodeapjs		\end{deluxetable}	\fi
\ifsubmodeastroph	\end{deluxetable*}	\fi

\subsection{Cosmic Ray Rejection}

With the relative shifts within each orbit having been corrected, the next step consists of creating a cosmic ray mask for all the exposures of a given filter, for each camera, during a given orbit, by carrying out another run of \multidrizzle, this time with the improved relative shifts. The cosmic rays are identified in the driz\_cr step of \multidrizzle\ using a process that first creates a series of separately drizzled images, one for each input exposure, which are subsequently used to create a median image using the ``minmed'' algorithm in \multidrizzle, which enables the minimum to be used instead of the median in cases where valid pixels from only two or three exposures are present, if one of them exceeds the others by $>\,$4$\sigma$. The clean median image is then transformed back to the distorted detector frame of each input exposure to carry out cosmic ray rejection using the following approach. The input counts in a given pixel in the original exposure, $I_{\rm exp}$ are compared with the counts from the median image $I_{\rm med}$ for the same pixel, together with the derivative of the median image $\Delta_{\rm med}$, defined as the steepest gradient from that pixel to its surrounding pixels (with all these quantities being in units of electrons). A pixel is flagged as a cosmic ray if it exceeds a threshold defined as follows:
	\begin{equation*}
	| I_{\rm exp} - I_{\rm med} |  >  S \Delta_{\rm med} + {\it SNR}  \sqrt{\sigma_{\rm read}^2 + | I_{\rm med} + B |}
	\end{equation*}
where $S$ and {\it SNR} are adjustable scaling factors and $B$ is the background sky value that has been measured for the exposure. The inclusion of the gradient term $\Delta_{\rm med}$ effectively ``softens'' the cosmic ray rejection in regions of relatively steep gradients such as bright cores of objects, where the pixel-to-pixel variation can exceed simple Poissonian statistics. For the data processing in the CANDELS pipelines, this rejection is performed over two iterations, with the first pass going through all the pixels in the image and using $S$$\,=\,$1.2 and {\it SNR}$\,=\,$3.5, followed by a second pass in a 1-pixel wide region around each of the pixels flagged in the first pass, but using more stringent criteria of $S$$\,=\,$0.7 and {\it SNR}$\,=\,$3.0. This ensures that fainter pixels around cosmic rays are also flagged.

In addition, the ACS/WFC exposures are significantly impacted by CTE degradation, which introduces substantial deferred-charge trails extending along columns away from cosmic ray hits. Although this has already been accounted for to some extent by means of the CTE algorithm described in Section~\ref{sec:cte}, some residual effects can remain in the images near the locations of bright cosmic rays. These are mitigated in \multidrizzle\ by a subsequent rejection iteration whereby the cosmic ray masks are convolved with a linear kernel in the direction opposite to the read-out direction, and with a length of 30~pixels, which is about the worst-case length of the CTE trails. The rejection is then carried out in this additional region, using the same parameters as were used in the second-pass iteration for pixels around the originally flagged ones.

For the WFC3/IR images, most of the cosmic rays are already rejected during the up-the-ramp sampling. However, there are occasional cosmic rays that are not fully removed, or warm pixels that are not accounted for in the dark file correction, so the WFC3 exposures for each filter are also passed through this step. For the ACS/WFC data, there are typically between two and four exposures per filter in a given orbit. Given the typical cosmic ray rate of $\sim\,$1$\,-\,$2\% during our exposure times, this means that for a 4-exposure depth, $\lesssim\,$1$\,-\,$2$\,$pixels can be expected to be hit by cosmic rays during all four exposure, while for two exposures this number increases to $\sim\,$2000$\,-\,$6000 pixels would be affected by cosmic rays during both exposures. However, this is still only $\sim\,$0.01$\,-\,$0.04\%, meaning that $\sim\,$1 out of every 100 small galaxies ($\sim\,$100$\,$pixels in area) would be affected, losing  $\sim\,$1$\,-\,$4\% of its pixels on average.

\subsection{Absolute Astrometric Calibration}\label{sec:absolute_astrometry}

The absolute astrometric accuracy of any {\it HST} observation is limited by the astrometric uncertainty of the primary guidestar, which for the GSC-2 is $\sim\,$0$\farcs$3$\,-\,$0$\farcs$5. In addition to a shift, these uncertainties also introduce errors in the knowledge of the orientation of the telescope, which is derived from the guidestar positions as well. All these uncertainties can be problematic for a program like CANDELS, where observations from many different visits and epochs, taken with potentially different guidestars, need to be combined into a single set of images, and where accuracy to levels better than $\sim\,$0.1$\,$pixel are demanded. We address this using a 2-stage technique: firstly, obtaining the best possible astrometric solution for all the visits relative to one another by using a deep external ground-based catalog of the field which provides sufficient source density to match enough objects on the {\it HST} images, and secondly by ensuring that the entire ground-based catalog / mosaic reference frame is registered to an absolute external astrometric reference coordinate system.

For each of the five different CANDELS fields, appropriate external astrometric reference catalogs are available, which we list in Table~\ref{tab:astrom_catalogs} and describe here briefly. For GOODS-North, the astrometry is based on the Subaru/Suprime-Cam R-band imaging
	\citep{2004AJ....127..180C}\note{Capak et al. 2004},
whose absolute astrometric reference frame has been registered to SDSS, 2MASS, and the new deep VLA 20~cm survey of the field
	\citep{2010ApJS..188..178M}\note{Morrison et al. 2010}.
For GOODS-South, we use an R-band mosaic from the ESO-MPI 2.2m / Wide Field Imager (WFI)  obtained as part of the ESO Imaging Survey (EIS) of the field
	\citep{2001A&A...379..740A}\note{Arnouts et al. 2001},
registered to the GSC-2
	\citep{2008AJ....136..735L}\note{Lasker \etal 2008}.
For the COSMOS field, the catalog used is based on CFHT/Megacam $i^*$ imaging, supplemented by deeper Subaru/Suprime-Cam $I^+$ imaging (both described in
	\citealt{2007ApJS..172...99C}\note{Capak et al. 2007}),
with absolute astrometry determined by registering to the VLA 20~cm survey of the COSMOS field
	\citep{2004AJ....128.1974S}\note{Schinnerer et al. 2004}.
For the EGS field, we use the Deep2 CFHT/12k mosaic imaging of the field
	\citep{2004ApJ...609..525C,2007ApJ...660L...1D}\note{Coil et al. 2004; Davis et al. 2007},
initially registered to the USNO-A2.0 catalog and subsequently improved by matching to the VLA 20~cm survey of this field
	\citep{2007ApJ...660L..77I}\note{Ivison et al. 2007}.
The absolute astrometric accuracy is generally $\lesssim\,$0$\farcs$1 for each of these reference catalogs, limited primarily by residual uncertainties in the underlying reference frame. In addition, since extensive {\it HST} imaging is available for all these fields, which has already been registered onto the astrometric systems provided by these catalogs and are much deeper with better resolution, we use these {\it HST} data in determining the astrometry for the new CANDELS data, and only use the ground-based catalogs in regions that are not covered by the existing {\it HST} data. Only the fifth field, UKIDSS/UDS, has insufficient existing {\it HST} coverage. Therefore our fundamental astrometric frame for this is the UKIDSS K-band catalog
	(Almaini et al., priv. comm.;
	\citealt{2007MNRAS.379.1599L}\note{Lawrence et al. 2007}),
which has been obtained by imaging the field with UKIRT/WFCAM and is registered to the VLA 20~cm survey of the field
	\citep{2006MNRAS.372..741S}\note{Simpson et al. 2006}
to a similar level of astrometric accuracy as the other surveys.

\begin{figure*}[h]
\begin{center}
\ifsubmodeastroph
  \includegraphics[width=3in]{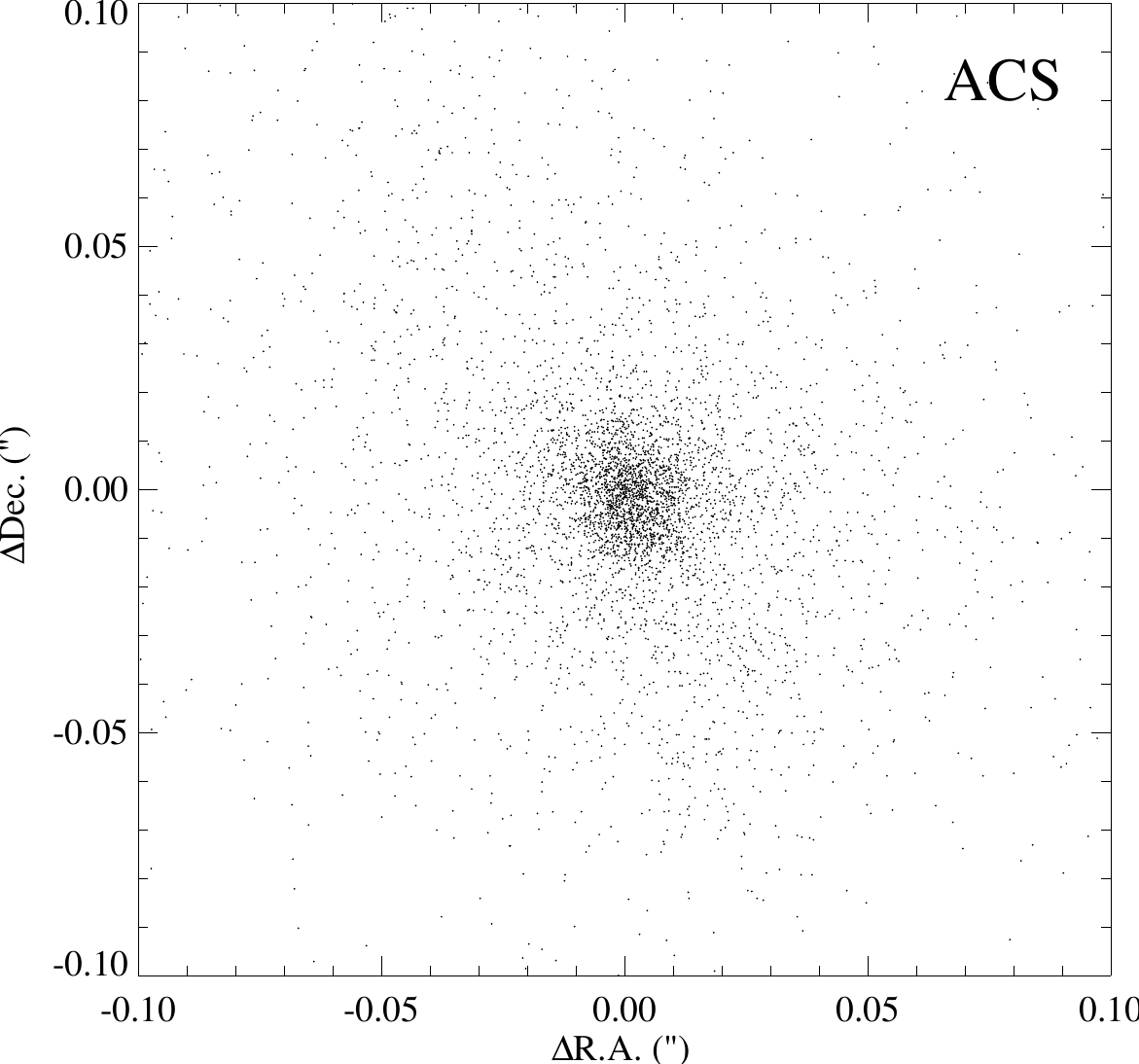} \hspace{0.25in}
  \includegraphics[width=3in]{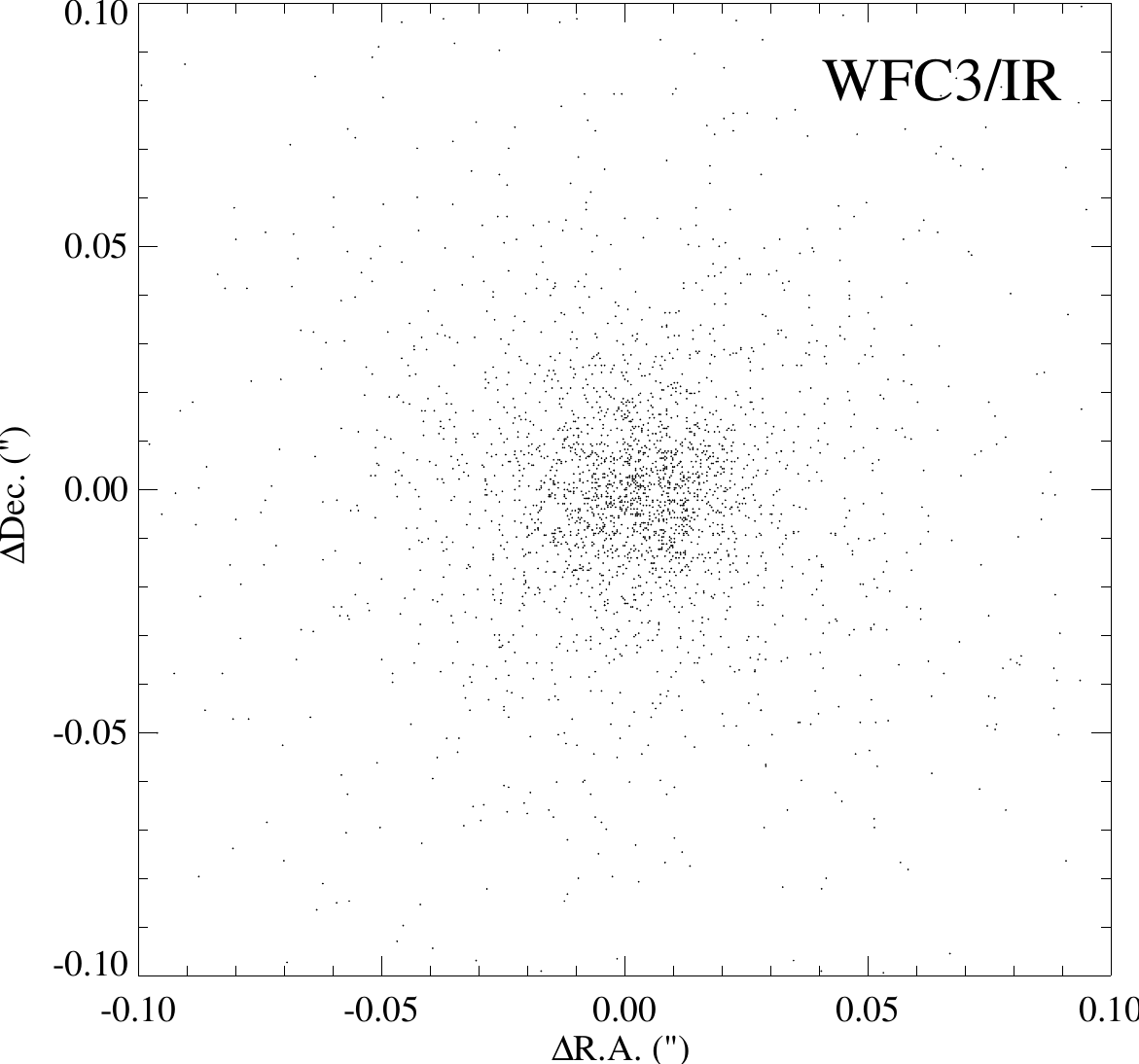}
\fi
\figcaption{\label{fig:astrom}%
Results showing the residual uncertainties in the positions of all the sources in the first epoch of GOODS-South as an example, where the plotted values indicate the difference in position for each source as measured from the new CANDELS data, compared with the position of the same source in the reference GOODS astrometric catalog, for the ACS/WFC images (left) and WFC3/IR images (right). For WFC3/IR the larger uncertainties are predominantly related to the broader PSF and larger pixel size, relative to ACS. The new CANDELS ACS data that overlap the previous GOODS-ACS data generally have smaller residuals than the WFC3/IR data. However, a subset of the new CANDELS ACS data lie outside the existing GOODS-ACS region, where we had to register the images to the ground-based WFI R-band data, and these sources have larger residuals (up to $\sim\,$0\farcs1 per source). Given $\sim\,$300$\,-\,$400 sources per tile, with the measurement errors on each source being drawn from this distribution, the overall accuracy on the absolute astrometric alignment of a given tile is therefore $\lesssim$0$\farcs$005, once the uncertainties on the positions of all the individual objects have been combined in quadrature. Therefore, for both instruments, the resulting absolute astrometry for each orbit, after combining the astrometric information from all the sources, is accurate to better than $\lesssim\,$0.1\,pixel, sufficient to enable robust combination of data in overlapping regions as well as permitting cosmic ray rejection across multiple epochs once the images have been placed onto this common astrometric grid.}
\end{center}
\begin{center}
\ifsubmodeastroph
  \includegraphics[width=3in]{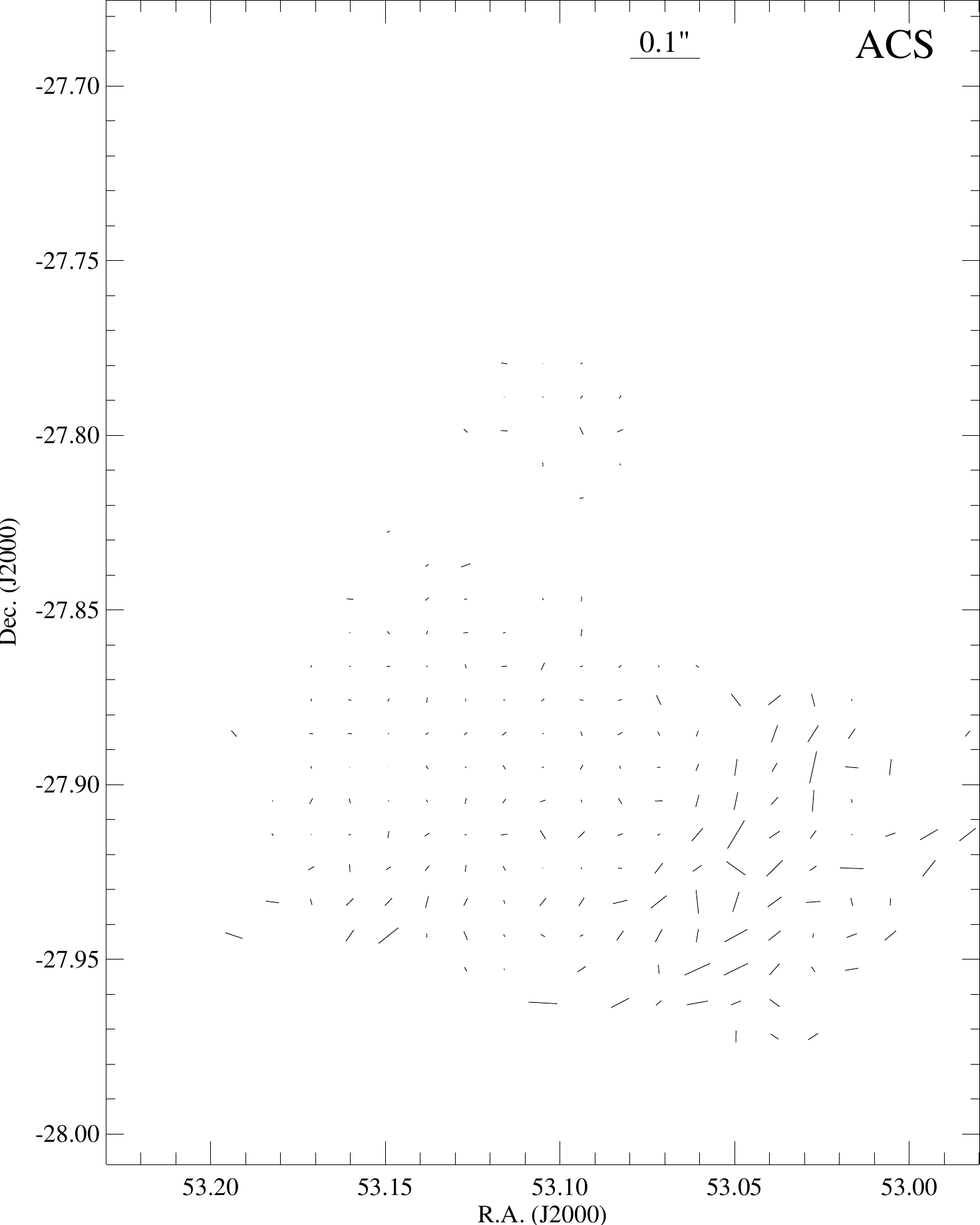} \hspace{0.25in}
  \includegraphics[width=3in]{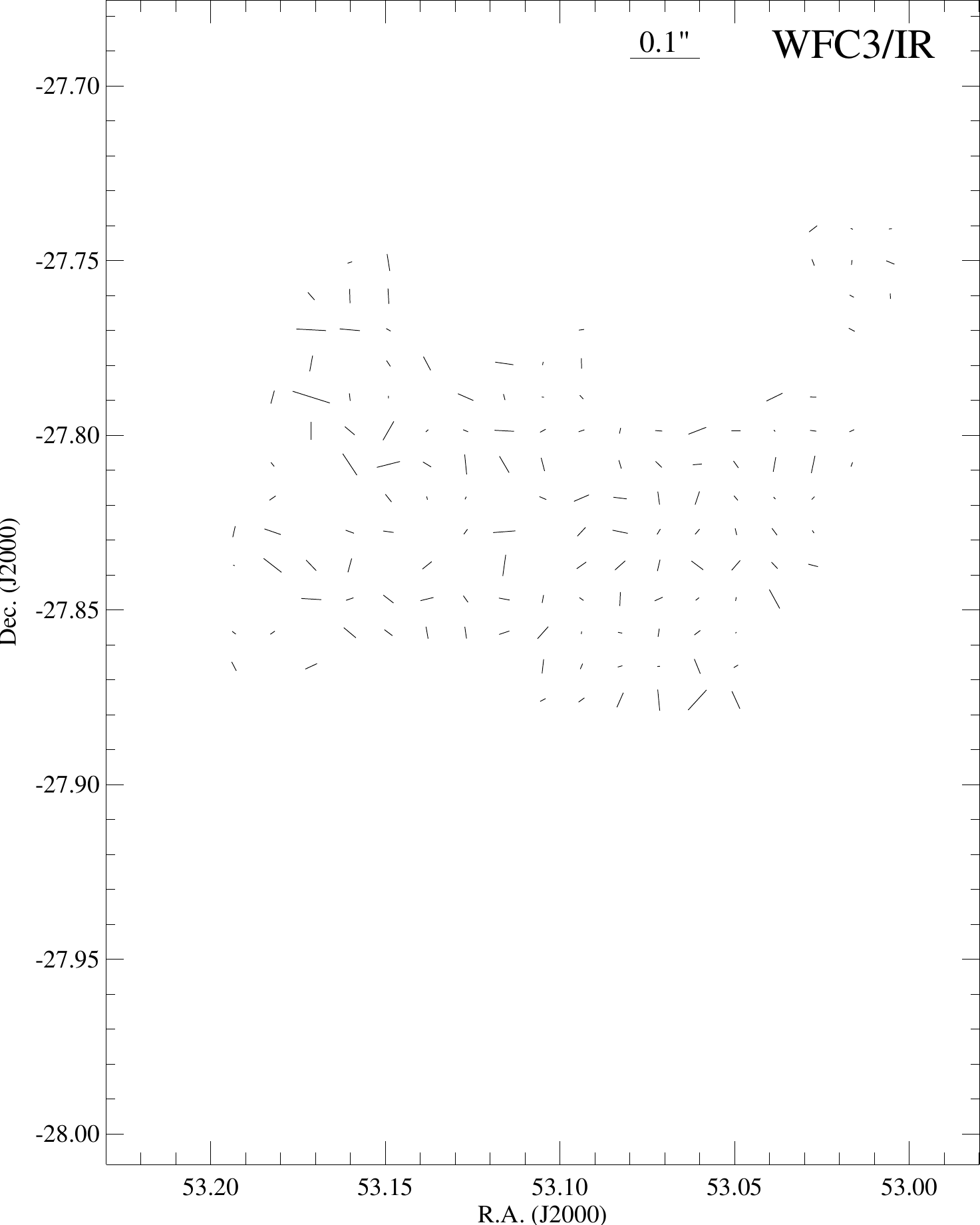}
\fi
\figcaption{\label{fig:vector}%
Binned vector residual plots showing the distribution of astrometric uncertainties across the GOODS-S field, once the images have all been registered onto the GOODS astrometric grid, again for the first epoch of GOODS-South, showing the ACS/WFC data (left) and WFC3/IR (right). The binned cells are 40\arcsec\ in extent, where the size is chosen for display purposes to ensure a sufficient number of objects per cell while also providing a sufficient number of cells across the mosaic to show the general structure of the residuals. The two cameras are offset from one another since the ACS/WFC data were obtained in parallel to the WFC3/IR data. Note also the sharp change in accuracy for the ACS data toward the south-west. This area lies outside the GOODS ACS v2.0 catalog, where we had to register the images to the ground-based WFI R-band data. However, even in this region the accuracy remains reliable to $\lesssim\,$0\farcs1 for each individual source, which means that when an overall shift is computed for these exposures using the full sample of $\sim\,$300$\,-\,$400 sources per exposure, the final shift is accurate to $\lesssim\,$0\farcs005$\,-\,$0\farcs01 once the measurements from all the different sources are combined, thereby enabling robust alignment of each exposure relative to those from different orbits.}
\end{center}
\end{figure*}

For all the exposures of a given instrument in each orbit, an ``astrometric detection image'' is then produced by doing another run of \multidrizzle\ to produce a single combined image (separately for each different instrument), for the tile covered during that orbit, containing all the exposures of all the filters for that instrument, and applying the cosmic ray masks that have been produced when combining the filters separately. This is motivated by the fact that the relative shifts for all the exposures, for all filters in a given orbit, have now already been determined to an accuracy of $\sim\,$0.5$\,-\,$1$\,$mas. Therefore nothing further would be gained by now attempting to solve for shifts separately for different filters in a given orbit. Therefore, the exposures for all the filters are combined into a single image (separately for each different instrument), yielding one image for WFC3/IR, another for WFC3/UVIS, and another for ACS/WFC, for each orbit. A catalog is then produced from this single multi-filter image, which also has the advantage of providing increased depth and reducing the impact from cosmic rays compared with the individual filter images.

All the sources in the multi-filter catalog for each orbit are then matched to the sources in the relevant portion of the external catalog, using a number of iterative steps. The first iteration uses a relatively large tolerance (up to a few arcseconds) and only the brightest $\sim\,$20$\,-\,$30 sources in each image, in order to determine the dominant terms in the shifts for right ascension and declination. Once these have been accounted for, several additional iterations are carried out using the full catalog of sources in each image, using progressively tighter matching tolerances down to 0$\farcs$1 and solving for the residual remaining shifts as well as the rotation errors due to the uncertainties in guidestar position. For all visits, $\sim\,$300$\,-\,$400 sources are typically then matched at the faintest levels and tightest tolerances between the {\it HST} \multidrizzle-combined image and the reference catalog.

The results of this procedure are shown in Figures~\ref{fig:astrom} and~\ref{fig:vector}, for the ACS/WFC and WFC3/IR images of the GOODS-S field. Figure~\ref{fig:astrom} shows the distribution in right ascension and declination offsets between the reference catalog positions of the sources and those measured on the new CANDELS data, after having solved for the astrometry as described above, for all sources in the first epoch of the GOODS-South field by way of example. In general, the offset between the new measured position and the catalog position is $\lesssim\,$0\farcs1 per object, where these residual differences are due to differences in sensitivity between the new and the old data, as well as morphological differences due to different filters, and astrometric uncertainties in the reference catalog. Since the ACS/WFC and WFC3/IR images for each orbit typically contain at least $\sim\,$300$\,-\,$400 sources, each with a residual $\lesssim\,$0\farcs1, this means that the overall absolute astrometric solutions for the exposures in each orbit are known to better than $\lesssim\,$0\farcs005 (i.e., $\sim\,$0.1\,pixel for ACS), once the uncertainties on the positions of all the individual objects have been combined in quadrature.

Figure~\ref{fig:vector} is based on the same data as in Figure~\ref{fig:astrom}, but this time showing the residuals as a function of position across the field, where each vector indicates the mean residual in a grid of cells, each 40\arcsec on a side, where the size is chosen for display purposes to ensure a sufficient number of objects per cell while also providing a sufficient number of cells across the mosaic to show the general structure of the residuals. The residuals are generally below $\sim],$0\farcs05 for most sources across the GOODS-South region, except for the CANDELS fields in the south-western corner which lie outside the previous GOODS-ACS coverage. For those fields, the only available reference catalog positions are those from the ground-based 2.2m ESO WFI R-band catalog previously discussed, and therefore the residuals per object are somewhat larger in that region. However, even for those tiles, the overall astrometric solutions for each orbit better than $\sim\,$0.1\,pixel for ACS once all the objects in all the exposures are taken into account, thereby enabling these exposures to be accurately aligned with the rest of the data.

We are continuing to investigate further improvements to these technique as we accumulate more data on the CANDELS fields, including the possibility of solving simultaneously for the positions of all sources, on all exposures, for all filters and all pointings, using the ground-based catalog as an external constraint. This may have the additional advantage of solving for the small residual scale changes from one exposure to the next, as well as any possible changes in rotation during an orbit, and perhaps even long-term changes in the skew terms of the detector which might not yet be captured in the distortion models. However, the current approach yields an absolute astrometric accuracy on the order of $\sim\,$5$\,$mas, or $\sim\,$0.1~pixel for ACS/WFC, between images obtained with different guidestars. This is sufficient to prevent registration problems between different {\it HST} observations of the same object in overlapping images from different visits or epochs and thereby preserve the morphological properties of galaxies in the survey.

\subsection{Satellite Trails, Optical Ghosts, and Crosstalk}

In addition to cosmic rays, several other artifacts are also present in the images and are best identified visually and masked by hand. These include trails from bright satellites, optical filter ghosts from bright stars, and anomalous persistence signals that might not have been identified in other ways. To this end, a data quality validation team has been assembled from across the CANDELS collaboration, who access all the individual exposures and \multidrizzle-combined images by means of a set of webpages that enable them to visually inspect graphical displays of the images, as well as to retrieve the images and submit data quality input results that are subsequently collected and used to identify and correct anomalous images.

Due to the relatively low altitude of the orbit of {\it HST}, several percent of exposures are affected by the passage of satellites across the field of view during the exposure. In addition, the WFC3/IR detector can exhibit persistence at the location of the satellite trail for up to several additional orbits if the satellite is bright enough. All exposures that are affected by this are identified visually, and a software script is then run to mask the satellite trail in the undistorted image and subsequently transform the mask back to the distorted frame of the input exposure. These masks are then included along with the cosmic ray masks in identifying pixels to be excluded from the combined mosaics for each individual epoch.

Optical ghosts from stars are also visually identified and masked on the affected exposures. In this case the masks are not used in creating the combined image for each individual epoch, since the dither pattern used is sufficiently small that the stellar ghosts do not move by much, which would then result in large holes in the resulting image. Instead, these masks are used when combining multiple epochs, since the locations of optical ghosts in one epoch are generally covered by pixels from another epoch that are not affected by ghosts.

Finally, an electronic effect that is present particularly in CCDs, and to a lesser extent in the WFC3/IR detector, is electronic crosstalk from bright sources, producing a region of somewhat lower flux that is located symmetrically in another quadrant. For ACS/WFC this effect is substantially mitigated by our use of GAIN=2 for all exposures and is not present at any significant level in the CANDELS data. Specifically, the inter-CCD cross-talk is negligible at GAIN=2, while the intra-CCD cross-talk of $\sim\,$0.007\% is corrected by the current version of the bias de-striping code. For WFC3/UVIS, however, it is quite noticeable particularly for stars that are bright enough to have a charge-bleeding column of bright pixels, leading to a corresponding low-valued section of columns on the opposite quadrant. The brightest galaxies in these fields also introduce this effect, with the affected region being more diffuse in that case. For WFC3/IR the effect is observed for only the very brightest stars and consists simply of a region with a relatively minor decrease in flux corresponding approximately to the area of saturation of the star. In our CANDELS imaging pipelines, we are able to flag the worst of the WFC3/UVIS crosstalk by identifying bleeding charge-trail columns in bright sources and then replicating a pixel mask for the corresponding area on the opposite side of the detector. These pixels are then excluded when creating the combined images for each individual epoch, since the dither offsets are large enough to ensure that such regions are generally sufficiently covered by other pixels that are not affected by this issue.

\subsection{Residual Background Subtraction}

Prior to combining all the exposures into a final mosaic, it is necessary to remove all background emission due to non-astronomical sources, for example the sky background from earthshine, which is relevant to all three cameras, as well as zodiacal light and low-level thermal background emission, which are more applicable to the WFC3/IR camera. Because these vary with time, their contribution to the pixel counts would lead to photometric errors in the final count-rates of sources if their relative differences between one exposure and the next are not removed. Moreover, the early versions of the WFC3/IR flatfields contained significant low-level large-scale residuals, $\sim\,$3$\,-\,$4\%, and we first embarked on an effort to improve these by constructing skyflats based on empty regions in all the CANDELS data that we had to date. However, new skyflats have now been released by the WFC3 team that provide improved flatfield corrections to $\sim$1\% for WFC3/IR, and after we incorporated these into the CANDELS products we found that they significantly reduced the large-scale structure across the images, as well as improving the consistency of photometric calibration across the detector.

In addition, some of the most severe instances of sky background emission can come from cases where the telescope is observing close to the limb of the bright earth, with backgrounds increasing significantly for bright earth limb angles below $\sim\,$35$\,-\,$40$\arcdeg$, and in some cases being non-uniform across the detector. For the WFC3/UVIS and ACS/WFC exposures not much can be done to mitigate this in any given exposure, since each CCD exposure is just a single integration. For WFC3/IR, however, we have been investigating the possibility of excluding specific reads near the beginning or end of the MULTIACCUM sequence, since the reads are sampled at either 50s or 100s and often only a few of them are affected by bright earth limb emission near the beginning or end of the orbit, while the remainder of the reads are obtained at higher limb angles with a more nominal background. Thus excluding the high-background ones can potentially improve the S/N (since the loss in S/N from excluding some reads can be compensated for by the increased S/N due to a lower background in the remaining ones). The exposures obtained so far generally do not show a sufficient degree of improvement to warrant this, and therefore we are currently retaining all the reads, but this technique remains available as an option for any future observations that may potentially be significantly affected by unusually bright sky backgrounds at low earth limb angles.

For all exposures, once all the low-level residual structure has been removed, the mean background level is then determined from a masked version of the exposure, which is constructed as follows. Firstly, the clean \multidrizzle-combined image for each visit is run through a source-detection step to create a mask of all sources in the images, which is done by smoothing the image and applying a sigma-clipping threshold for pixels that exceed the mean background level. This is necessary because many sources display faint isophotes that reach into the noise of the image and are not excluded when doing a simple sigma-clip on the image, thereby contributing to the pixel statistics at faint levels and biasing the resulting sky estimate. Smoothing the image effectively increases the significance of these faint outer isophotes and thereby enables these pixels to be flagged to much fainter levels than in the unsmoothed image, thereby reducing their impact on the final sky estimates. Typically a total of $\sim\,$20$\,-\,$30\% of all the pixels in the images are excluded in this way, leading to a significant reduction in the impact of these pixels on the final pixel statistics and a much more stable background sky estimate.

The statistics of the remaining pixels are still slightly biased toward positive values, which is simply a consequence of the much larger number of faint sources below the detection thresholds of the images, whose combined signal adds a slight net positive bias to the pixels. To these are added other faint positive signals such as faint tails from cosmic rays and deferred-charge trails from CTE which in both cases may be below the detection thresholds for the algorithms that are responsible for identifying and masking these. Ultimately, however, these residual effects are approximately uniform from one exposure to the next and represent a net DC component, and all that changes between exposures is the varying sky, zodiacal and thermal background. Therefore, the background level on all the remaining unmasked pixels on each image ($\sim\,$70$\,-\,$80\% of the total number of pixels) is determined by doing an iterative sigma-clipped fit to these pixels and determining its mean, which is then used as the final background offset value and subtracted from each individual exposure to yield an image containing only the counts from the sources.

\ifsubmodeapjs		\begin{deluxetable}{llllcccc}\rotate	\fi
\ifsubmodeastroph	\begin{deluxetable*}{llllcccc}[t!]	\fi
\tablecaption{\label{tab:wcs}
	World Coordinate System Information for each CANDELS {\it HST} Mosaic}
\tablehead{%
Field	& \multicolumn{2}{c}{Mosaic Tangent Point}		& Instrument/Camera	& Pixel Scale	& \multicolumn{2}{c}{Mosaic size}	& Reference Pixel	\\
	&  R.A.($\arcdeg$ J2000)&  Dec.($\arcdeg$ J2000)		&		&		& x (pix)	& y(pix)		& (x,y)			}
\startdata
GOODS-N	& 189.228621		& $+$62.238572			& WFC3/IR		& 0$\farcs$06	& 18600		& 18600			& (9900.5, 9600.5)	\\
	&			&				& WFC3/UVIS, ACS/WFC	& 0$\farcs$03	& 37200		& 37200			& (19800.5, 19200.5)	\vspace{6pt}	\\
GOODS-S	& \hspace{4pt}53.122751	& $-$27.805089			& WFC3/IR		& 0$\farcs$06	& 18600		& 18600			& (9900.5, 9600.5)	\\
	&			&				& WFC3/UVIS, ACS/WFC	& 0$\farcs$03	& 37200		& 37200			& (19800.5, 19200.5)	\vspace{6pt}	\\
COSMOS	& 150.116321  		& $+$2.2009731	 		& WFC3/IR		& 0$\farcs$06	& 12800		& 30720			& (6400.5, 12000.5)	\\
	&			&				& WFC3/UVIS, ACS/WFC	& 0$\farcs$03	& 25600		& 61440			& (12800.5, 24000.5)	\vspace{6pt}	\\
EGS	& 214.825000 		& $+$52.825000			& WFC3/IR		& 0$\farcs$06	& 40800		& 12600			& (29640.5, 7020.5)	\\
	&			&				& WFC3/UVIS, ACS/WFC	& 0$\farcs$03	& 81600		& 25200			& (59280.5, 14040.5)	\vspace{6pt}	\\
UDS	& \hspace{4pt}34.406250 & $-$5.2000000			& WFC3/IR		& 0$\farcs$06	& 30720		& 12800			& (12000.5, 6400.5)	\\
	&			&				& WFC3/UVIS, ACS/WFC	& 0$\farcs$03	& 61440		& 25600			& (24000.5, 12800.5)	
\vspace{3pt}
\enddata
\ifsubmodeapjs		\end{deluxetable}	\fi
\ifsubmodeastroph	\end{deluxetable*}	\fi

\subsection{Final \multidrizzle\ Mosaic Combination}\label{sec:multidrizzle}

\subsubsection{Inverse Variance Images}

In preparation for the final step of combining all the exposures for each filter into a single mosaic using \multidrizzle, our pipelines first create for each exposure a corresponding inverse variance image, which contains all the ``intrinsic'' error terms associated with each pixel (including noise from accumulated dark current, detector read-out, and photon noise from the background as modulated multiplicatively by the flatfield and the detector gain), but not the additional ``extrinsic'' Poisson noise from the astronomical sources in the image. This approach was described by
	\cite{2000AJ....120.2747C}	
for the Hubble Deep Field South and has also been used in
	GOODS \citep{2004ApJ...600L..93G},	
	COSMOS \citep{2007ApJS..172....1S,2007ApJS..172..196K},	
	AEGIS (Davis et al. 2007; Newman et al. 2011, in prep.),
	UDF \citep{2006AJ....132.1729B},	
and other projects, and is implemented as a routine option in \multidrizzle\
	(see \citealt{2002hstc.conf..337K}).	
We provide here a brief description of it as applied to the new instruments that are used to observe the CANDELS fields, but refer the reader to the above references for further details.

\begin{figure*}[t]
\begin{center}
\ifsubmodeastroph
  \includegraphics[width=5in]{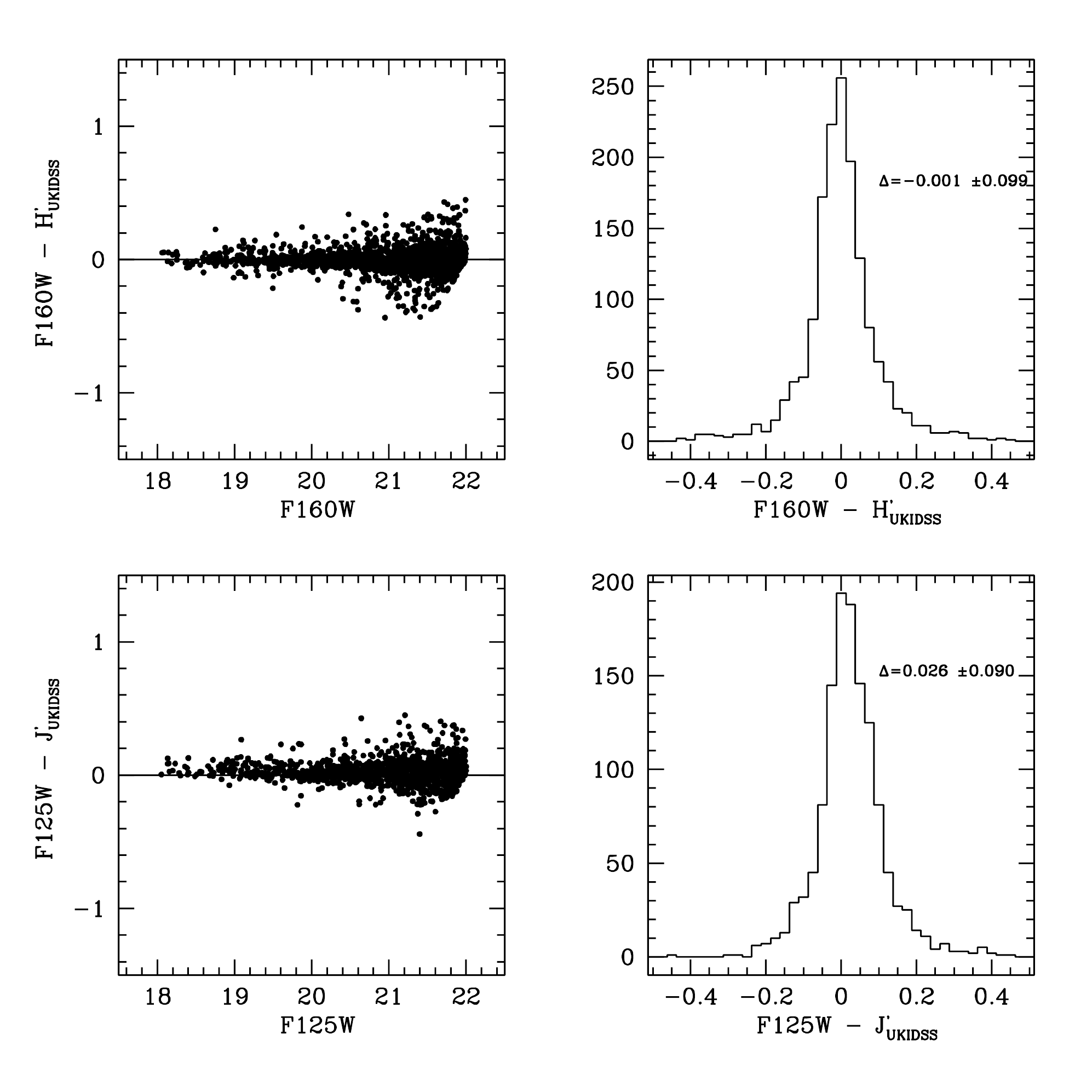}
\fi
\figcaption{\label{fig:photom}%
Photometric comparison of magnitudes measured from the CANDELS WFC3/IR imaging of the UKIDSS/UDS field, relative to the photometry of the same objects from the publicly released UKIDSS J and H data. Since the filter bandpasses are somewhat different, the UDS photometry has been transformed to the WFC3/IR system using stellar spectral libraries
	\citep{1998PASP..110..863P,		
		1983ApJS...52..121G},		
thereby enabling a direct comparison of any remaining possible offsets. The agreement is generally robust to better than a few percent, in agreement with the current published photometric performance of WFC3.}
\end{center}
\end{figure*}

When combining images where each pixel value has noise associated with it, the optimal approach involves weighting each pixel according to the inverse square of its noise. In particular, it can be demonstrated that the appropriate quantity to use consists of the background noise only; it should not include the noise associated with the emission from the object since that would lead to a biased estimation of the true flux in the pixel and result in photometric errors. Our CANDELS pipeline constructs inverse variance images for all the exposures obtained with WFC3/IR, WFC3/UVIS and ACS/WFC, using essentially the same formalism as described in Casertano et al. (2000) for the original WFPC2 Hubble Deep Field South observations, but modified to take into account the differences in the data formats for the ACS and WFC3 detectors. In particular, these detectors are calibrated to electrons per second and have been corrected for the detector gain, while the WFPC2 data are in DN and therefore still need to have the gain included in the calculation. Specifically, the WFC3/IR images are in electrons per second while WFC3/UVIS and ACS/WFC are in units of total electrons; however, the input inverse variance images for all of them are calculated to be in units of counts per second, since the ACS images are scaled to be in electrons/second while drizzling and since that is also what the output units of the mosaics are in all cases. The formula that is used to calculate the inverse variance images, in units of (e/s)$^-2$, is given by:
	\begin{equation*}
	{\it Inverse\,Var.} = \frac {(f\,t)^2} {(D+ fB) + \sigma_{\rm read}^2}
	\end{equation*}
where $f$ is the inverse flatfield (as defined in the conventions for the flatfield reference files used in calibration), $t$ is the exposure time (in seconds), $D$ is the total accumulated dark current signal during the exposure, $B$ is the total accumulated background level during the exposure, and $\sigma_{\rm read}$ is the read-out noise, with all three of the latter quantities being in units of electrons.

\subsubsection{Pixel Scale and Final \drizzle\ Parameters}

The pixel scale for the output mosaics is driven by the detector plate scale and pixel size, together with the full width half max (FWHM) of the point spread function (PSF) produced by the telescope optics. At the wavelengths of the WFC3/IR F105W to F160W observations, the {\it HST} PSF has a FWHM $\sim\,$0$\farcs$12$\,-\,$0$\farcs$18, which is subsequently convolved by the 0$\farcs$128 WFC3/IR detector pixel scale. Hence, the best PSF that could be recovered (without deconvolution), even in the ideal scenario of combining images using interlacing, which would minimize additional convolutions, still has a FWHM $\sim\,$0$\farcs$17$\,-\,$0$\farcs$19 in the final images. We choose an output pixel scale of 0$\farcs$06/pixel for the final WFC3/IR mosaics, providing adequate sampling of the PSF.

Having determined an output pixel scale facilitates the choice of the other relevant \drizzle\ parameter, namely \pixfrac, which defines how much the input pixels are reduced in linear size before being mapped onto the output grid (see the basic description of \drizzle\ parameters in
	\citealt{2002hstc.conf..337K}).	
The initial PSF produced by the {\it HST} optics is first convolved by the detector pixel size when imaging the sky, then a second time by the rescaled detector pixel size when mapping onto the output grid (thus applying the \pixfrac\ parameter), then is convolved a final time by the output pixel scale. The impact of the second convolution can thus be minimized by setting \pixfrac\ to a sufficiently small value; setting it to 0 would remove this convolution, corresponding to pure interlacing (as was done for the UDF as described in
	Beckwith et al. 2006,
where up to 144 exposures were available), but for a program like CANDELS with a limited number of exposures, we set \pixfrac\,=\,0.8 in order to avoid introducing too much variation between pixels in the corresponding weight images.

For the ACS/WFC and WFC3/UVIS exposures, the PSF is considerably sharper due to both the shorter wavelengths and the smaller pixel scale on the detector; thus we choose an output mosaic pixel scale of 0$\farcs$03/pixel for these (which has become the standard for GOODS, UDF, and most other large {\it HST} ACS surveys), and also set \pixfrac\,=\,0.8 based on similar considerations as with the WFC3/IR data. The resulting PSF, after taking into account the relevant convolutions, is $\sim\,$0$\farcs$07$\,-\,$0$\farcs$11 in the final images, across the range of UV/optical filters that we are using and is very well sampled by our chosen 0$\farcs$03/pixel scale.

The final pass of \multidrizzle\ is then run using the above output pixel scale and \pixfrac\ settings and applying the inverse variance weight image associated with each exposure, which contains the full set of masks from cosmic rays, bad pixels, satellite trails, and other blemishes in the detector. In each case the images for all filters are drizzled onto a common tangent plane projection on the sky, which is defined to match the existing ones where known, to facilitate a direct comparison with pre-existing data on these fields. Four of the fields (GOODS-North, GOODS-South, COSMOS, and EGS) all have an existing tangent plane point already defined, which we adopt for the CANDELS mosaics as well. For the UDS field, we adopt a common tangent plane point designed to satisfy the current surveys on that field. The World Coordinate System (WCS) properties of all the CANDELS mosaics are shown in Table~\ref{tab:wcs}.

\begin{figure*}[h]
\begin{center}
\ifsubmodeastroph
  \includegraphics[width=5in]{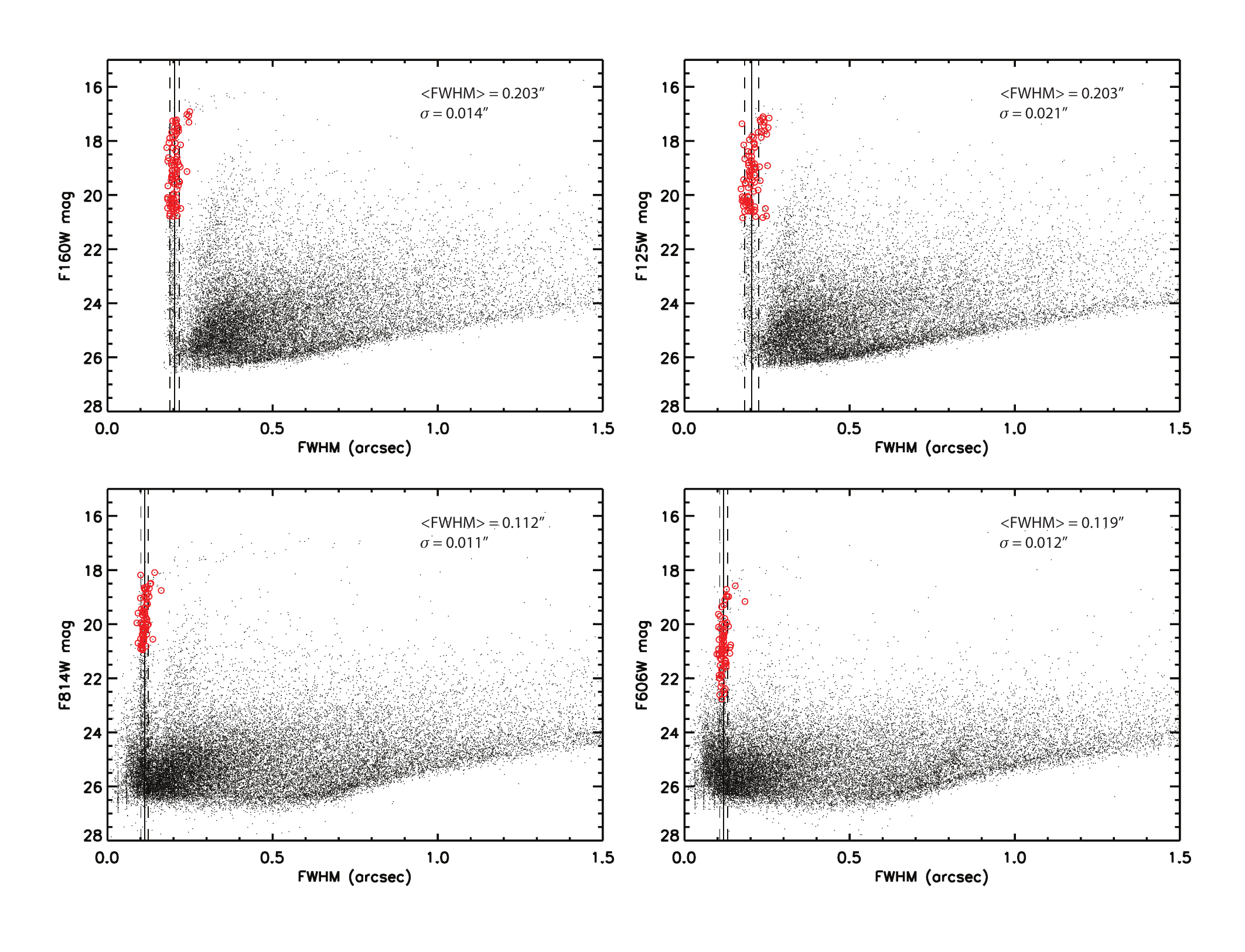}
\fi
\figcaption{\label{fig:psf_fwhm_mag}%
Plot of measured FWHM values for all the sources in the CANDELS UDS field, for all four filters (ACS/WFC F606W and F814W, and WFC3/IR F125W and F160W). Stars that have been manually inspected and verified to be valid, isolated point sources with no significant defects are identified in red. The mean FWHM for these stars is shown as a solid line, with dashed lines indicatring the 1-sigma standard deviation boundaries in each case.}
\end{center}
\begin{center}
\ifsubmodeastroph
  \includegraphics[width=5in]{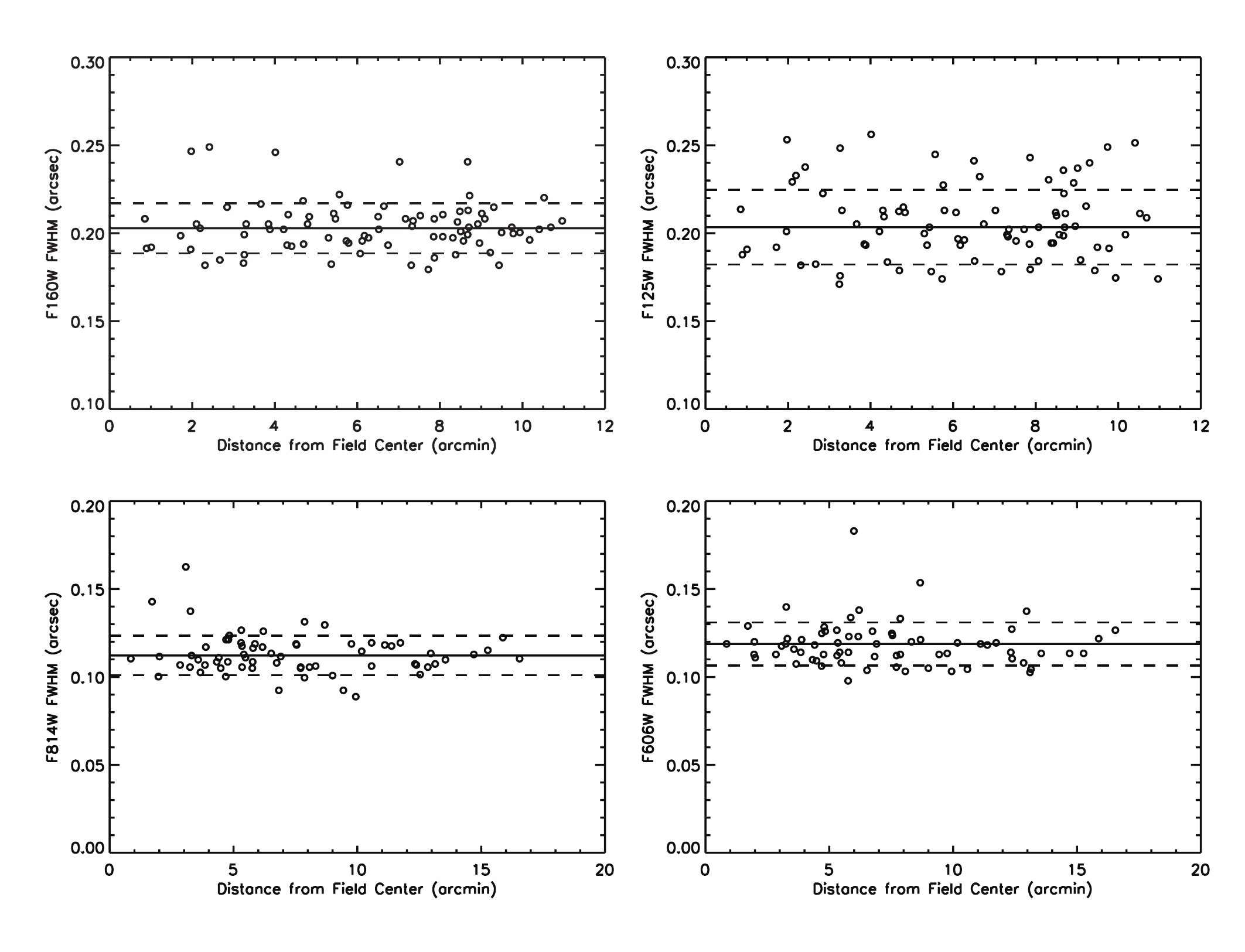}
\fi
\figcaption{\label{fig:psf_fwhm_pos}%
Plot of measured FWHM values for the subset of manually verified unresolved stars in CANDELS UDS field, for all four filters (ACS/WFC F606W and F814W, and WFC3/IR F125W and F160W), as a function of position across the field. The plots indicate the mean FWHM with a solid line and the standard deviation boundaries using dashed lines. Generally the stars are well behaved across the field, with no significant trends evident as a function of field position. We note that the somewhat higher scatter of the WFC3/IR points is due primarily to the increased importance of pixellation due to its larger detector pixel size, thus the ACS measurements are our primary diagnostic and these do not reveal any significant trends as a function of position across the field.}
\end{center}
\end{figure*}

Figures~\ref{fig:gsd01_wfc3} to \ref{fig:uds_full_acs} show the final combined mosaics obtained to date for the GOODS-S and UDS fields, including the WFC3/IR and UVIS observations, which are all obtained as primary exposures. Also shown are the ACS/WFC exposures, which are obtained in parallel and are therefore offset from the WFC3 pointings. In each case we show a color mosaic representing the various filters obtained in each given instrument (either WFC3/IR or WFC3/UVIS+ACS), together with the inverse variance weight images which show the total extent of the coverage and degree of overlap between the WFC3/IR pointings as well as the WFC3/UVIS and ACS/WFC images. The WFC3/IR exposures successfully adjoin one another with minimal overlap while maximizing the area, as designed. However, the larger ACS/WFC exposures overlap each other significantly, given that the offsets from one pointing to another are governed by the size of the WFC3/IR detector. The ACS/WFC exposures are therefore effectively doubled at most locations across the field.

Three months after the observations of a given CANDELS epoch are completed, we release the calibrated mosaics to the public via the STScI archive\footnote{http://archive.stsci.edu/prepds/candels/}, including the drizzled science mosaics as well as the inverse variance weight files that describe the noise associated with each pixel. Updates on the data obtained will be provided at the primary CANDELS project website\footnote{http://candels.ucolick.org/} as the survey progresses. Subsequent papers will present details on the catalogs and other measurements derived from the CANDELS data.

\subsection{Photometric Validation}\label{sec:photom}

We have carried out a series of photometric tests between the HST and ground-based imaging in fields for which both types of datasets exist, in order to quantify the level of agreement between photometry derived from the two types of data, with emphasis on the new WFC3/IR imaging. In particular we examined the UKIDSS/UDS field since this is the first CANDELS field to have been completed, and extensive ground-based imaging exists in the near-IR which we can compare with our new data. The total WFC3/IR system throughput in the F125W and F160W filters, including the filter transmission curves, the detector sensitivity, and the optical telescope assembly response, is somewhat different from the ground-based system, therefore we carried out a rigorous conversion between the two bandpasses using the following procedure.

The first step involved identifying all the unresolved objects in the UDS field that were covered by CANDELS F125W and F160W as well as UKIDSS J and H, also ensuring that the objects were unsaturated in all cases and in a magnitude range that is well covered by the UDS data. The UKIDSS magnitudes were then converted to the WFC3/IR system using a set of stellar spectral libraries
	\citep{1998PASP..110..863P,		
		1983ApJS...52..121G}		
to obtain the required filter conversions, using detailed existing information about the filters and total system throughput in each case, following a methodology similar to previous multi-band catalog studies that have involved observations from several different instruments
	(e.g., \citealt{2006A&A...449..951G,	
			2010A&A...511A..20C}).	
We plot the resulting comparison in Figure~\ref{fig:photom}. The agreement is generally tight and shows no significant systematics, with the zeropoints agreeing to better than a few percent, which is also in good agreement with what has been reported for the performance of WFC3
	\citep{2010SPIE.7731E..27M}\footnote{See\,also\,http://www.stsci.edu/hst/wfc3}.
We have also verified that initial checks on the data obtained so far for the other partially completed CANDELS fields yield similar results.

\subsection{Point-Spread Function Validation}\label{sec:psf}

The quality of the PSF is important to verify, both in terms of the scientific utility of the data (including morphological measurements and signal-to-noise limits for compact or unresolved sources) as well as the quality of the data (providing a good diagnostic of the alignment accuracy between exposures and the cosmic ray rejection quality for compact and unresolved sources).

Within the CANDELS team, we have compiled an extensive list of unresolved sources in each pointing obtained to date (verified by manual inspection), and we use these here to diagnose the PSF quality of the UKIDSS/UDS dataset, since this is the first field to be completed and has the most extensive CANDELS  WFC3/IR and ACS/WFC imaging. In Figure~\ref{fig:psf_fwhm_mag} we show the measured PSF of all the sources in this field, for all four filters (ACS/WFC F606W and F814W, and WFC3/IR F125W and F160W), and identifying in red the stars that have been manually verified as being valid point sources that are not saturated and are clearly isolated.

Generally the unresolved sources are very well behaved, until reaching $\sim\,$18th magnitude where saturation becomes an issue. We note that the ACS FWHM are slightly broader than might be expected, because all data in this particular dataset have been drizzled to a 0$\farcs$06/pixel scale, primarily to facilitate cross-image cataloging and related work. However we have verified that images drizzled to a finer scale (0$\farcs$03/pixel) recover the $\sim\,$0$\farcs$08$\,-\,$0$\farcs$09 FWHM generally achieved for ACS imaging. The WFC3/IR data display a broader FWHM primarily as a result of the broader HST PSF in the IR as well as the larger $\sim\,$0$\farcs$13 detector pixel scale, which results in more dither-related pixellation due to the relatively small number of dither offset positions used.

We also examine the measured PSF for this subset of unresolved sources as a function of position across the field, and show the result in Figure~\ref{fig:psf_fwhm_pos}. This test is useful for identifying regions of the image where misalignment or cosmic ray rejection problems might cause a net upward shift of the FWHM by a significant amount. However, it is evident that across the field the PSF values remain well behaved, especially for ACS which is more sensitive to CR rejection issues resulting from misalignment, and that no significant systematics are present across the field.

\section{Summary}\label{sec:summary}

We have described the {\it HST} observational imaging data products and processing pipelines for the Cosmic Assembly Near-infrared Deep Extragalactic Legacy Survey (CANDELS), a 902-orbit {\it HST} Multi-Cycle Treasury program aimed at documenting the first third of galactic evolution from redshift $z \sim 8$ to 1.5 via deep imaging of more than 250,000 galaxies with the {\it Hubble Space Telescope}, together with new constraints on the use of SNe~Ia as tracers of dark energy to higher redshifts than previously studied. The survey covers five well-studied extragalactic fields, namely GOODS-North, GOODS-South, COSMOS, EGS and UKIDSS/UDS, targeting each one predominantly with WFC3/IR as prime and ACS/WFC in parallel, together with some WFC3/UVIS observations where necessary for the science goals. The data are all recalibrated and combined using our own pipelines as described here. Three months after the observations of a given CANDELS epoch are completed, we release the calibrated mosaics to the public via the STScI archive. At the time of writing, we have already released the first few GOODS-South epochs, and the full UDS campaign. We strongly encourage the astronomical community to make use of the CANDELS data to advance their own research.

\section{Acknowledgements}

\acknowledgments

We would like to thank our Program Coordinators, Tricia Royle and Beth Perriello, along with the rest of the Hubble planning team, for their efforts to schedule this challenging program. The WFC3 team has made substantial contributions to the program by calibrating and characterizing the instrument and have provided much useful advice. Rychard Bouwens provided helpful input on the observing strategy for the CANDELS/Deep survey. John Mackenty suggested using 2$\times$2 on-chip binning for the UV observations, which will significantly improve the signal-to-noise ratio of those observations. The CANDELS observations would not have been possible without the contributions of hundreds of other individuals to the Hubble missions and the development and installation of new instruments. Support for program number HST-GO-12060 is provided by NASA through a grant from the Space Telescope Science Institute, which is operated by the Association of Universities for Research in Astronomy, Incorporated, under NASA contract NAS5-26555.

Facilities: \facility{HST(ACS,WFC3)}



\bibliographystyle{apj}
\bibliography{apjmnemonic,manuscript}

\begin{thebibliography}{}

\bibitem[\protect\citeauthoryear{{Aird} et~al.}{{Aird}
  et~al.}{2008}]{2008MNRAS.387..883A}
{Aird}, J., {Nandra}, K., {Georgakakis}, A., {Laird}, E.~S., {Steidel}, C.~C.,
  \& {Sharon}, C. 2008, \mnras, 387, 883

\bibitem[\protect\citeauthoryear{{Anderson}}{{Anderson}}{2007}]{2007.Anderson.%
ISR}
{Anderson}, J. 2007, ISR-ACS-2007-08

\bibitem[\protect\citeauthoryear{{Anderson} \& {Bedin}}{{Anderson} \&
  {Bedin}}{2010}]{2010PASP..122.1035A}
{Anderson}, J.,  \& {Bedin}, L.~R. 2010, \pasp, 122, 1035

\bibitem[\protect\citeauthoryear{{Arnouts} et~al.}{{Arnouts}
  et~al.}{2001}]{2001A&A...379..740A}
{Arnouts}, S., et~al. 2001, \aap, 379, 740

\bibitem[\protect\citeauthoryear{{Babul} \& {Rees}}{{Babul} \&
  {Rees}}{1992}]{1992MNRAS.255..346B}
{Babul}, A.,  \& {Rees}, M.~J. 1992, \mnras, 255, 346

\bibitem[\protect\citeauthoryear{{Beckwith} et~al.}{{Beckwith}
  et~al.}{2006}]{2006AJ....132.1729B}
{Beckwith}, S.~V.~W., et~al. 2006, \aj, 132, 1729

\bibitem[\protect\citeauthoryear{{Bournaud}, {Elmegreen}, \&
  {Elmegreen}}{{Bournaud} et~al.}{2007}]{2007ApJ...670..237B}
{Bournaud}, F., {Elmegreen}, B.~G.,  \& {Elmegreen}, D.~M. 2007, \apj, 670, 237

\bibitem[\protect\citeauthoryear{{Bouwens} et~al.}{{Bouwens}
  et~al.}{2007}]{2007ApJ...670..928B}
{Bouwens}, R.~J., {Illingworth}, G.~D., {Franx}, M.,  \& {Ford}, H. 2007, \apj,
  670, 928

\bibitem[\protect\citeauthoryear{{Bouwens} et~al.}{{Bouwens}
  et~al.}{2008}]{2008ApJ...686..230B}
{Bouwens}, R.~J., {Illingworth}, G.~D., {Franx}, M.,  \& {Ford}, H. 2008, \apj,
  686, 230

\bibitem[\protect\citeauthoryear{{Bouwens} et~al.}{{Bouwens}
  et~al.}{2010}]{2010ApJ...708L..69B}
{Bouwens}, R.~J., et~al. 2010, \apjl, 708, L69

\bibitem[\protect\citeauthoryear{{Brusa} et~al.}{{Brusa}
  et~al.}{2009}]{2009A&A...507.1277B}
{Brusa}, M., et~al. 2009, \aap, 507, 1277

\bibitem[\protect\citeauthoryear{{Bullock}, {Kravtsov}, \&
  {Weinberg}}{{Bullock} et~al.}{2000}]{2000ApJ...539..517B}
{Bullock}, J.~S., {Kravtsov}, A.~V.,  \& {Weinberg}, D.~H. 2000, \apj, 539, 517

\bibitem[\protect\citeauthoryear{{Capak} et~al.}{{Capak}
  et~al.}{2007}]{2007ApJS..172...99C}
{Capak}, P., et~al. 2007, \apjs, 172, 99

\bibitem[\protect\citeauthoryear{{Capak} et~al.}{{Capak}
  et~al.}{2004}]{2004AJ....127..180C}
{Capak}, P., et~al. 2004, \aj, 127, 180

\bibitem[\protect\citeauthoryear{{Casertano} et~al.}{{Casertano}
  et~al.}{2000}]{2000AJ....120.2747C}
{Casertano}, S., et~al. 2000, \aj, 120, 2747

\bibitem[\protect\citeauthoryear{{Castellano} et~al.}{{Castellano}
  et~al.}{2010}]{2010A&A...511A..20C}
{Castellano}, M., et~al. 2010, \aap, 511, A20

\bibitem[\protect\citeauthoryear{{Chary} et~al.}{{Chary}
  et~al.}{2007}]{2007ApJ...665..257C}
{Chary}, R., {Teplitz}, H.~I., {Dickinson}, M.~E., {Koo}, D.~C., {Le Floc'h},
  E., {Marcillac}, D., {Papovich}, C.,  \& {Stern}, D. 2007, \apj, 665, 257

\bibitem[\protect\citeauthoryear{{Cimatti} et~al.}{{Cimatti}
  et~al.}{2008}]{2008A&A...482...21C}
{Cimatti}, A., et~al. 2008, \aap, 482, 21

\bibitem[\protect\citeauthoryear{{Cirasuolo} et~al.}{{Cirasuolo}
  et~al.}{2007}]{2007MNRAS.380..585C}
{Cirasuolo}, M., et~al. 2007, \mnras, 380, 585

\bibitem[\protect\citeauthoryear{{Clarke} \& {Oey}}{{Clarke} \&
  {Oey}}{2002}]{2002MNRAS.337.1299C}
{Clarke}, C.,  \& {Oey}, M.~S. 2002, \mnras, 337, 1299

\bibitem[\protect\citeauthoryear{{Coil} et~al.}{{Coil}
  et~al.}{2004}]{2004ApJ...609..525C}
{Coil}, A.~L., et~al. 2004, \apj, 609, 525

\bibitem[\protect\citeauthoryear{{Conroy}, {Wechsler}, \& {Kravtsov}}{{Conroy}
  et~al.}{2006}]{2006ApJ...647..201C}
{Conroy}, C., {Wechsler}, R.~H.,  \& {Kravtsov}, A.~V. 2006, \apj, 647, 201

\bibitem[\protect\citeauthoryear{{Cooray} et~al.}{{Cooray}
  et~al.}{2004}]{2004ApJ...606..611C}
{Cooray}, A., {Bock}, J.~J., {Keatin}, B., {Lange}, A.~E.,  \& {Matsumoto}, T.
  2004, \apj, 606, 611

\bibitem[\protect\citeauthoryear{{Daddi} et~al.}{{Daddi}
  et~al.}{2007}]{2007ApJ...670..156D}
{Daddi}, E., et~al. 2007, \apj, 670, 156

\bibitem[\protect\citeauthoryear{{Daddi} et~al.}{{Daddi}
  et~al.}{2005}]{2005ApJ...626..680D}
{Daddi}, E., et~al. 2005, \apj, 626, 680

\bibitem[\protect\citeauthoryear{{Davis} et~al.}{{Davis}
  et~al.}{2007}]{2007ApJ...660L...1D}
{Davis}, M., et~al. 2007, \apjl, 660, L1

\bibitem[\protect\citeauthoryear{{Dickinson} et~al.}{{Dickinson}
  et~al.}{2004}]{2004ApJ...600L..99D}
{Dickinson}, M., et~al. 2004, \apjl, 600, L99

\bibitem[\protect\citeauthoryear{{Dunlop}, {Cirasuolo}, \& {McLure}}{{Dunlop}
  et~al.}{2007}]{2007MNRAS.376.1054D}
{Dunlop}, J.~S., {Cirasuolo}, M.,  \& {McLure}, R.~J. 2007, \mnras, 376, 1054

\bibitem[\protect\citeauthoryear{{Elbaz} et~al.}{{Elbaz}
  et~al.}{2007}]{2007A&A...468...33E}
{Elbaz}, D., et~al. 2007, \aap, 468, 33

\bibitem[\protect\citeauthoryear{{Elmegreen}, {Bournaud}, \&
  {Elmegreen}}{{Elmegreen} et~al.}{2008}]{2008ApJ...688...67E}
{Elmegreen}, B.~G., {Bournaud}, F.,  \& {Elmegreen}, D.~M. 2008, \apj, 688, 67

\bibitem[\protect\citeauthoryear{{Fan}, {Carilli}, \& {Keating}}{{Fan}
  et~al.}{2006}]{2006ARA&A..44..415F}
{Fan}, X., {Carilli}, C.~L.,  \& {Keating}, B. 2006, \araa, 44, 415

\bibitem[\protect\citeauthoryear{{Fan} et~al.}{{Fan}
  et~al.}{2003}]{2003AJ....125.1649F}
{Fan}, X., et~al. 2003, \aj, 125, 1649

\bibitem[\protect\citeauthoryear{{Fernandez} et~al.}{{Fernandez}
  et~al.}{2010}]{2010ApJ...710.1089F}
{Fernandez}, E.~R., {Komatsu}, E., {Iliev}, I.~T.,  \& {Shapiro}, P.~R. 2010,
  \apj, 710, 1089

\bibitem[\protect\citeauthoryear{{F{\"o}rster Schreiber} et~al.}{{F{\"o}rster
  Schreiber} et~al.}{2009}]{2009ApJ...706.1364F}
{F{\"o}rster Schreiber}, N.~M., et~al. 2009, \apj, 706, 1364

\bibitem[\protect\citeauthoryear{{F{\"o}rster Schreiber} et~al.}{{F{\"o}rster
  Schreiber} et~al.}{2006}]{2006ApJ...645.1062F}
{F{\"o}rster Schreiber}, N.~M., et~al. 2006, \apj, 645, 1062

\bibitem[\protect\citeauthoryear{{Fruchter} \& {Hook}}{{Fruchter} \&
  {Hook}}{2002}]{2002PASP..114..144F}
{Fruchter}, A.~S.,  \& {Hook}, R.~N. 2002, \pasp, 114, 144

\bibitem[\protect\citeauthoryear{{Fujita} et~al.}{{Fujita}
  et~al.}{2002}]{2002ApJ...577...11F}
{Fujita}, Y., {Sarazin}, C.~L., {Nagashima}, M.,  \& {Yano}, T. 2002, \apj,
  577, 11

\bibitem[\protect\citeauthoryear{{Genzel} et~al.}{{Genzel}
  et~al.}{2008}]{2008ApJ...687...59G}
{Genzel}, R., et~al. 2008, \apj, 687, 59

\bibitem[\protect\citeauthoryear{{Genzel} et~al.}{{Genzel}
  et~al.}{2006}]{2006Natur.442..786G}
{Genzel}, R., et~al. 2006, \nat, 442, 786

\bibitem[\protect\citeauthoryear{{Giavalisco} et~al.}{{Giavalisco}
  et~al.}{2004}]{2004ApJ...600L..93G}
{Giavalisco}, M., et~al. 2004, \apjl, 600, L93

\bibitem[\protect\citeauthoryear{{Gnedin}, {Kravtsov}, \& {Chen}}{{Gnedin}
  et~al.}{2008}]{2008ApJ...672..765G}
{Gnedin}, N.~Y., {Kravtsov}, A.~V.,  \& {Chen}, H. 2008, \apj, 672, 765

\bibitem[\protect\citeauthoryear{{Grazian} et~al.}{{Grazian}
  et~al.}{2006}]{2006A&A...449..951G}
{Grazian}, A., et~al. 2006, \aap, 449, 951

\bibitem[\protect\citeauthoryear{{Greggio}, {Renzini}, \& {Daddi}}{{Greggio}
  et~al.}{2008}]{2008MNRAS.388..829G}
{Greggio}, L., {Renzini}, A.,  \& {Daddi}, E. 2008, \mnras, 388, 829

\bibitem[\protect\citeauthoryear{{Grogin} et~al.}{{Grogin}
  et~al.}{2011}]{2011.Grogin}
{Grogin}, N.~A., et~al. 2011, in prep

\bibitem[\protect\citeauthoryear{{Grogin} et~al.}{{Grogin}
  et~al.}{2010}]{2010.Grogin.bias}
{Grogin}, N.~A., {Lim}, P.~L., {Maybhate}, A., {Hook}, R.~N.,  \& {Loose}, M.
  2010, in The 2010 HST Calibration Workshop, ed. {S.~Deustua and Cristina
  Oliveira (Baltimore: STScI)}

\bibitem[\protect\citeauthoryear{{Gunn} \& {Stryker}}{{Gunn} \&
  {Stryker}}{1983}]{1983ApJS...52..121G}
{Gunn}, J.~E.,  \& {Stryker}, L.~L. 1983, \apjs, 52, 121

\bibitem[\protect\citeauthoryear{{Immeli} et~al.}{{Immeli}
  et~al.}{2004}]{2004ApJ...611...20I}
{Immeli}, A., {Samland}, M., {Westera}, P.,  \& {Gerhard}, O. 2004, \apj, 611,
  20

\bibitem[\protect\citeauthoryear{{Ivison} et~al.}{{Ivison}
  et~al.}{2007}]{2007ApJ...660L..77I}
{Ivison}, R.~J., et~al. 2007, \apjl, 660, L77

\bibitem[\protect\citeauthoryear{{Iwata} et~al.}{{Iwata}
  et~al.}{2009}]{2009ApJ...692.1287I}
{Iwata}, I., et~al. 2009, \apj, 692, 1287

\bibitem[\protect\citeauthoryear{{Kobayashi} \& {Nomoto}}{{Kobayashi} \&
  {Nomoto}}{2009}]{2009ApJ...707.1466K}
{Kobayashi}, C.,  \& {Nomoto}, K. 2009, \apj, 707, 1466

\bibitem[\protect\citeauthoryear{{Koekemoer} et~al.}{{Koekemoer}
  et~al.}{2004}]{2004ApJ...600L.123K}
{Koekemoer}, A.~M., et~al. 2004, \apjl, 600, L123

\bibitem[\protect\citeauthoryear{{Koekemoer} et~al.}{{Koekemoer}
  et~al.}{2007}]{2007ApJS..172..196K}
{Koekemoer}, A.~M., et~al. 2007, \apjs, 172, 196

\bibitem[\protect\citeauthoryear{{Koekemoer} et~al.}{{Koekemoer}
  et~al.}{2002}]{2002hstc.conf..337K}
{Koekemoer}, A.~M., {Fruchter}, A.~S., {Hook}, R.~N.,  \& {Hack}, W. 2002, in
  The 2002 HST Calibration Workshop, ed. {S.~Arribas, A.~Koekemoer, \&
  B.~Whitmore (Baltimore: STScI)}, 337

\bibitem[\protect\citeauthoryear{{Kozhurina-Platais}
  et~al.}{{Kozhurina-Platais} et~al.}{2009}]{2009.Kozhurina.ISR}
{Kozhurina-Platais}, V., {Cox}, C., {McLean}, B., {Petro}, L., {Dressel}, L.,
  \& {Bushouse}, H. 2009, ISR-WFC3-2009-34

\bibitem[\protect\citeauthoryear{{Lasker} et~al.}{{Lasker}
  et~al.}{2008}]{2008AJ....136..735L}
{Lasker}, B.~M., et~al. 2008, \aj, 136, 735

\bibitem[\protect\citeauthoryear{{Lawrence} et~al.}{{Lawrence}
  et~al.}{2007}]{2007MNRAS.379.1599L}
{Lawrence}, A., et~al. 2007, \mnras, 379, 1599

\bibitem[\protect\citeauthoryear{{Lee} et~al.}{{Lee}
  et~al.}{2006}]{2006ApJ...642...63L}
{Lee}, K., {Giavalisco}, M., {Gnedin}, O.~Y., {Somerville}, R.~S., {Ferguson},
  H.~C., {Dickinson}, M.,  \& {Ouchi}, M. 2006, \apj, 642, 63

\bibitem[\protect\citeauthoryear{{Lee} et~al.}{{Lee}
  et~al.}{2009}]{2009ApJ...695..368L}
{Lee}, K.-S., {Giavalisco}, M., {Conroy}, C., {Wechsler}, R.~H., {Ferguson},
  H.~C., {Somerville}, R.~S., {Dickinson}, M.~E.,  \& {Urry}, C.~M. 2009, \apj,
  695, 368

\bibitem[\protect\citeauthoryear{{MacKenty} et~al.}{{MacKenty}
  et~al.}{2010}]{2010SPIE.7731E..27M}
{MacKenty}, J.~W., {Kimble}, R.~A., {O'Connell}, R.~W.,  \& {Townsend}, J.~A.
  2010, in Society of Photo-Optical Instrumentation Engineers (SPIE) Conference
  Series, Vol. 7731, Society of Photo-Optical Instrumentation Engineers (SPIE)
  Conference Series

\bibitem[\protect\citeauthoryear{{Mannucci} et~al.}{{Mannucci}
  et~al.}{2005}]{2005A&A...433..807M}
{Mannucci}, F., {Della Valle}, M., {Panagia}, N., {Cappellaro}, E., {Cresci},
  G., {Maiolino}, R., {Petrosian}, A.,  \& {Turatto}, M. 2005, \aap, 433, 807

\bibitem[\protect\citeauthoryear{{Massey}}{{Massey}}{2010}]{2010MNRAS.409L.109%
M}
{Massey}, R. 2010, \mnras, 409, L109

\bibitem[\protect\citeauthoryear{{Mobasher} et~al.}{{Mobasher}
  et~al.}{2005}]{2005ApJ...635..832M}
{Mobasher}, B., et~al. 2005, \apj, 635, 832

\bibitem[\protect\citeauthoryear{{Morrison} et~al.}{{Morrison}
  et~al.}{2010}]{2010ApJS..188..178M}
{Morrison}, G.~E., {Owen}, F.~N., {Dickinson}, M., {Ivison}, R.~J.,  \& {Ibar},
  E. 2010, \apjs, 188, 178

\bibitem[\protect\citeauthoryear{{Newman} et~al.}{{Newman}
  et~al.}{2011}]{2011.Newman}
{Newman}, J.~A., et~al. 2011, in prep

\bibitem[\protect\citeauthoryear{{Noeske} et~al.}{{Noeske}
  et~al.}{2007}]{2007ApJ...660L..43N}
{Noeske}, K.~G., et~al. 2007, \apjl, 660, L43

\bibitem[\protect\citeauthoryear{{Noguchi}}{{Noguchi}}{1999}]{1999ApJ...514...%
77N}
{Noguchi}, M. 1999, \apj, 514, 77

\bibitem[\protect\citeauthoryear{{Page} et~al.}{{Page}
  et~al.}{2007}]{2007ApJS..170..335P}
{Page}, L., et~al. 2007, \apjs, 170, 335

\bibitem[\protect\citeauthoryear{{Perlmutter} et~al.}{{Perlmutter}
  et~al.}{1999}]{1999ApJ...517..565P}
{Perlmutter}, S., et~al. 1999, \apj, 517, 565

\bibitem[\protect\citeauthoryear{{Pickles}}{{Pickles}}{1998}]{1998PASP..110..8%
63P}
{Pickles}, A.~J. 1998, \pasp, 110, 863

\bibitem[\protect\citeauthoryear{{Ricotti}}{{Ricotti}}{2002}]{2002MNRAS.336L..%
33R}
{Ricotti}, M. 2002, \mnras, 336, L33

\bibitem[\protect\citeauthoryear{{Riess} et~al.}{{Riess}
  et~al.}{1998}]{1998AJ....116.1009R}
{Riess}, A.~G., et~al. 1998, \aj, 116, 1009

\bibitem[\protect\citeauthoryear{{Riess} et~al.}{{Riess}
  et~al.}{2007}]{2007ApJ...659...98R}
{Riess}, A.~G., et~al. 2007, \apj, 659, 98

\bibitem[\protect\citeauthoryear{{Riess} et~al.}{{Riess}
  et~al.}{2004}]{2004ApJ...607..665R}
{Riess}, A.~G., et~al. 2004, \apj, 607, 665

\bibitem[\protect\citeauthoryear{{Rix} et~al.}{{Rix}
  et~al.}{2004}]{2004ApJS..152..163R}
{Rix}, H.-W., et~al. 2004, \apjs, 152, 163

\bibitem[\protect\citeauthoryear{{Ruiz-Lapuente} \& {Canal}}{{Ruiz-Lapuente} \&
  {Canal}}{1998}]{1998ApJ...497L..57R}
{Ruiz-Lapuente}, P.,  \& {Canal}, R. 1998, \apjl, 497, L57

\bibitem[\protect\citeauthoryear{{Ryan-Weber}, {Pettini}, \&
  {Madau}}{{Ryan-Weber} et~al.}{2006}]{2006MNRAS.371L..78R}
{Ryan-Weber}, E.~V., {Pettini}, M.,  \& {Madau}, P. 2006, \mnras, 371, L78

\bibitem[\protect\citeauthoryear{{Schinnerer} et~al.}{{Schinnerer}
  et~al.}{2004}]{2004AJ....128.1974S}
{Schinnerer}, E., et~al. 2004, \aj, 128, 1974

\bibitem[\protect\citeauthoryear{{Scoville} et~al.}{{Scoville}
  et~al.}{2007}]{2007ApJS..172....1S}
{Scoville}, N., et~al. 2007, \apjs, 172, 1

\bibitem[\protect\citeauthoryear{{Shapley} et~al.}{{Shapley}
  et~al.}{2006}]{2006ApJ...651..688S}
{Shapley}, A.~E., {Steidel}, C.~C., {Pettini}, M., {Adelberger}, K.~L.,  \&
  {Erb}, D.~K. 2006, \apj, 651, 688

\bibitem[\protect\citeauthoryear{{Simpson} et~al.}{{Simpson}
  et~al.}{2006}]{2006MNRAS.372..741S}
{Simpson}, C., et~al. 2006, \mnras, 372, 741

\bibitem[\protect\citeauthoryear{{Songaila}}{{Songaila}}{2001}]{2001ApJ...561L%
.153S}
{Songaila}, A. 2001, \apjl, 561, L153

\bibitem[\protect\citeauthoryear{{Spergel} et~al.}{{Spergel}
  et~al.}{2007}]{2007ApJS..170..377S}
{Spergel}, D.~N., et~al. 2007, \apjs, 170, 377

\bibitem[\protect\citeauthoryear{{Trujillo} et~al.}{{Trujillo}
  et~al.}{2006}]{2006MNRAS.373L..36T}
{Trujillo}, I., et~al. 2006, \mnras, 373, L36

\bibitem[\protect\citeauthoryear{{van Dokkum} et~al.}{{van Dokkum}
  et~al.}{2008}]{2008ApJ...677L...5V}
{van Dokkum}, P.~G., et~al. 2008, \apjl, 677, L5

\bibitem[\protect\citeauthoryear{{Wiklind} et~al.}{{Wiklind}
  et~al.}{2008}]{2008ApJ...676..781W}
{Wiklind}, T., {Dickinson}, M., {Ferguson}, H.~C., {Giavalisco}, M.,
  {Mobasher}, B., {Grogin}, N.~A.,  \& {Panagia}, N. 2008, \apj, 676, 781

\bibitem[\protect\citeauthoryear{{Windhorst} et~al.}{{Windhorst}
  et~al.}{2011}]{2011ApJS..193...27W}
{Windhorst}, R.~A., et~al. 2011, \apjs, 193, 27

\end{thebibliography}

\clearpage

\ifsubmodeastroph
  $\,$
  \vspace{-6in}
\fi
\appendix

\section{Additional HST Data in the CANDELS Fields}

In addition to the new CANDELS data described in this paper, several of these survey fields also contain significant previous {\it HST} investment, in many cases with instrument/filter combinations that provide further scientific leverage when combined with the CANDELS data. These surveys include
	GOODS	(Giavalisco et al. 2004),
	GEMS	(Rix et al. 2004),
	UDF	(Beckwith et al. 2006; Oesch et al, 2007; Bouwens et al. 2010),
	WFC3-ERS2 (Windhorst et al. 2011),
	COSMOS	(Scoville et al. 2007),
and
	EGS/AEGIS (Davis et al. 2007; Newman et al. 2011),
as well as a variety of other programs that have obtained data in various locations on these fields. We note that the {\it HST} datasets for these programs are all already public, in many cases with high-level science products already released, although in most cases the calibrations have improved since the data products were originally released. Thus we are reprocessing these data, combining them with newer data using the same CANDELS filters where appropriate and creating summed mosaic images that have been uniformly processed using updated calibration files and CANDELS methods as described in this paper. We note also that the exposure maps and sensitivity estimates of
	Grogin et al. (2011)
take these prior {\it HST} programs into account. These datasets are summarized in Table~\ref{tab:ancillary_hst_data}, showing the existing (non-CANDELS) broadband filter coverage in ACS/WFC and WFC3/IR and UVIS, as well as the relevant references where available.

The datasets are all retrieved from the archive and are recalibrated using the current generation of reference files, including all the new flatfield, dark current, bias and geometric distortion files, as well as improved photometric calibrations. We note that some of the earlier ACS data do not suffer from the same electronic issues as the post-SM4 ACS data, most notably the bias striping effect, and CTE degradation is also much less for the first year or two of ACS operations (since CTE degradation gradually worsens over time). However, the flatfields, darks, bias and distortion reference files are generally much improved, both for the older ACS data and for the newer WFC3 data, and are therefore applied to these reprocessed datasets. These reprocessed datasets will be delivered and served out along with the final combined CANDELS datasets once all the data have been obtained for the survey.

\ifsubmodeapjs		\begin{deluxetable}{llll}\rotate	\fi
\ifsubmodeastroph	\begin{deluxetable*}{llll}		\fi
\vspace{3.5in}
\tablecaption{\label{tab:ancillary_hst_data}
	Additional {\it HST} ACS/WFC and WFC3/IR,UVIS Broad-band Filter Coverage in the CANDELS Fields}
\tablehead{%
Field		& ACS/WFC Filters			& WFC3/UVIS Filters	& WFC3/IR Filters			}
\startdata
GOODS-N$^1$	& F435W F475W F606W F775W F814W	F850LP	& $-$			& F140W 				\\
GOODS-S$^2$	& F435W F475W F606W F775W F814W F850LP	& F225W F275W F336W	& F098M F105W F125W F140W F160W		\\	
COSMOS$^3$	& F814W					& F300W			& F140W F160W				\\
EGS$^4$		& F606W F814W				& $-$			& F140W					\\	
UDS		& $-$					& $-$			& $-$
\enddata
\tablenotetext{1}{Giavalisco et al. (2004); Malhotra et al. (2005); Rhodes et al. (2005); Riess et al. (2004)}		
\tablenotetext{2}{Giavalisco et al. (2004); Beckwith et al. (2006); Bouwens et al (2010); Malhotra et al. (2005); Rhoads et al. (2005); Riess et al. (2004); Rix et al. (2004); Oesch et al. (2007); Thompson et al. (2005); Windhorst et al. (2011)}	
\tablenotetext{3}{Scoville et al. (2007);  Franx et al. (2008); Scarlata et al. (2007); van Dokkum et al. (2006)}
\tablenotetext{4}{Davis et al. (2007); Newman et al. (2011, in prep.); Franx et al. (2008); van Dokkum et al. (2006)}
\ifsubmodeapjs		\end{deluxetable}	\fi
\ifsubmodeastroph	\end{deluxetable*}	\fi

\clearpage
\begin{figure*}[h]
\begin{center}
\ifsubmodeastroph
  \includegraphics[height=4in]{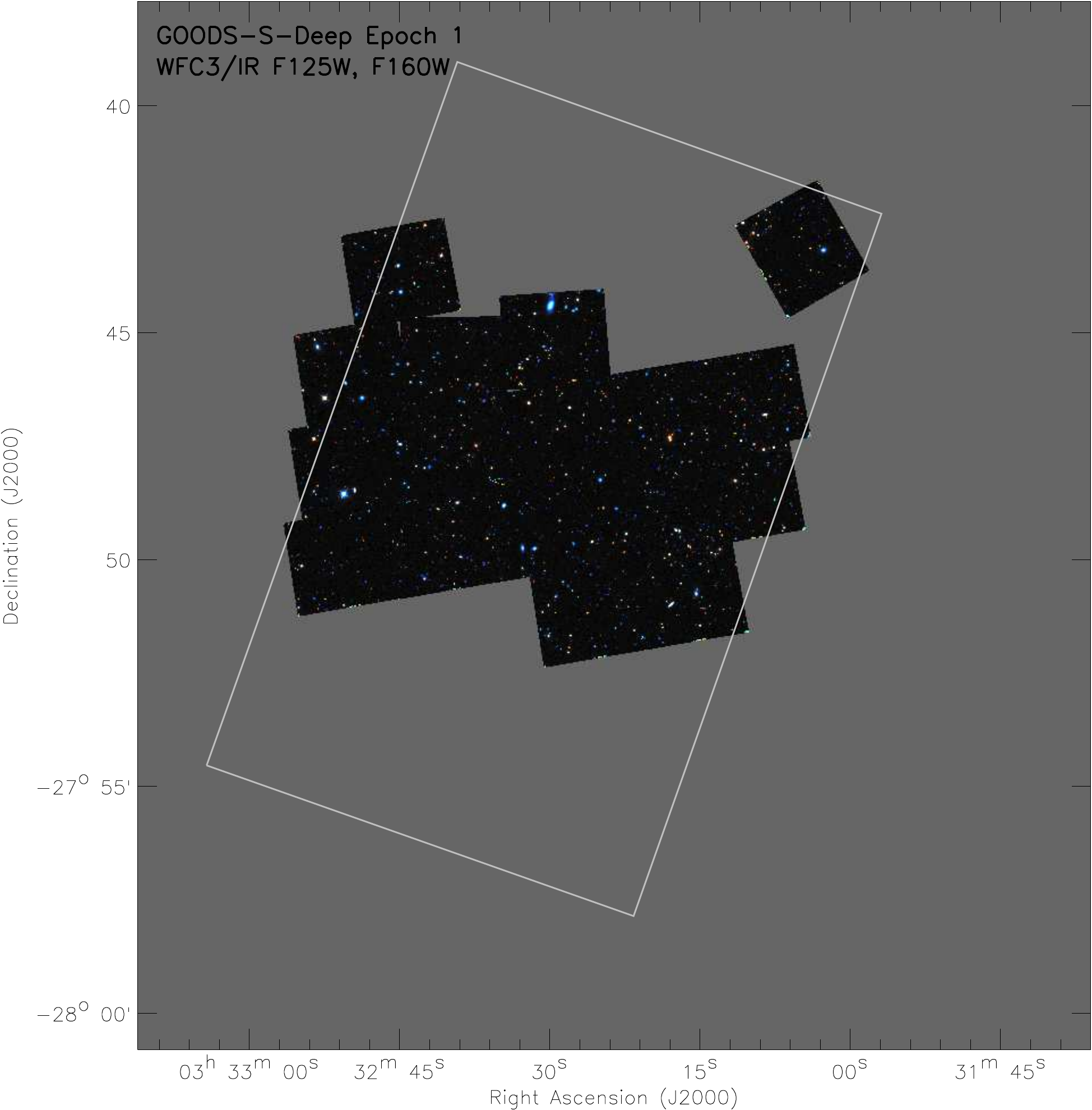}
  \includegraphics[height=4in]{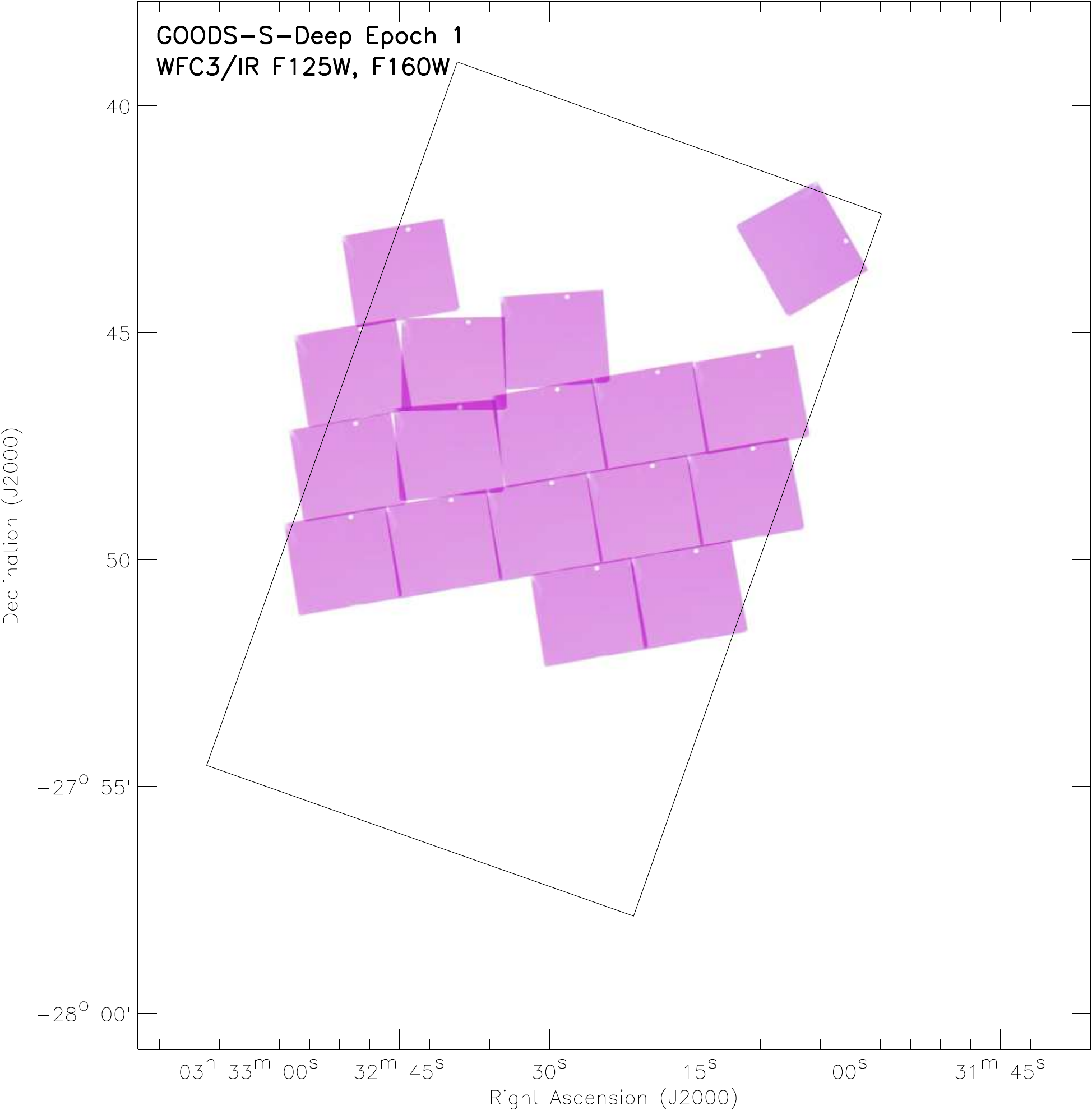}
\fi
\figcaption{\label{fig:gsd01_wfc3}%
Images showing the prime CANDELS WFC3/IR dataset for the first epoch obtained on the GOODS-S field (GOODS-S-Deep Epoch 1). This dataset also includes the first CANDELS test orbit (obtained in April 2010), to the north-west. The top panel shows a color composite of the WFC3/IR F125W and F160W images after combined mosaics were created for each filter separately using \multidrizzle, while the bottom panel shows the corresponding weight images, which are in units of inverse variance. The F125W data are shown in blue and the F160W data are shown in red. Regions containing bad pixels (such as the circular ``death star'' region) are set to 0 and thus have no weight in this single epoch dataset, in which the dither offsets were not yet large enough to move over such features. The overlap between pointings was chosen to be just large enough to provide contiguous coverage while also maximizing total area covered. Occasional tiles are intentionally tilted or offset to enable appropriate guide stars to be selected. The rectangular outlines indicate the nominal boundaries of the existing GOODS coverage.}
\end{center}
\end{figure*}

\clearpage
\begin{figure*}[h]
\begin{center}
\ifsubmodeastroph
  \includegraphics[height=4in]{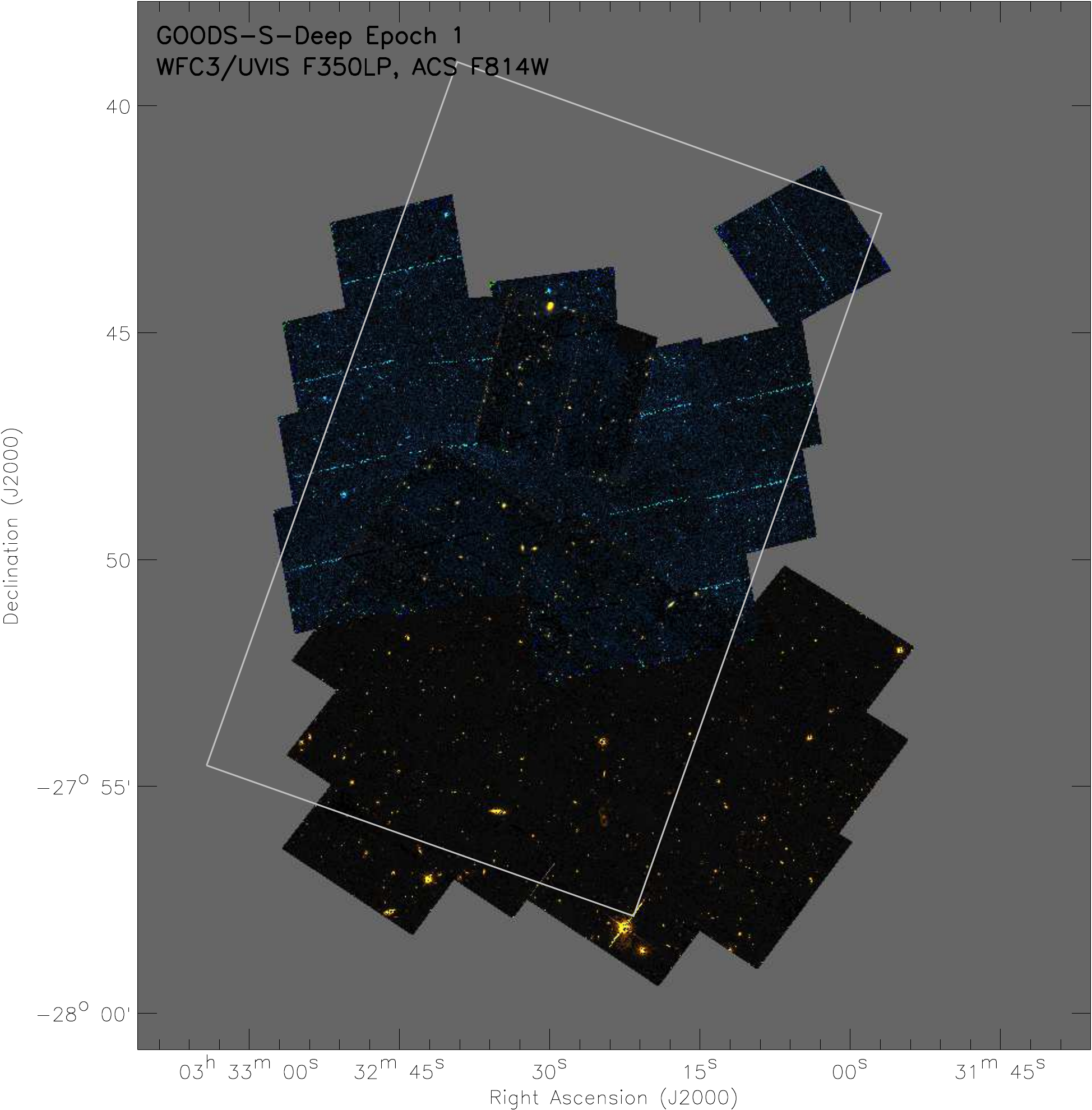}
  \includegraphics[height=4in]{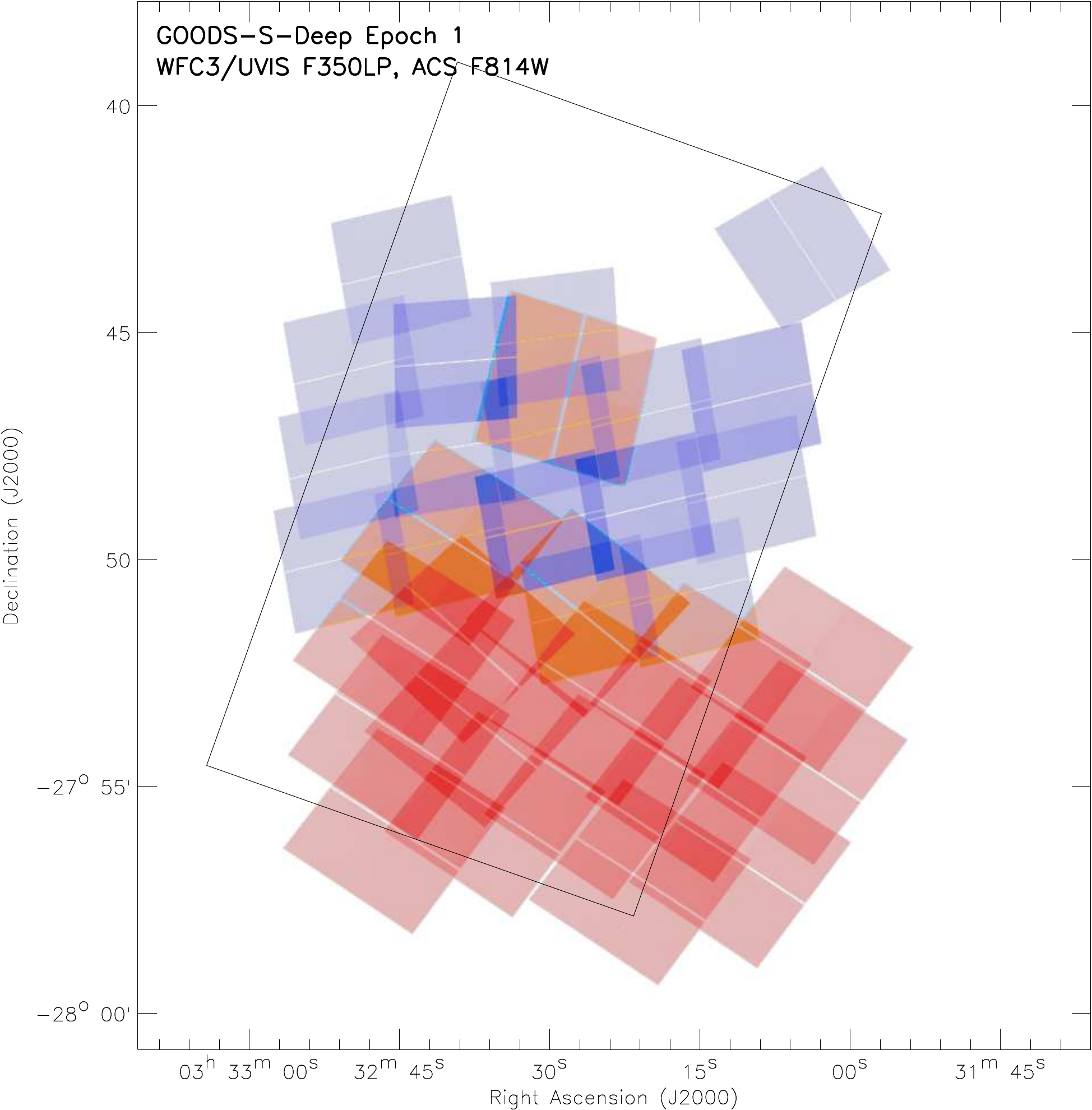}
\fi
\figcaption{\label{fig:gsd01_acs}%
Images showing the prime CANDELS WFC3/UVIS and parallel ACS/WFC datasets for the first epoch obtained on the GOODS-S field (GOODS-S-Deep Epoch 1). This dataset also includes the first CANDELS test orbit (obtained in April 2010), to the north-west. The top panel shows a color composite of the WFC3/UVIS F350LP and ACS/WFC F814W images after combined mosaics were created for each filter separately using \multidrizzle, while the bottom panel shows the corresponding weight images, which are in units of inverse variance. The F350LP data are shown in blue and the F814W data are shown in red. The WFC3/UVIS data consist of only one exposure per pointing and therefore still contain a large number of cosmic rays and other defects. Regions containing bad pixels are masked where necessary. Note also that, in general, the overlap between ACS pointings is sufficient to provide approximately twice the depth of a single pointing across much of the ACS area. The rectangular outlines indicate the nominal boundaries of the existing GOODS coverage.}
\end{center}
\end{figure*}

\clearpage
\begin{figure*}[h]
\begin{center}
\ifsubmodeastroph
  \includegraphics[height=4in]{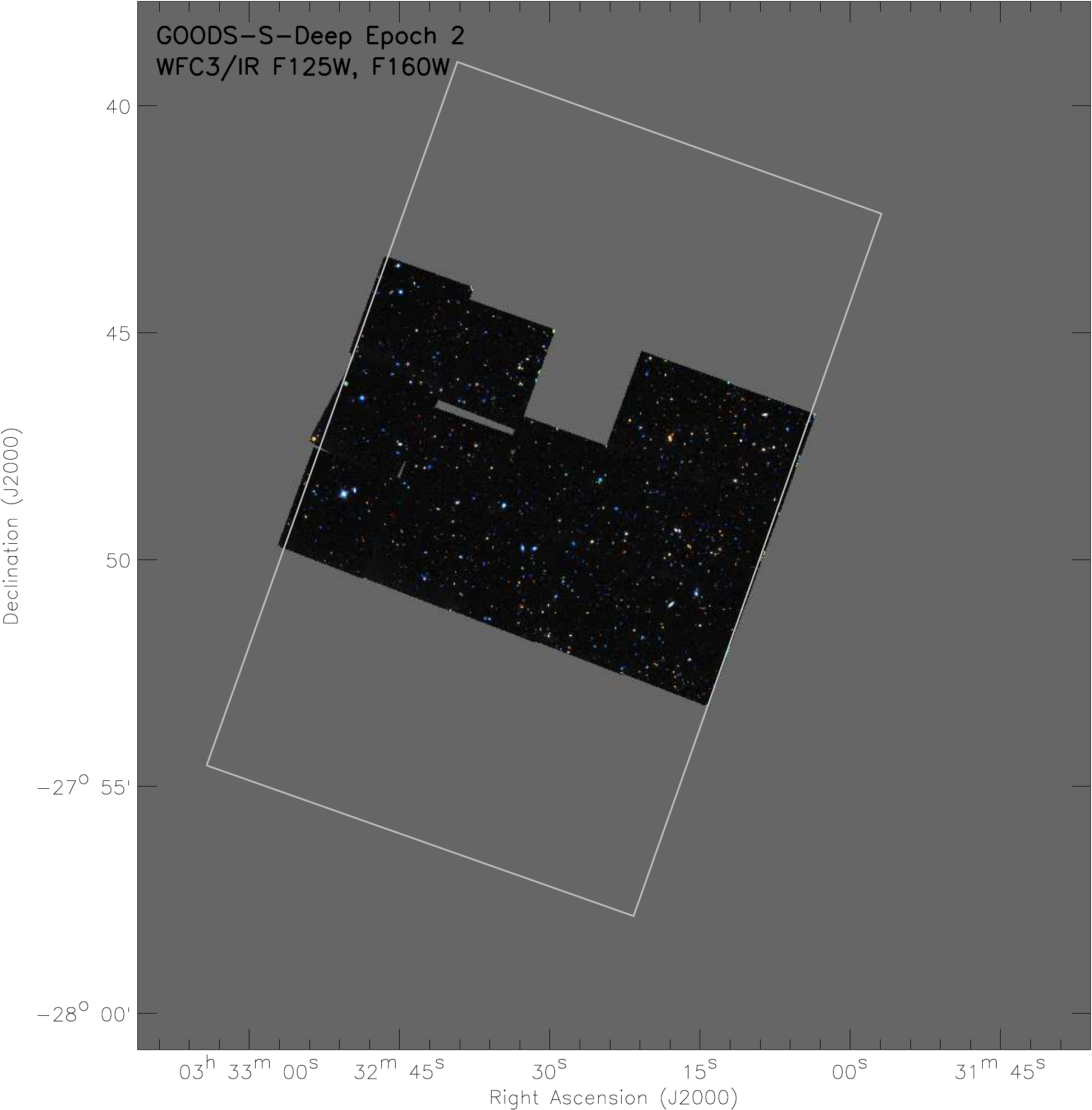}
  \includegraphics[height=4in]{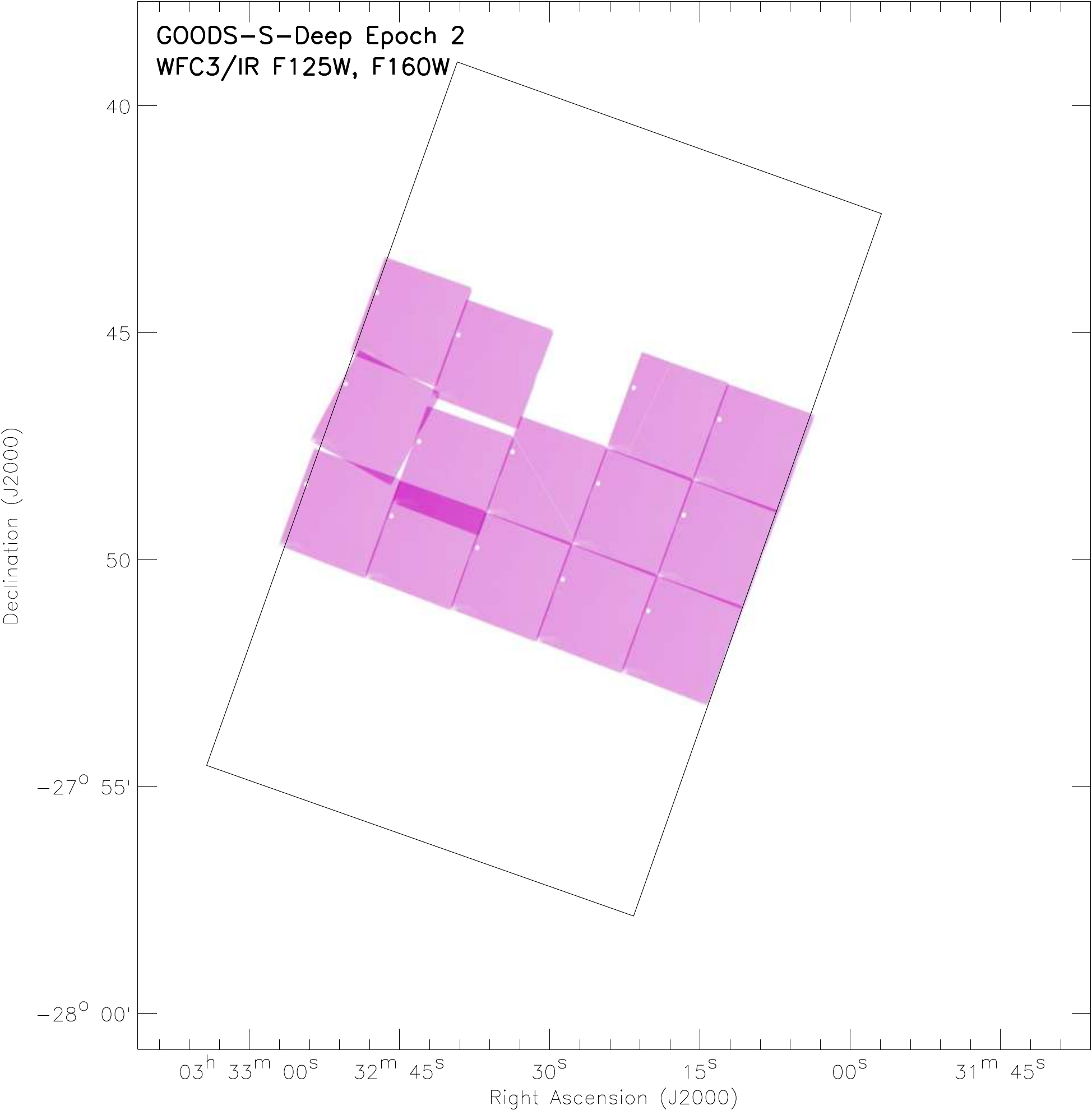}
\fi
\figcaption{\label{fig:gsd02_wfc3}%
As for Figure~\ref{fig:gsd01_wfc3}, but for GOODS-S-Deep Epoch 2.}
\end{center}
\end{figure*}

\clearpage
\begin{figure*}[h]
\begin{center}
\ifsubmodeastroph
  \includegraphics[height=4in]{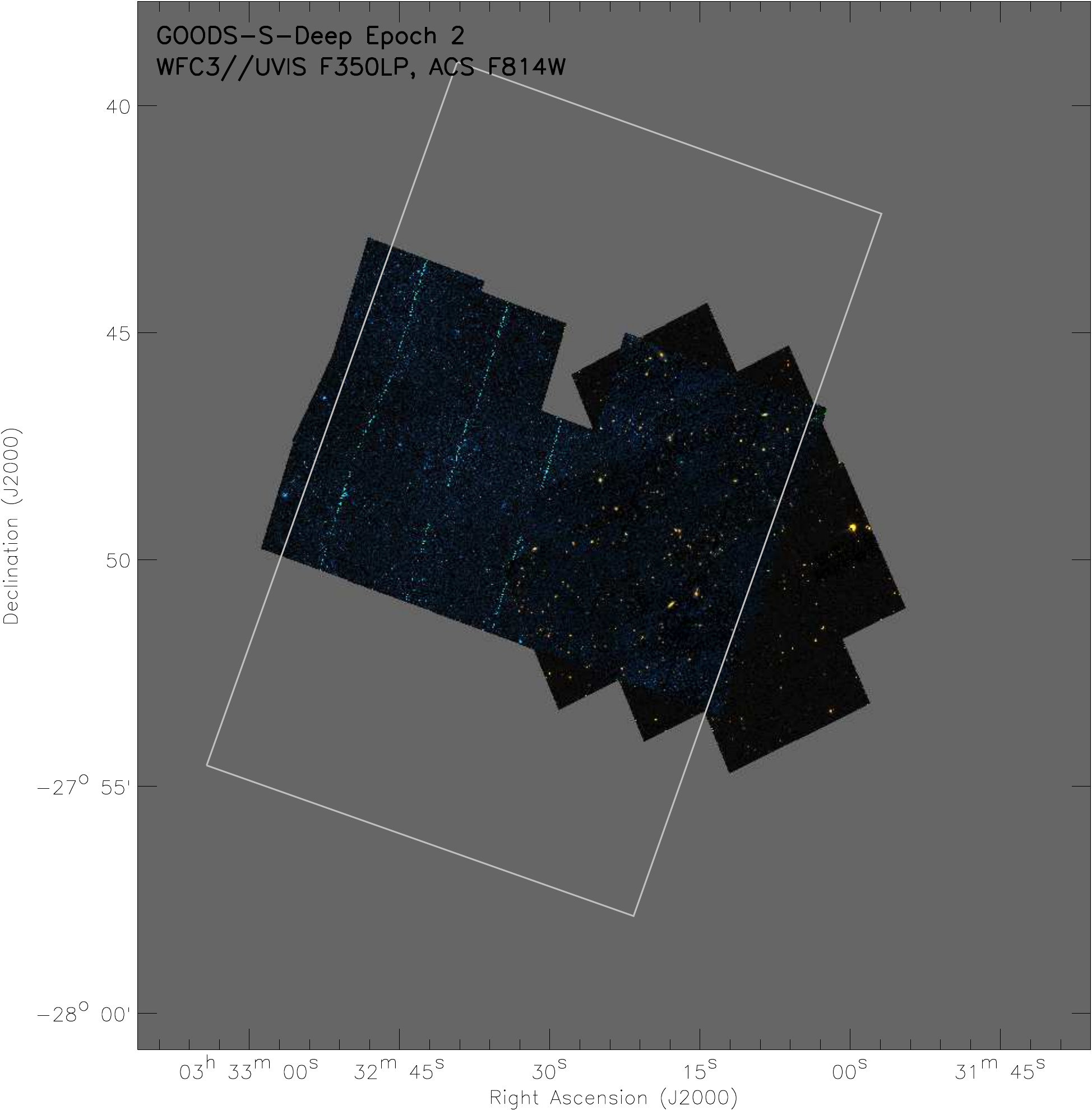}
  \includegraphics[height=4in]{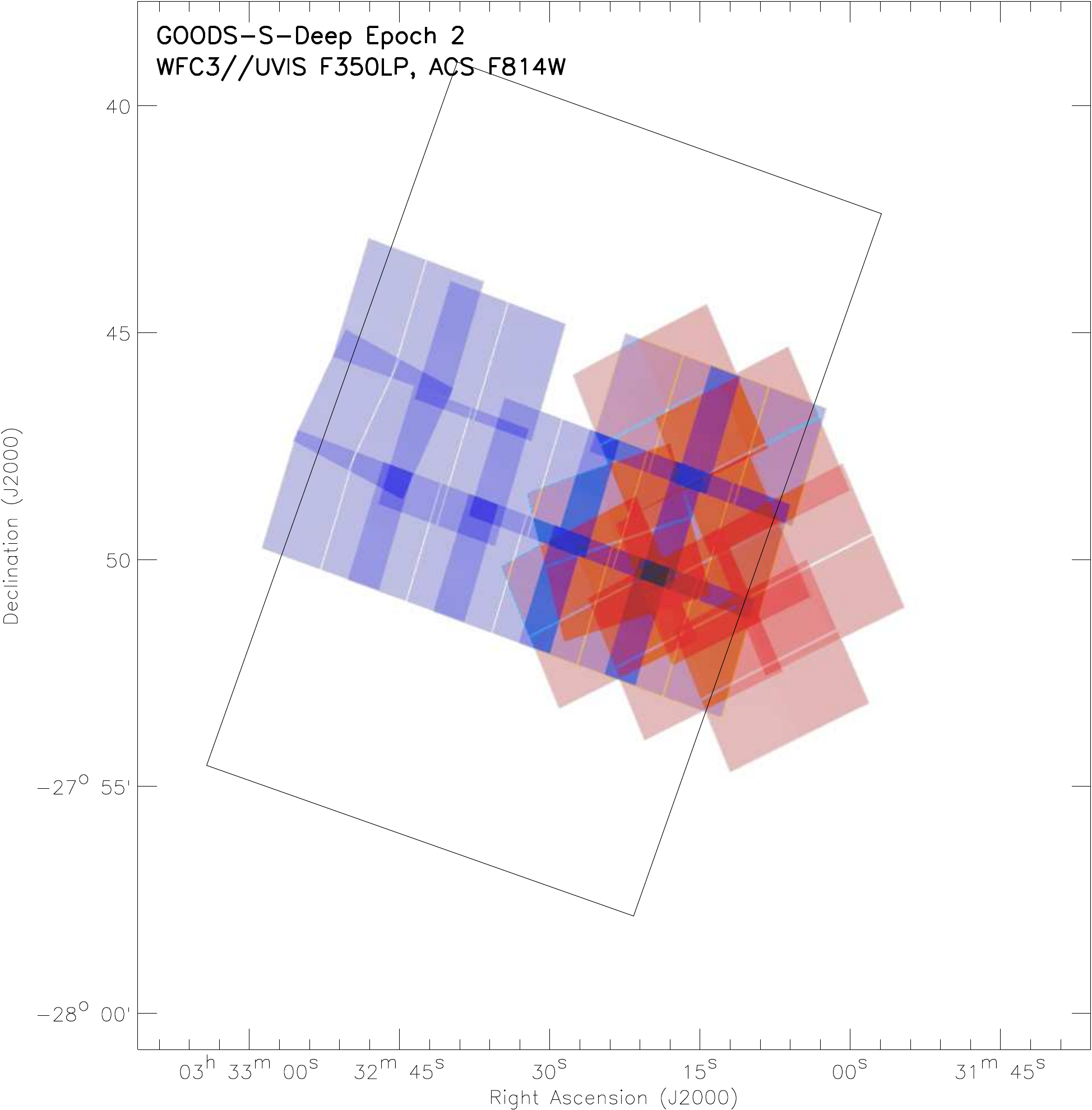}
\fi
\figcaption{\label{fig:gsd02_acs}%
As for Figure~\ref{fig:gsd01_acs}, but for GOODS-S-Deep Epoch 2.}
\end{center}
\end{figure*}

\clearpage
\begin{figure*}[h]
\begin{center}
\ifsubmodeastroph
  \includegraphics[height=4in]{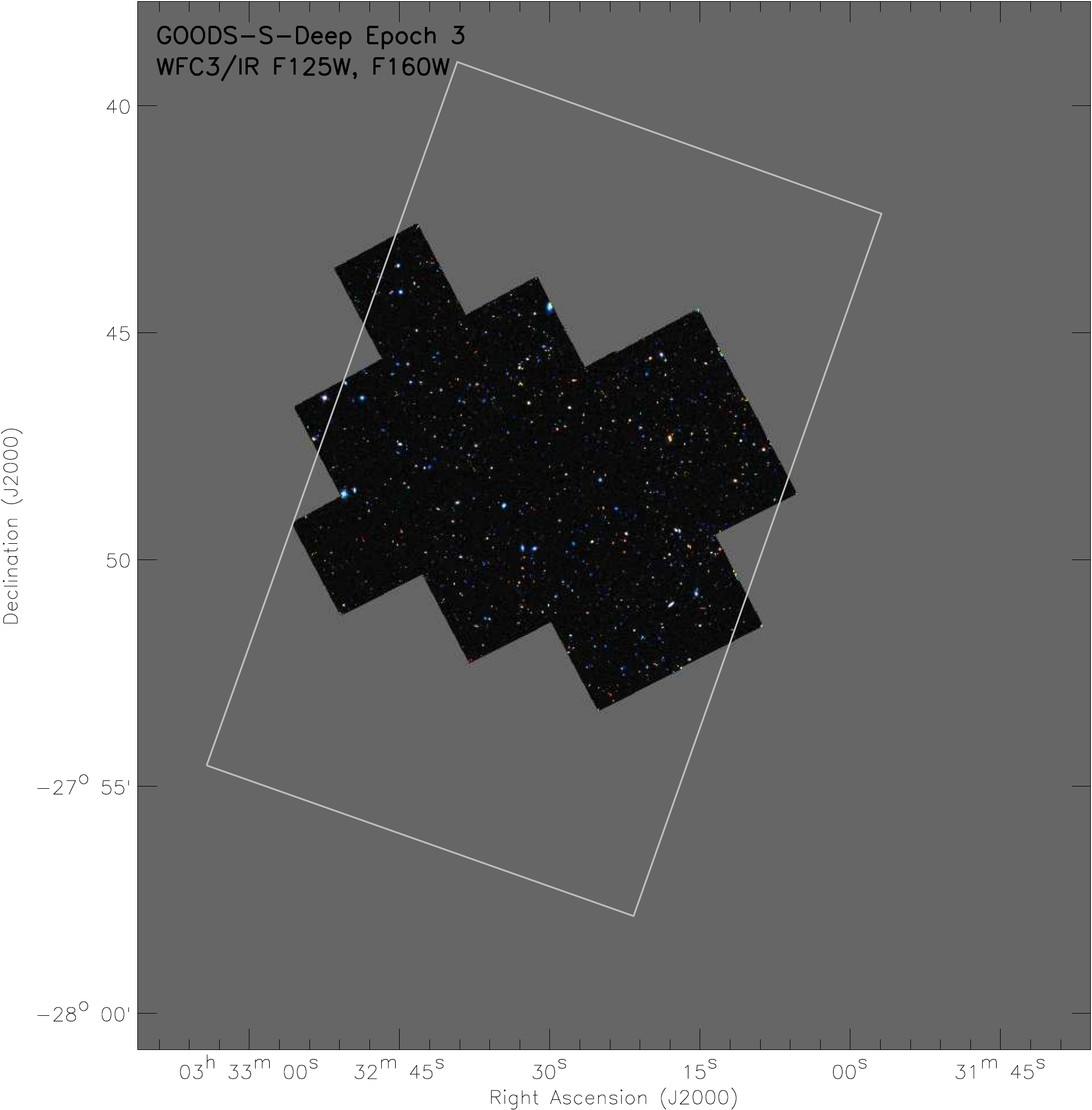}
  \includegraphics[height=4in]{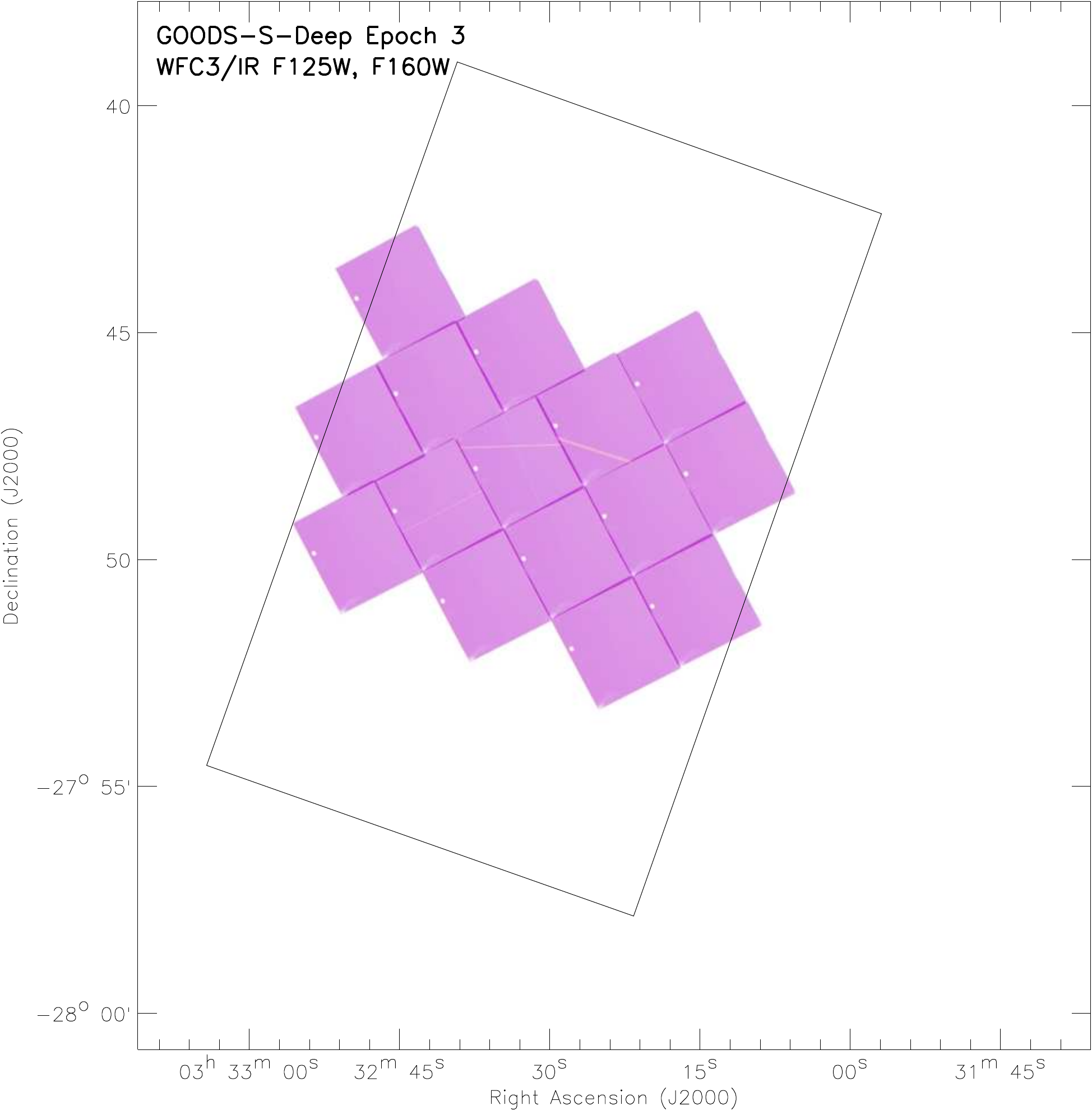}
\fi
\figcaption{\label{fig:gsd03_wfc3}%
As for Figure~\ref{fig:gsd01_wfc3}, but for GOODS-S-Deep Epoch 3.}
\end{center}
\end{figure*}

\clearpage
\begin{figure*}[h]
\begin{center}
\ifsubmodeastroph
  \includegraphics[height=4in]{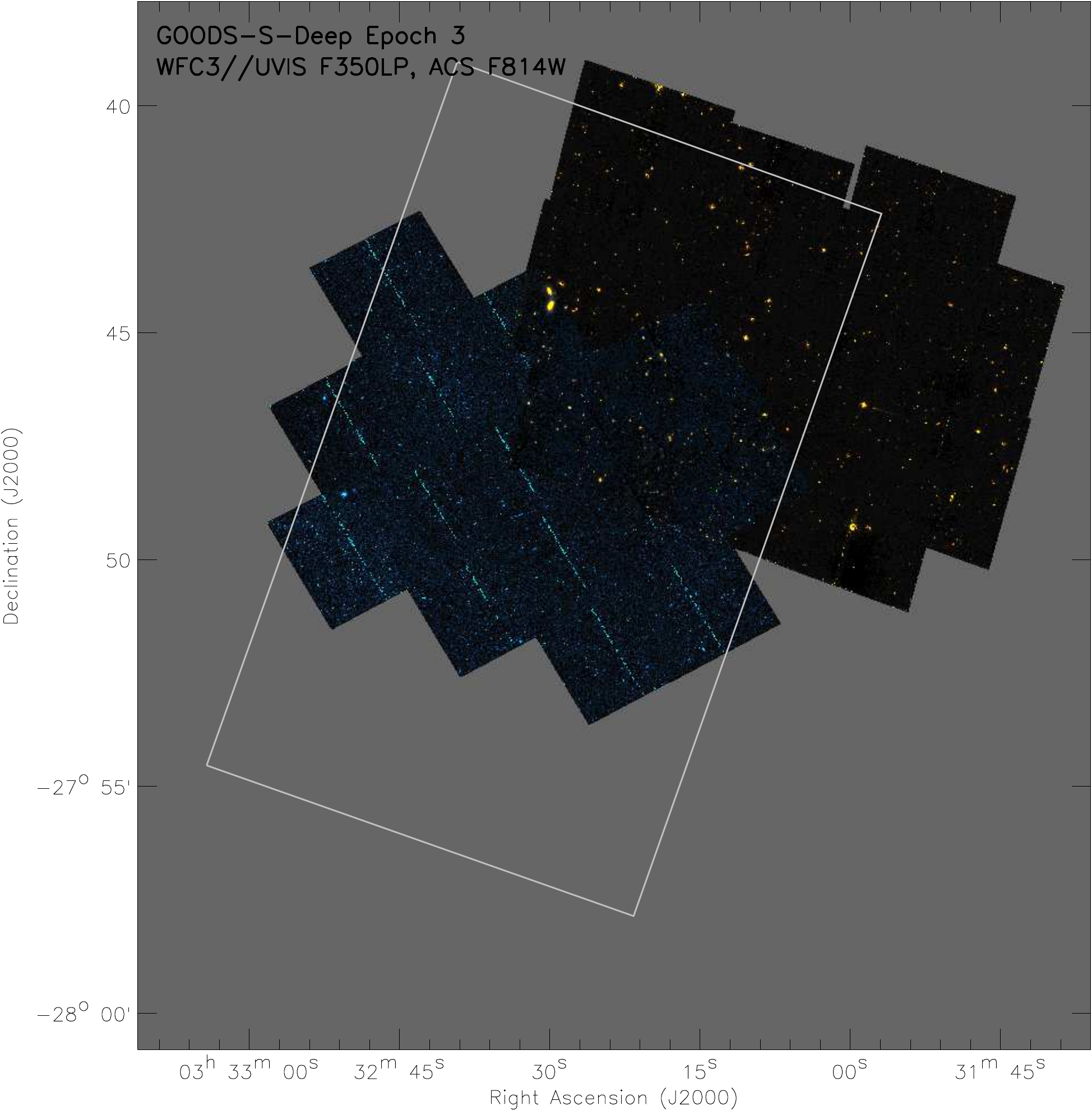}
  \includegraphics[height=4in]{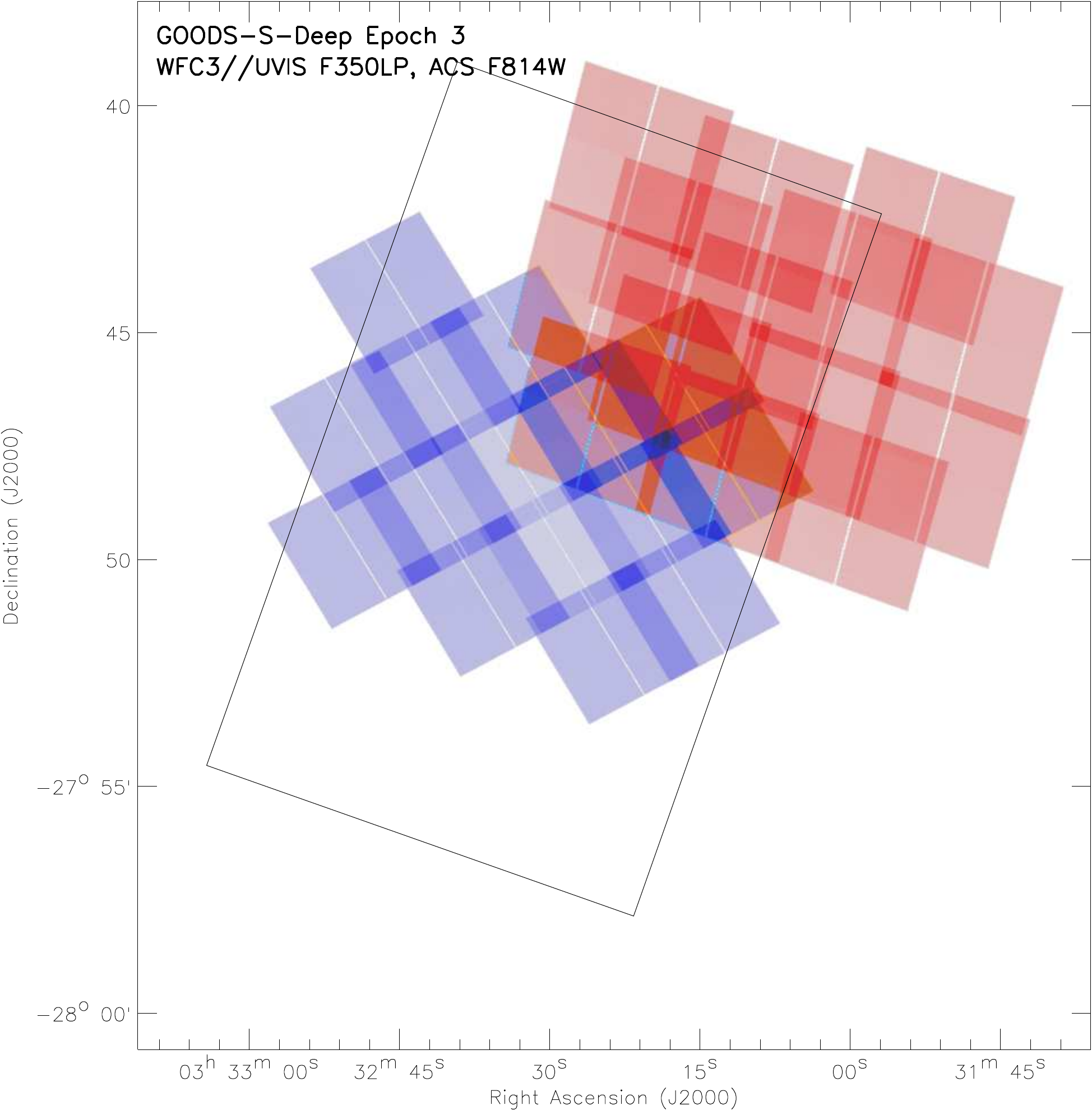}
\fi
\figcaption{\label{fig:gsd03_acs}%
As for Figure~\ref{fig:gsd01_acs}, but for GOODS-S-Deep Epoch 3.}
\end{center}
\end{figure*}

\clearpage
\begin{figure*}[h]
\begin{center}
\ifsubmodeastroph
  \includegraphics[height=4in]{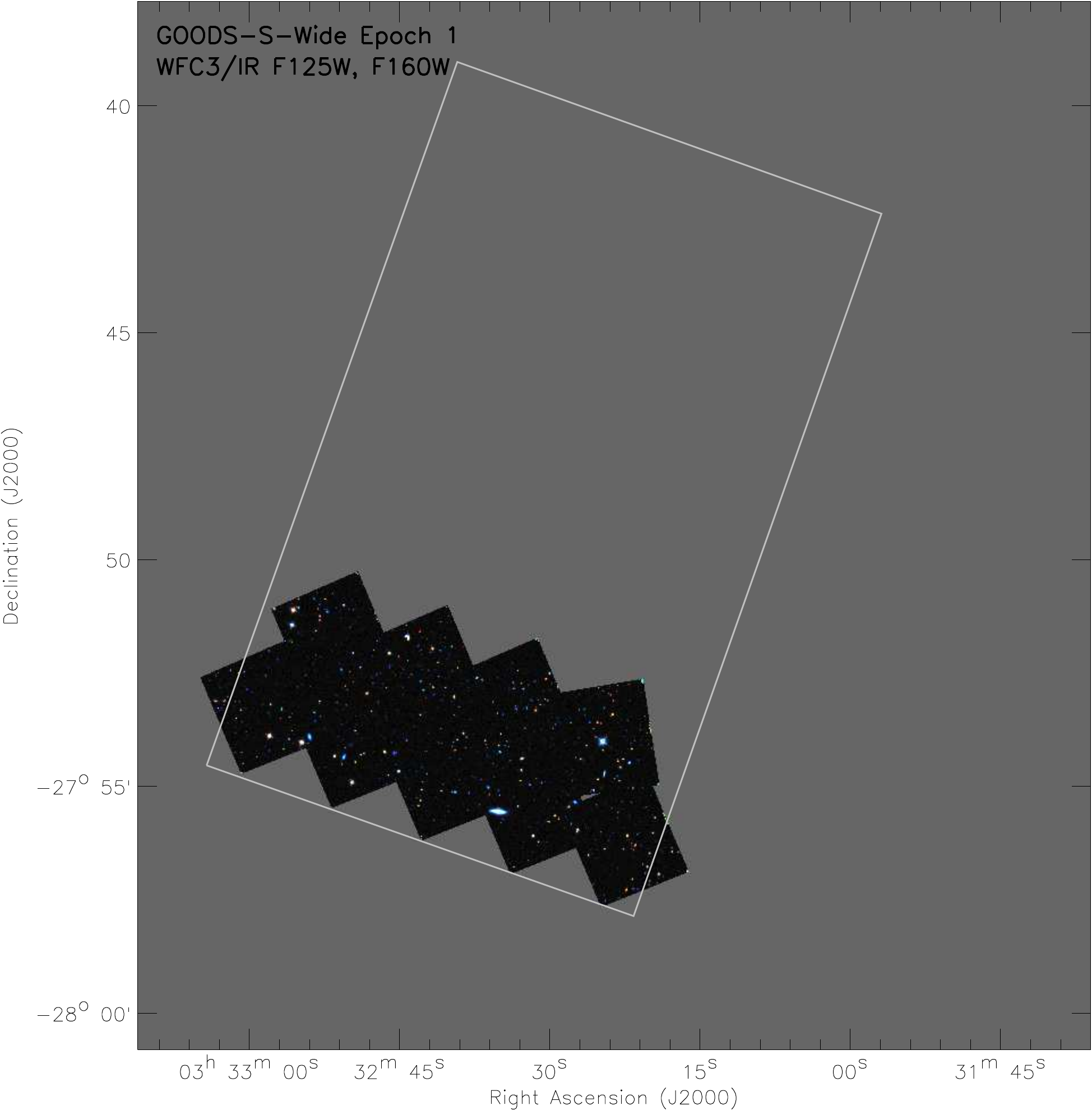}
  \includegraphics[height=4in]{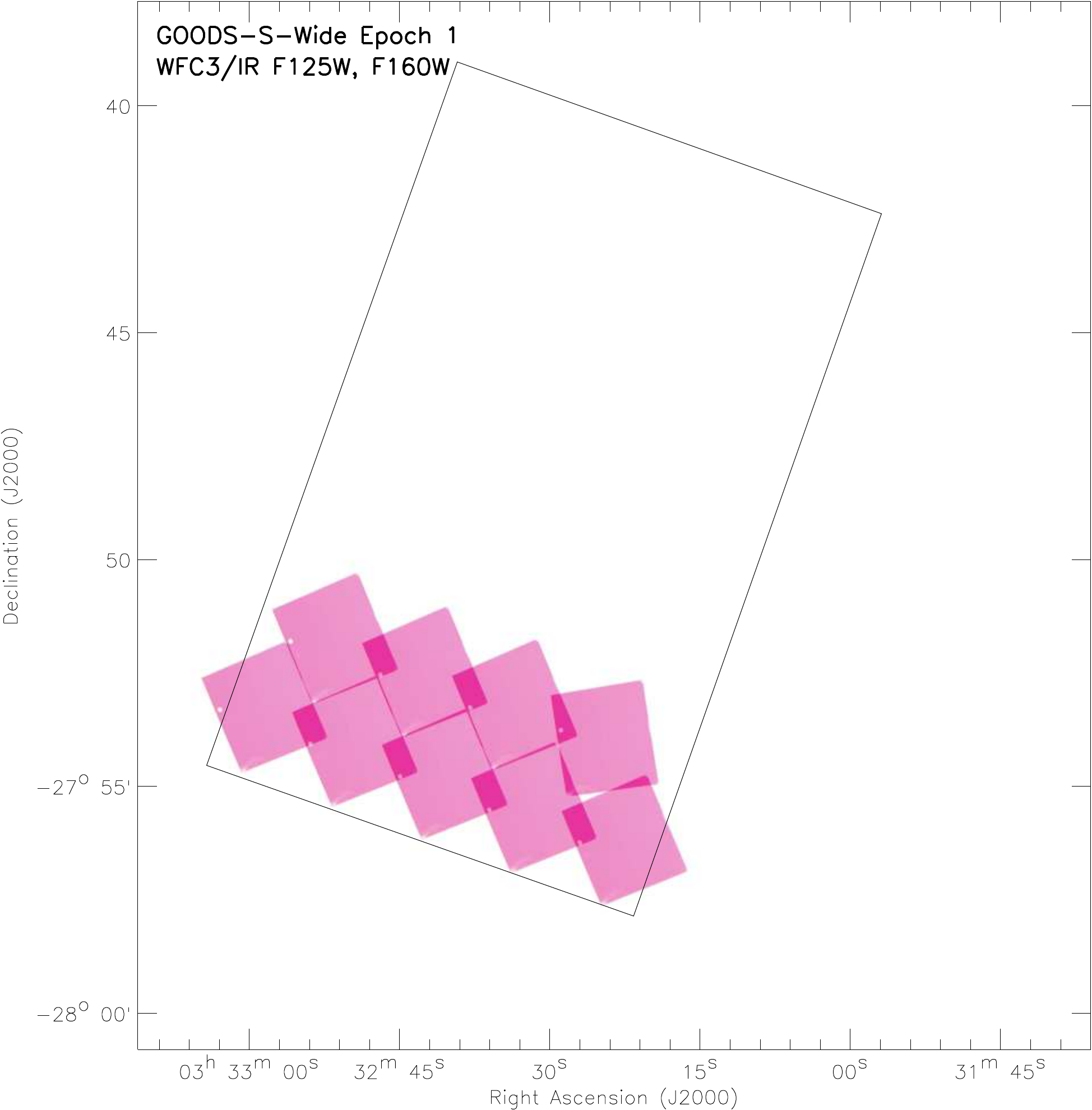}
\fi
\figcaption{\label{fig:gsw01_wfc3}%
As for Figure~\ref{fig:gsd01_wfc3}, but for GOODS-S-Wide Epoch 1.}
\end{center}
\end{figure*}

\clearpage
\begin{figure*}[h]
\begin{center}
\ifsubmodeastroph
  \includegraphics[height=4in]{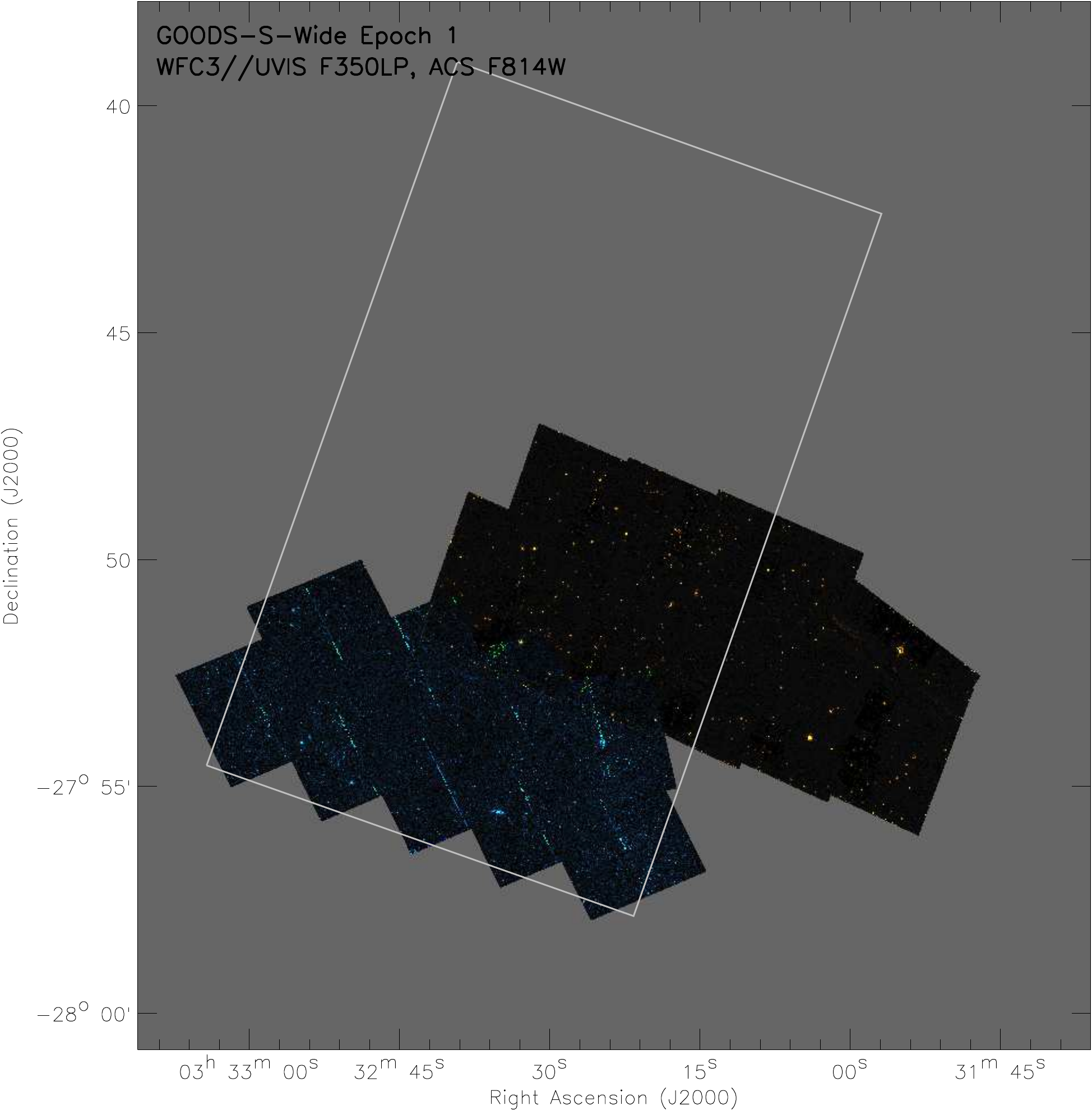}
  \includegraphics[height=4in]{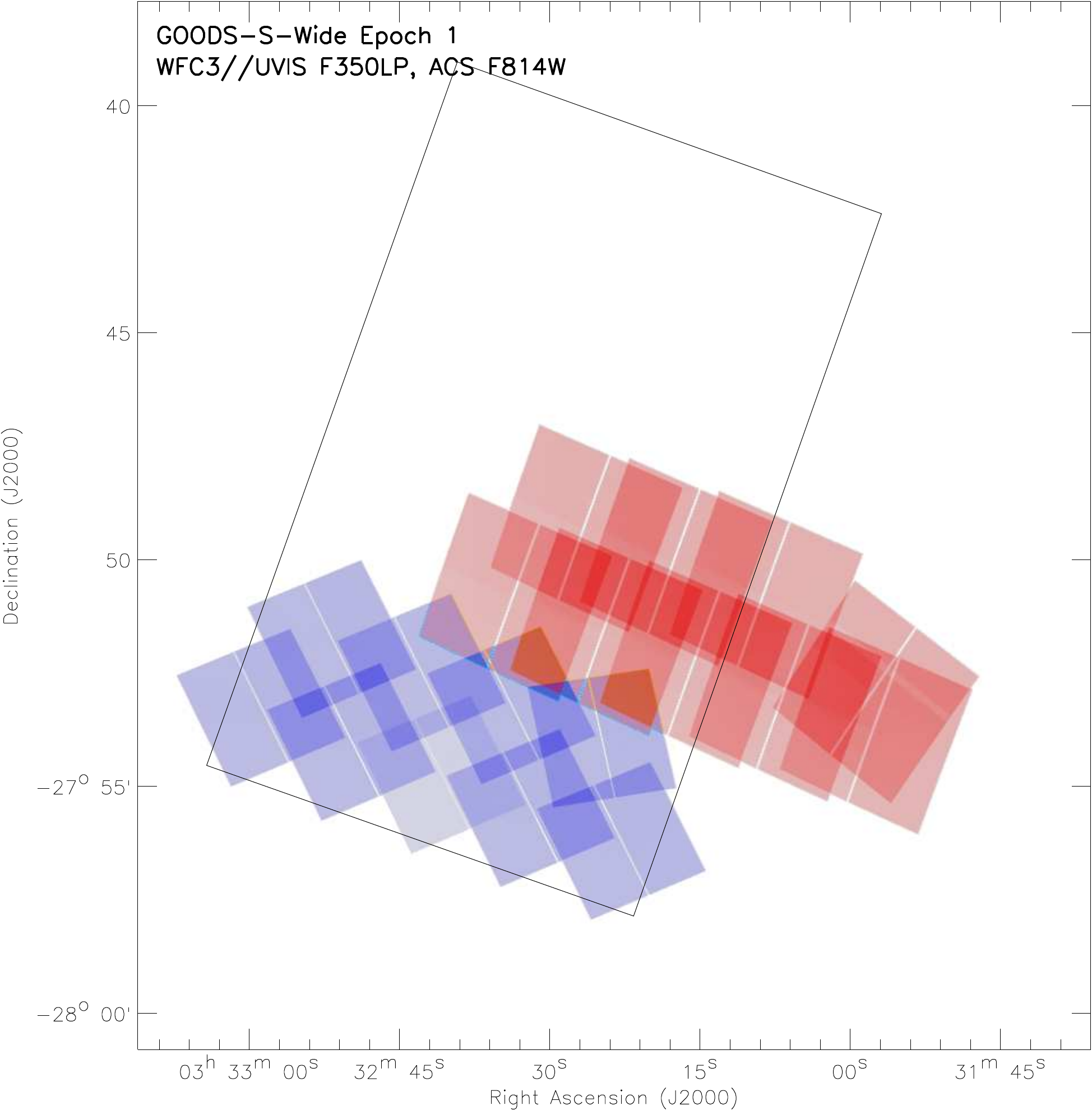}
\fi
\figcaption{\label{fig:gsw01_acs}%
As for Figure~\ref{fig:gsd01_acs}, but for GOODS-S-Wide Epoch 1.}
\end{center}
\end{figure*}

\clearpage
\begin{figure*}[h]
\begin{center}
\ifsubmodeastroph
  \includegraphics[height=4in]{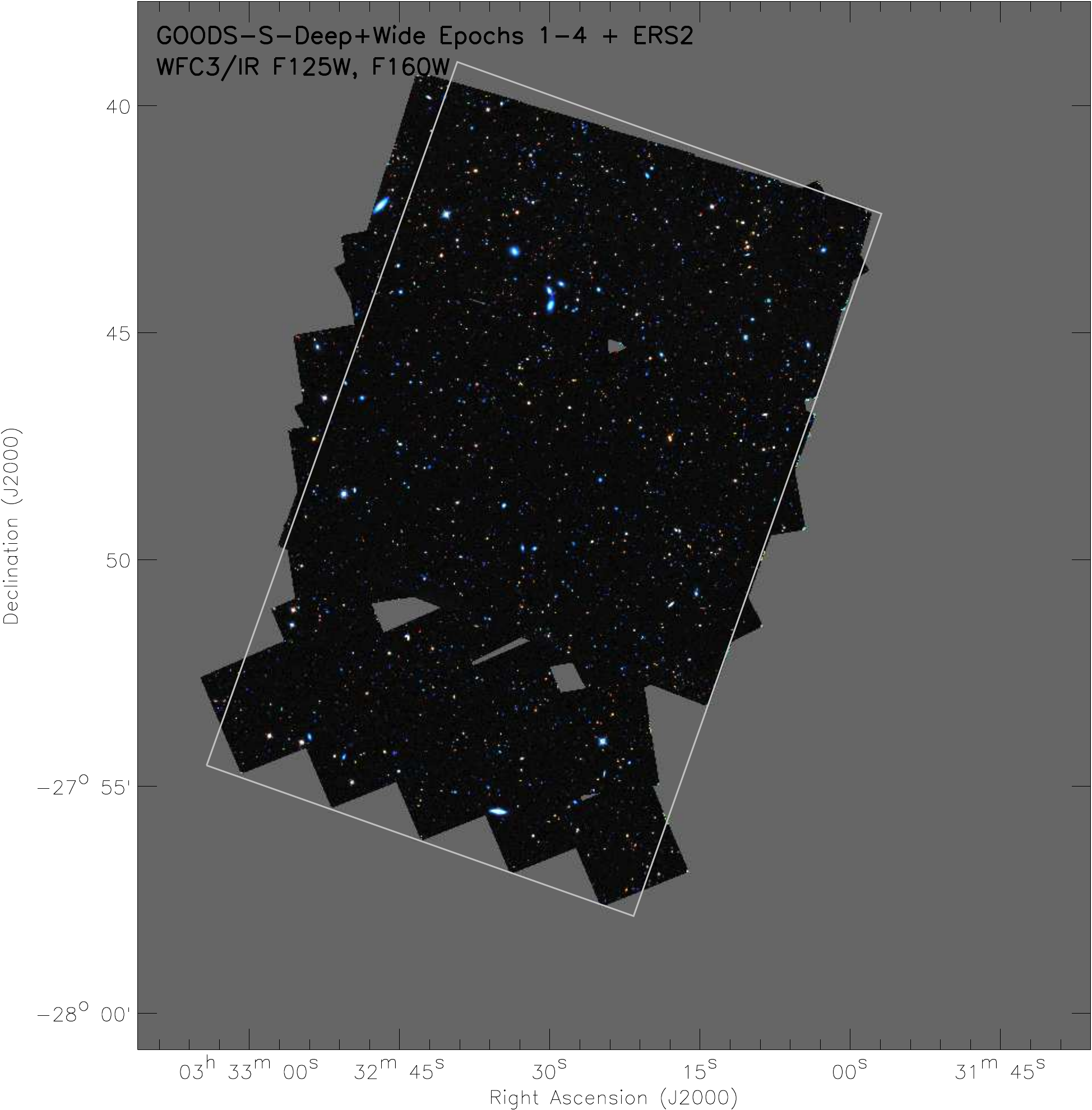}
  \includegraphics[height=4in]{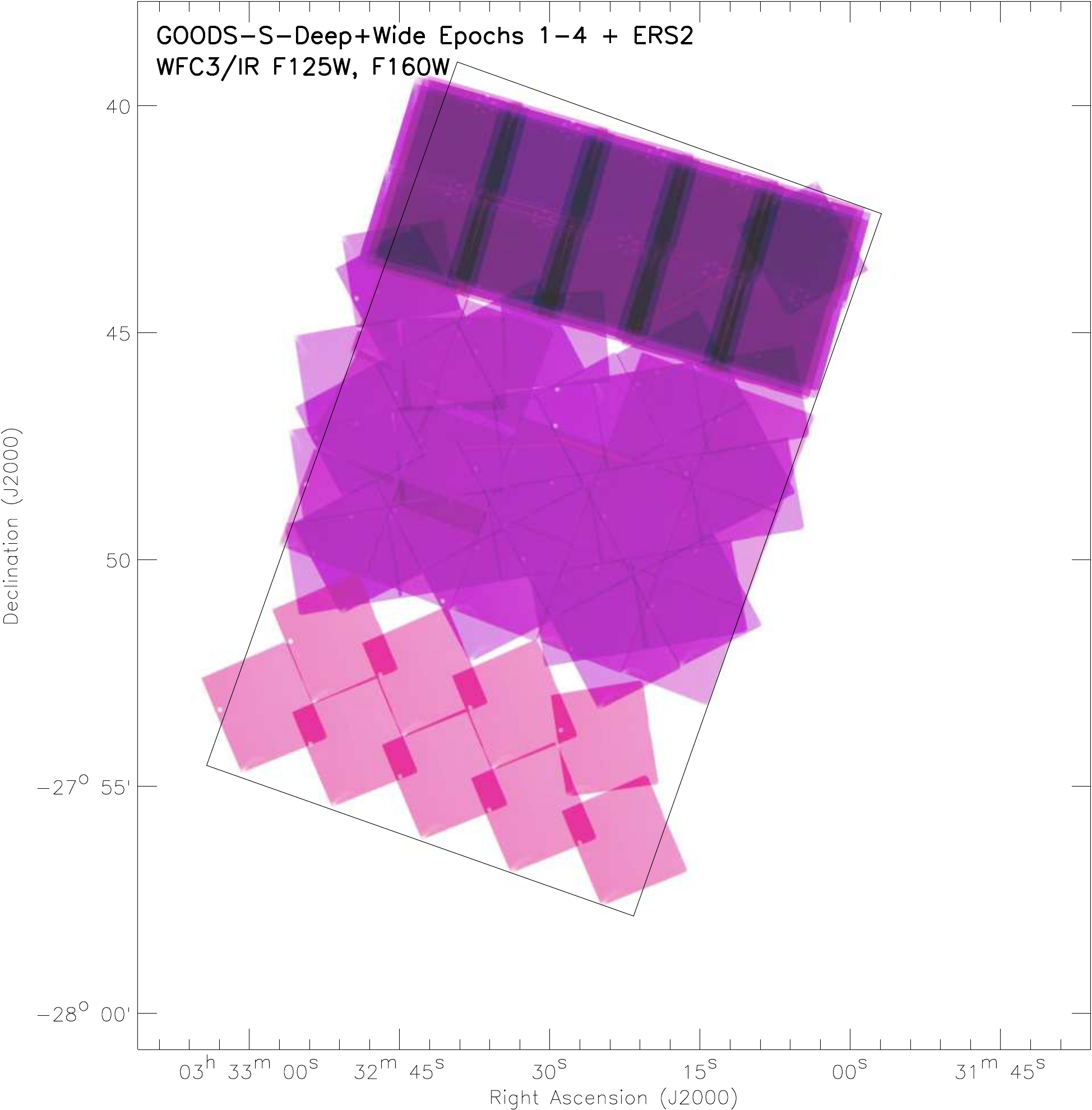}
\fi
\figcaption{\label{fig:gsa_wfc3}%
As for Figure~\ref{fig:gsd01_wfc3}, but this time showing the full accumulated CANDELS dataset on the GOODS-S field so far (including GOODS-S-Deep Epochs 1, 2, 3 as well as GOODS-S-Wide Epoch 1), together also with the WFC3/IR ERS2 F125W and F160W dataset, since the CANDELS observations were designed to abut this dataset.}
\end{center}
\end{figure*}

\clearpage
\begin{figure*}[h]
\begin{center}
\ifsubmodeastroph
  \includegraphics[height=4in]{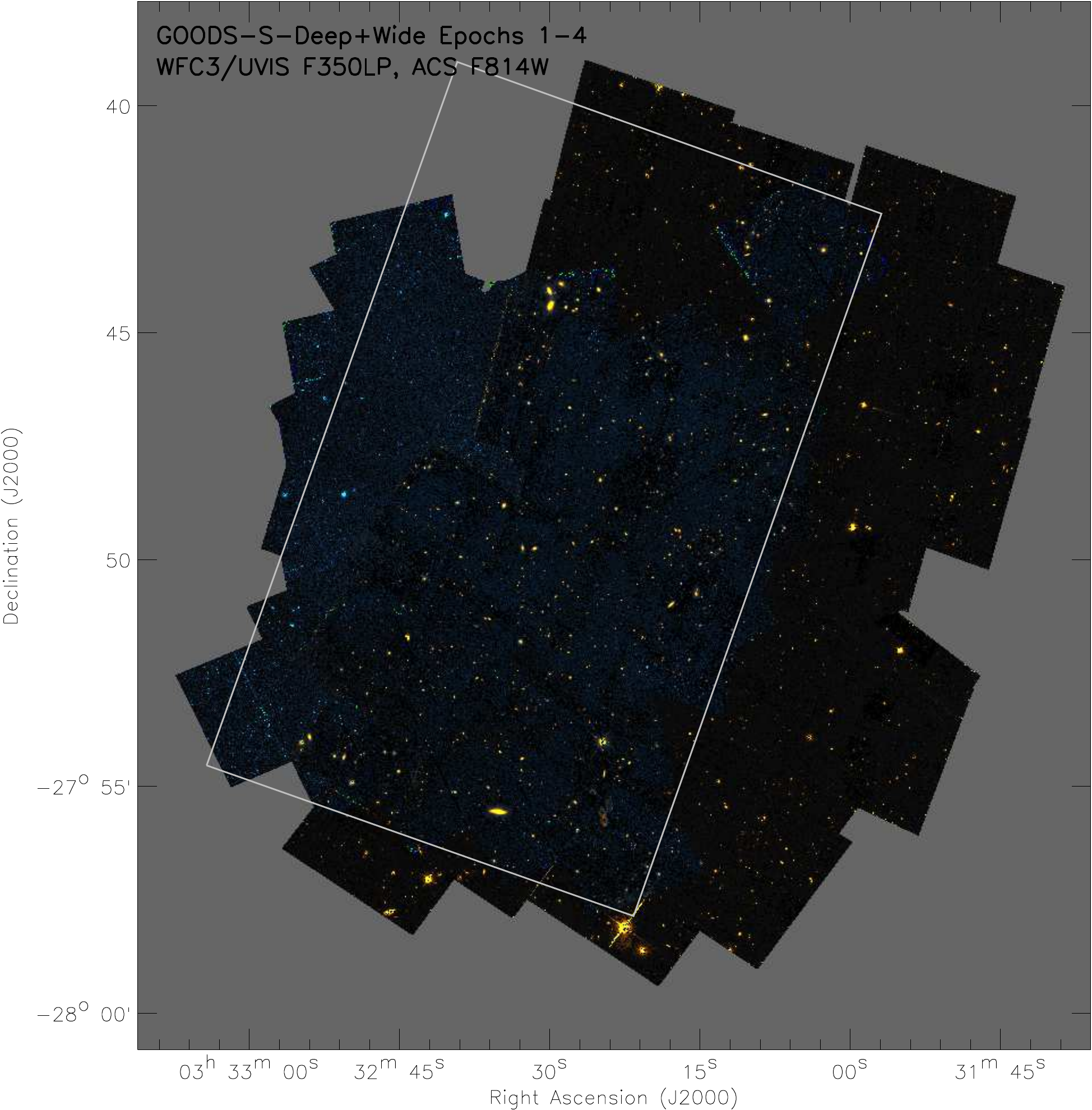}
  \includegraphics[height=4in]{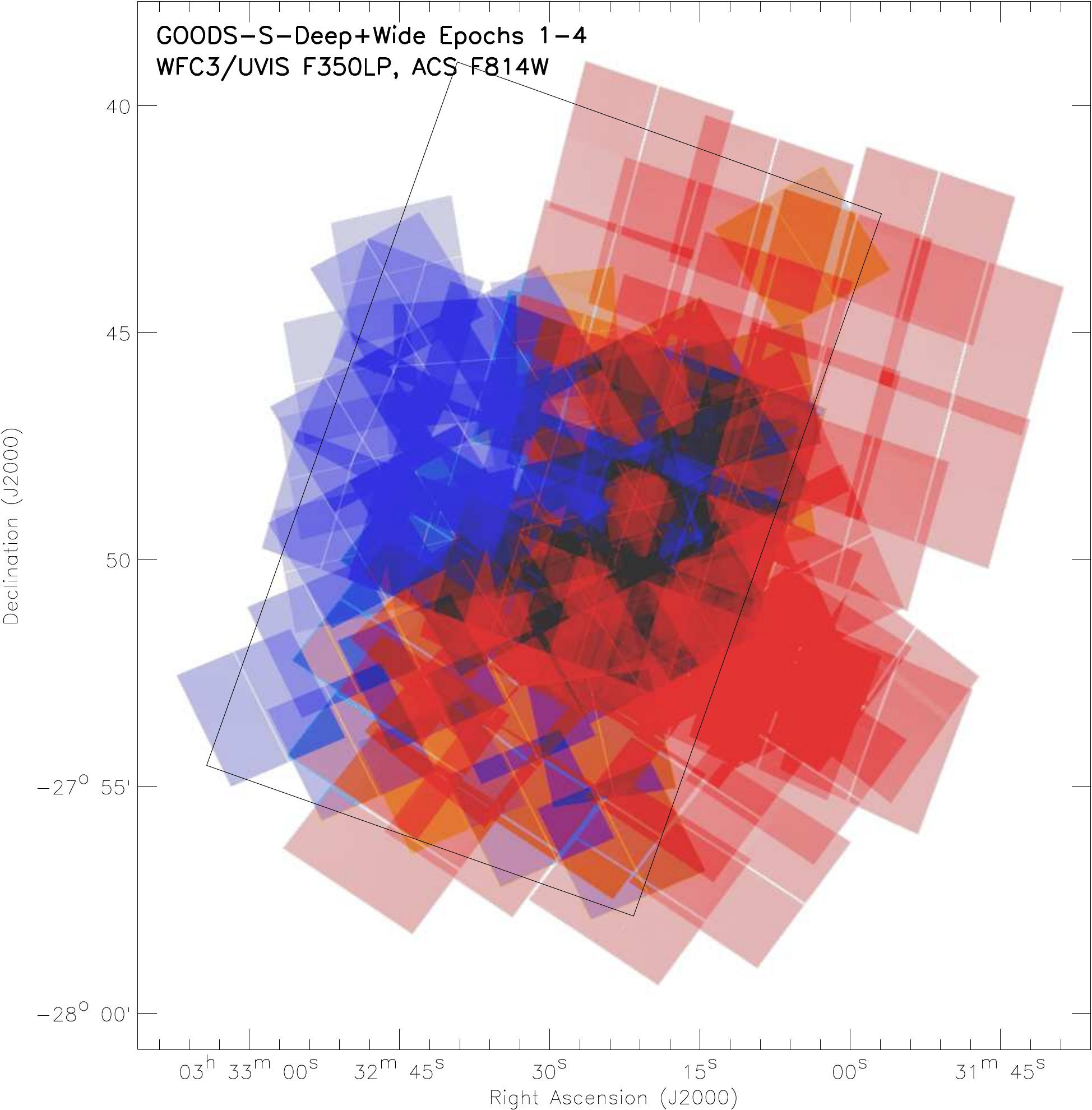}
\fi
\figcaption{\label{fig:gsa_acs}%
As for Figure~\ref{fig:gsd01_acs}, but this time showing the full accumulated CANDELS dataset on the GOODS-S field so far (including GOODS-S-Deep Epochs 1, 2, 3 as well as GOODS-S-Wide Epoch 1).}
\end{center}
\end{figure*}

\clearpage
\begin{figure*}[h]
\begin{center}
\ifsubmodeastroph
  \vspace{1in}%
  \includegraphics[height=3in]{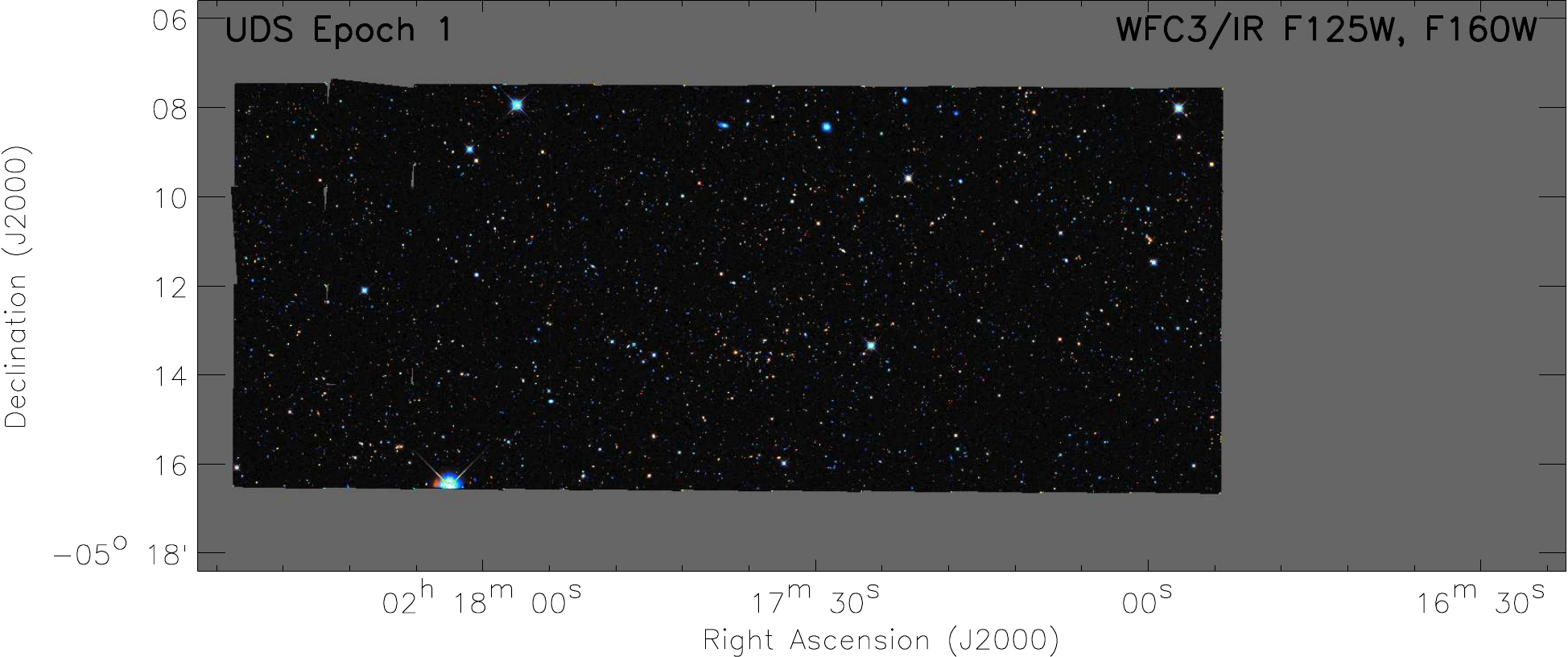}\vspace{0.45in}
  \includegraphics[height=3in]{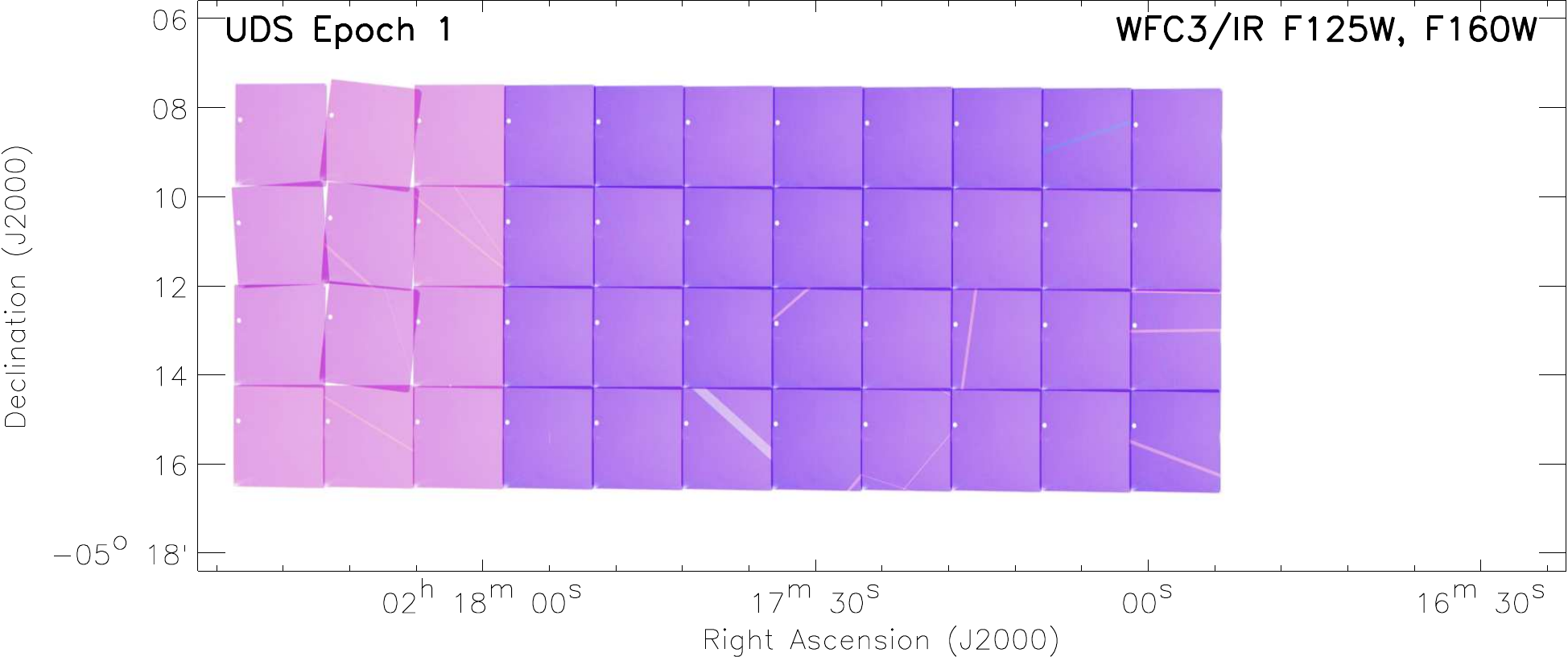}\vspace{0.5in}
\fi
\figcaption{\label{fig:uds01_wfc3}%
Images showing the prime CANDELS WFC3/IR dataset for the first epoch obtained on the UKIDSS/UDS field (UDS Epoch 1). The top panel shows a color composite of the WFC3/IR F125W and F160W images after combined mosaics were created for each filter separately using \multidrizzle, while the bottom panel shows the corresponding weight images, which are in units of inverse variance. The F125W data are shown in blue and the F160W data are shown in red. Regions containing bad pixels (such as the circular ``death star'' region) are set to 0 and thus have no weight in this single epoch dataset, in which the dither offsets were not yet large enough to move over such features. Other masked features are largely satellite trails. The overlap between pointings was chosen to be just large enough to provide contiguous coverage while also maximizing total area covered. Occasional tiles are intentionally tilted or offset to enable appropriate guide stars to be selected. Exposures are shorter at the east end of the mosaic in order to accommodate F350LP exposures during the same orbit.}
\end{center}
\end{figure*}

\clearpage
\begin{figure*}[h]
\begin{center}
\ifsubmodeastroph
  \vspace{1in}%
  \includegraphics[height=3in]{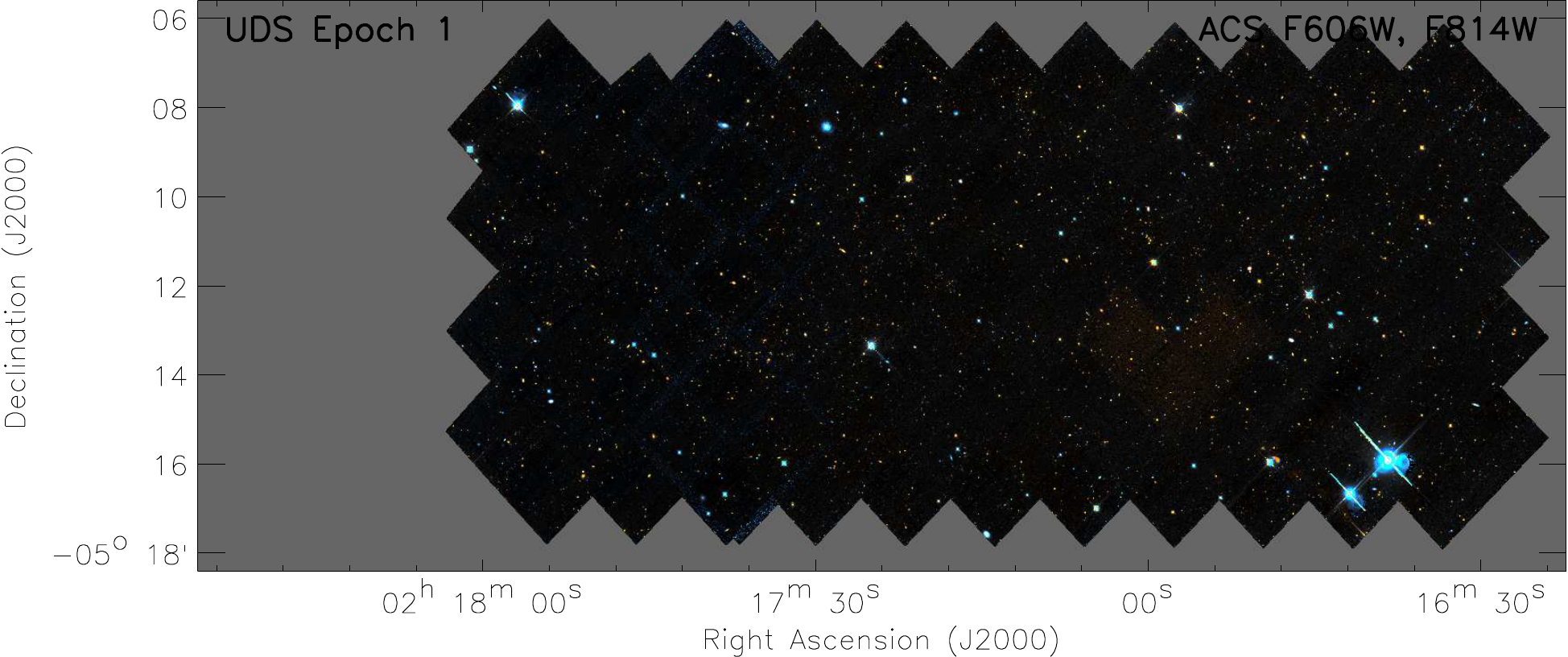}\vspace{0.5in}
  \includegraphics[height=3in]{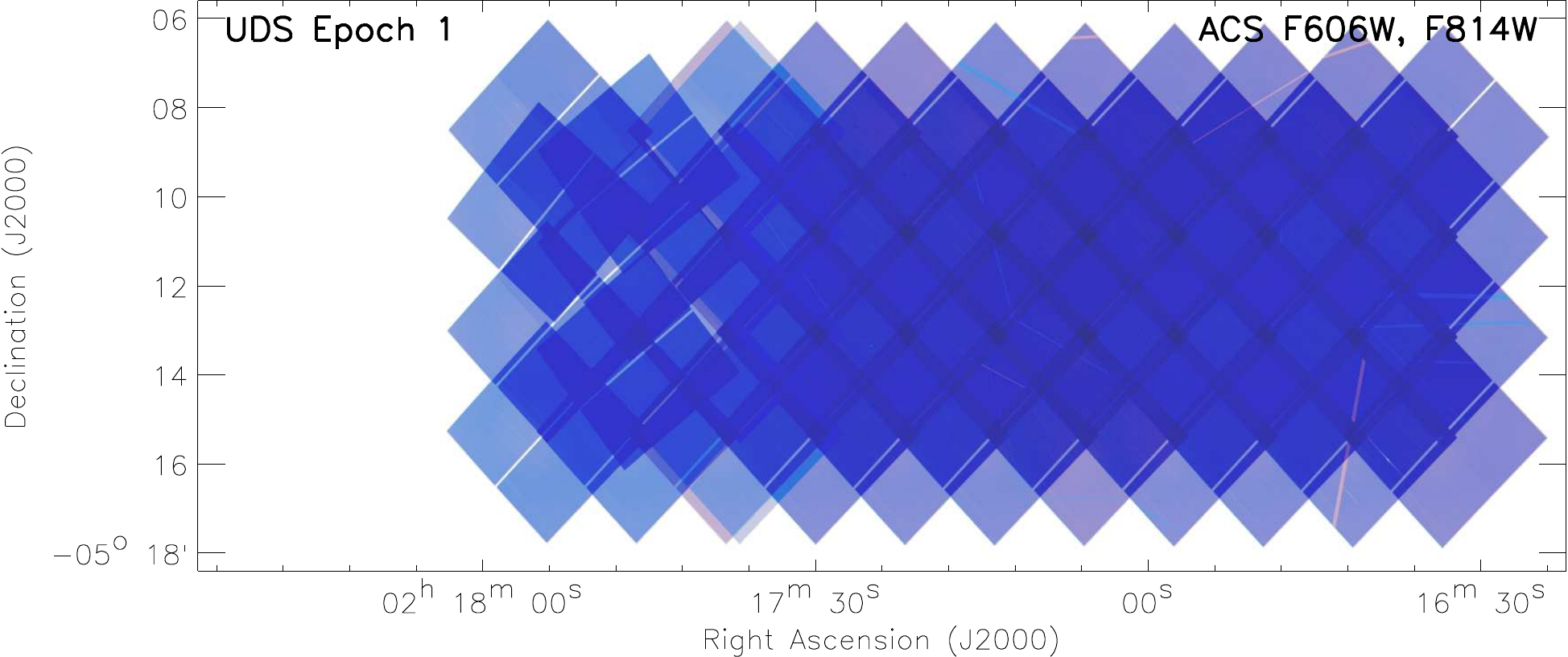}\vspace{0.5in}
\fi
\figcaption{\label{fig:uds01_acs}%
Images showing the parallel CANDELS ACS/WFC dataset for the first epoch obtained on the UKIDSS/UDS field (UDS Epoch 1). The top panel shows a color composite of the ACS/WFC F606W and F814W images after combined mosaics were created for each filter separately using \multidrizzle, while the bottom panel shows the corresponding weight images, which are in units of inverse variance. The F606W data are shown in blue and the F814W data are shown in red. Regions containing bad pixels (such as satellite trails) are masked where necessary. Note also that, in general, the overlap between ACS pointings is sufficient to provide approximately twice the depth of a single pointing across much of the ACS area.}
\end{center}
\end{figure*}

\clearpage
\begin{figure*}[h]
\begin{center}
\ifsubmodeastroph
  \vspace{1in}
  \includegraphics[height=3in]{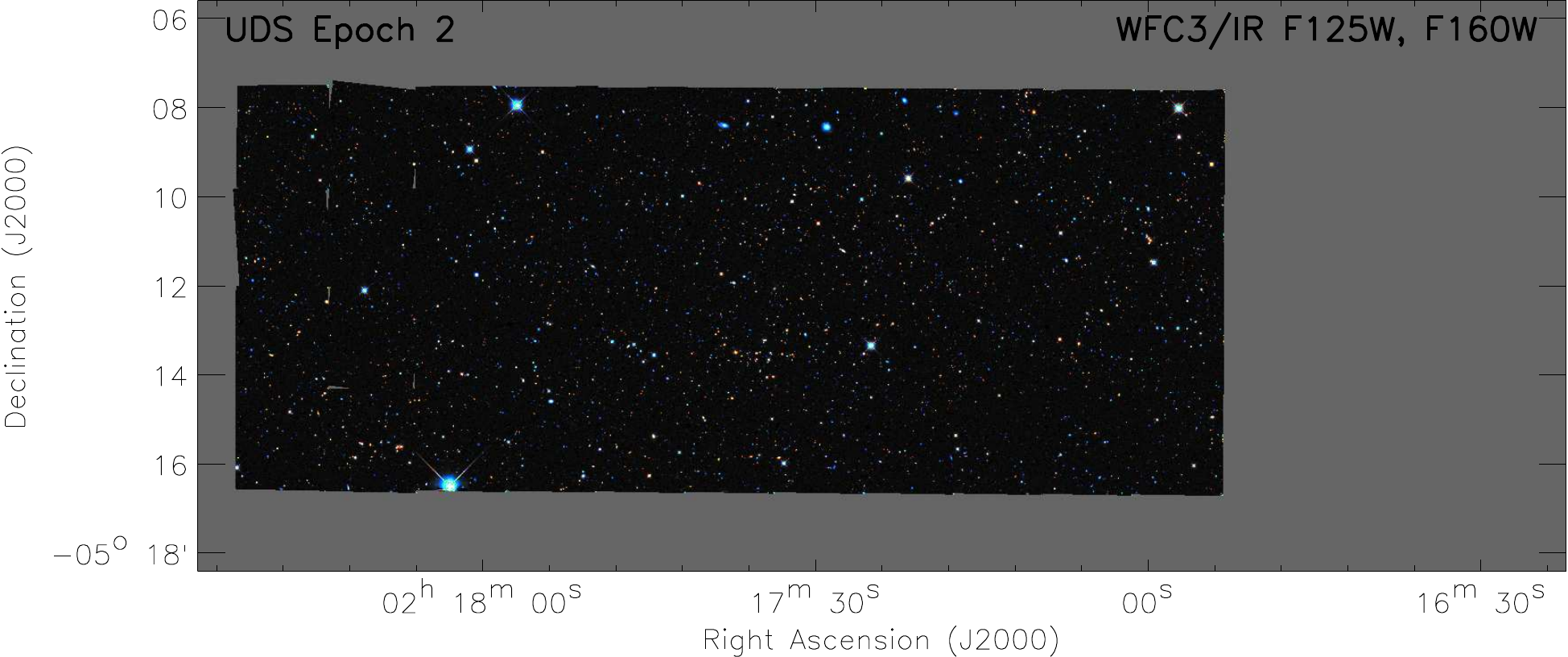}\vspace{0.5in}
  \includegraphics[height=3in]{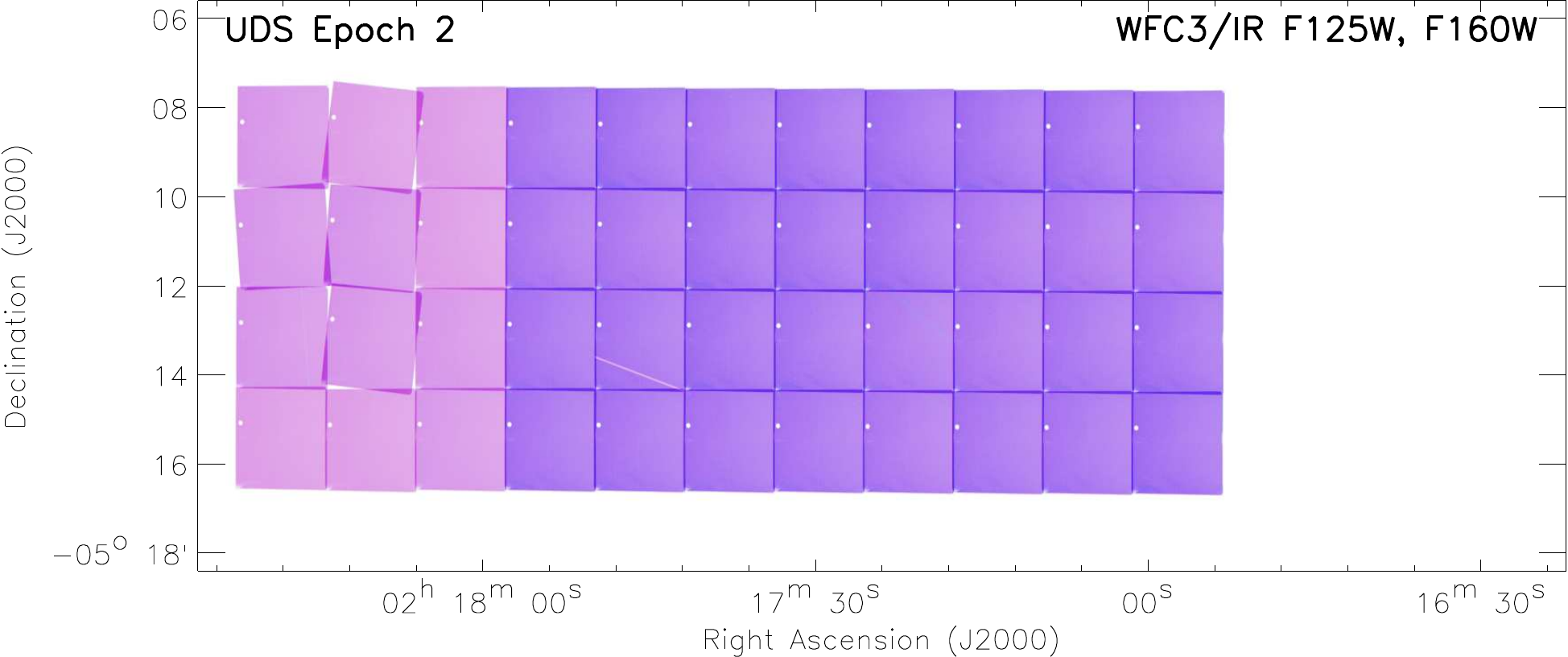}\vspace{0.5in}
\fi
\figcaption{\label{fig:uds02_wfc3}%
As for Figure~\ref{fig:uds01_wfc3}, but for UDS Epoch 2.}
\end{center}
\end{figure*}

\clearpage
\begin{figure*}[h]
\begin{center}
\ifsubmodeastroph
  \vspace{1in}
  \includegraphics[height=3in]{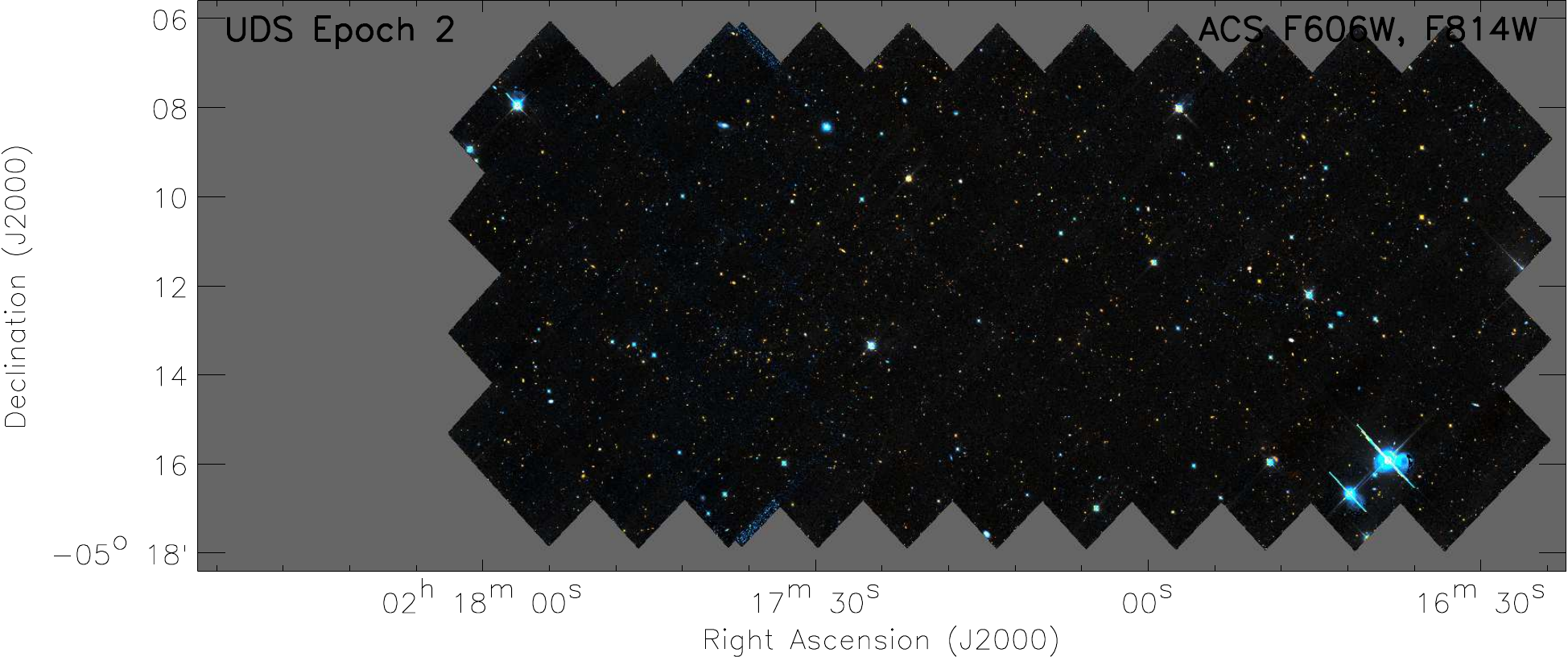}\vspace{0.5in}
  \includegraphics[height=3in]{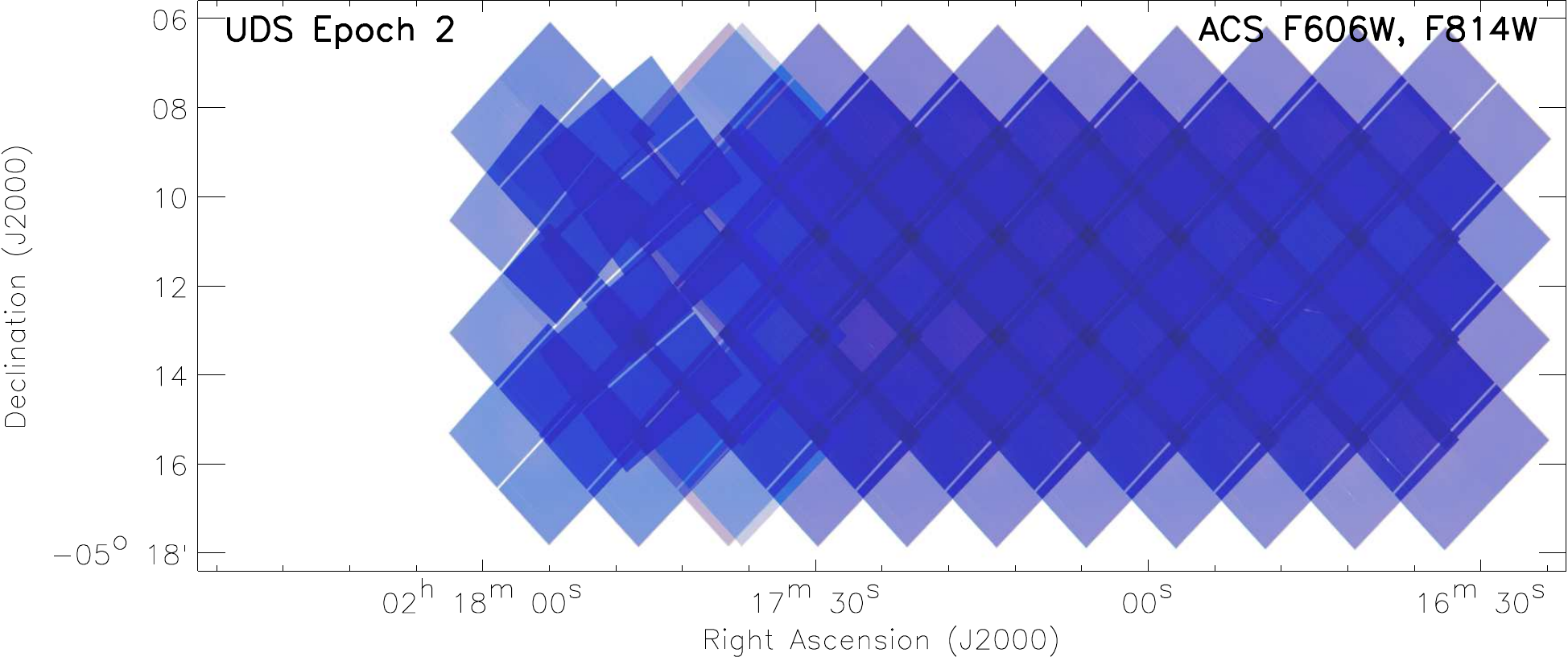}\vspace{0.5in}
\fi
\figcaption{\label{fig:uds02_acs}%
As for Figure~\ref{fig:uds01_acs}, but for UDS Epoch 2.}
\end{center}
\end{figure*}

\clearpage
\begin{figure*}[h]
\begin{center}
\ifsubmodeastroph
  \vspace{1in}
  \includegraphics[height=3in]{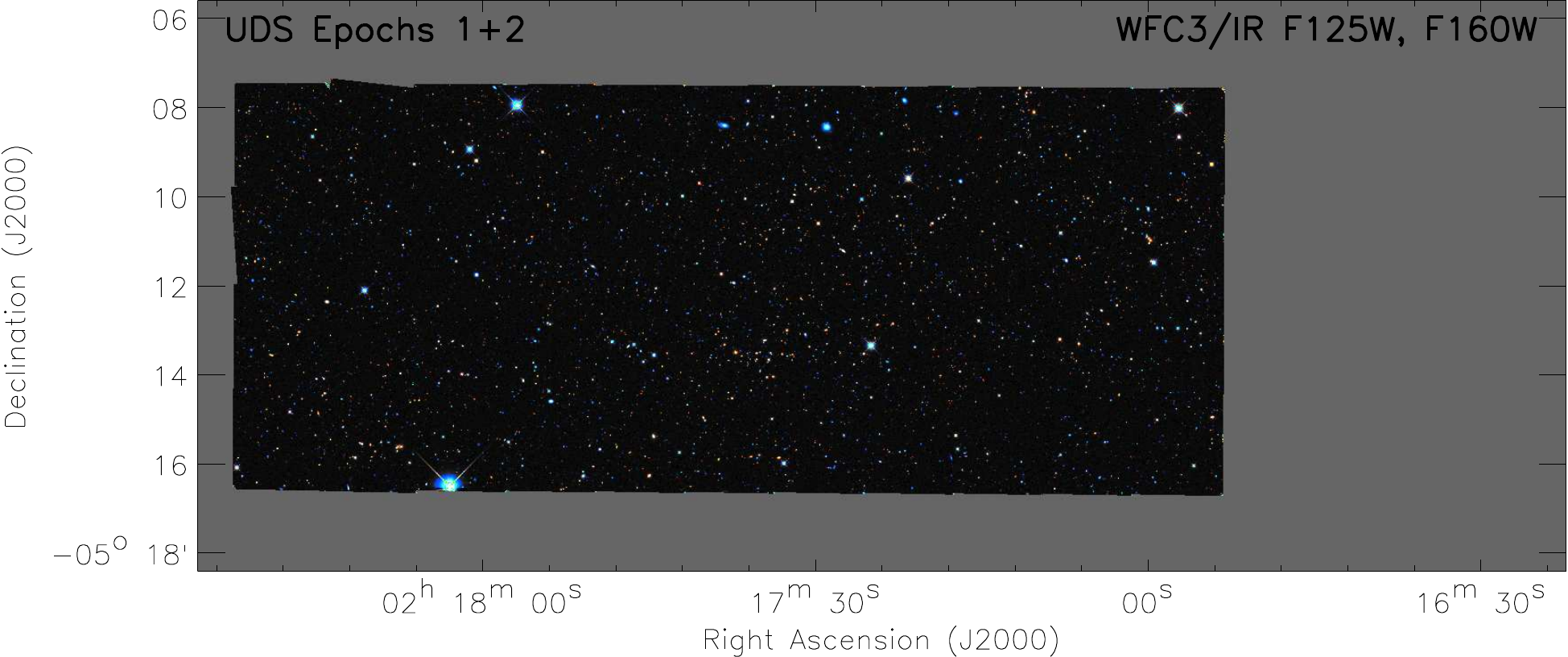}\vspace{0.5in}
  \includegraphics[height=3in]{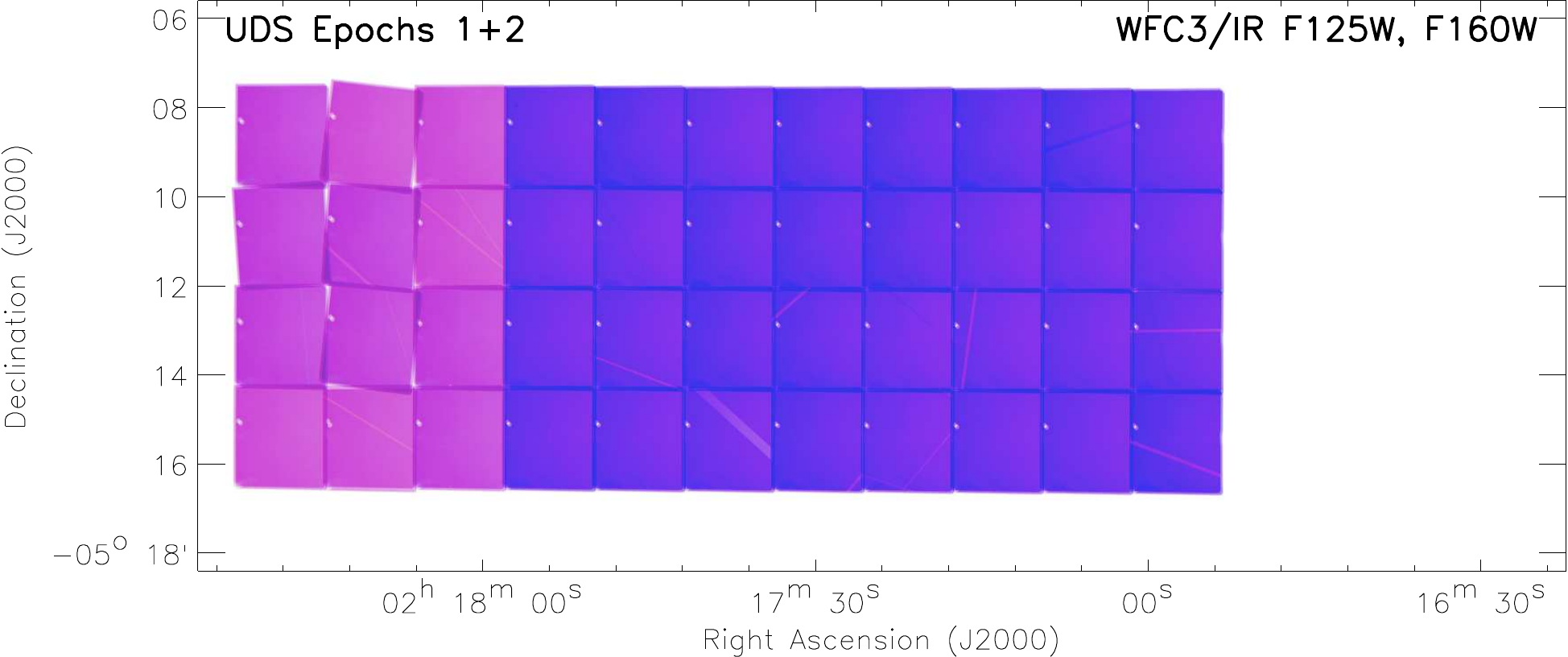}\vspace{0.5in}
\fi
\figcaption{\label{fig:uds_full_wfc3}%
As for Figure~\ref{fig:gsd01_wfc3}, but showing the full, final accumulated CANDELS dataset on the UDS field (including epochs 1 and 2).}
\end{center}
\end{figure*}

\clearpage
\begin{figure*}[h]
\begin{center}
\ifsubmodeastroph
  \vspace{1in}
  \includegraphics[height=3in]{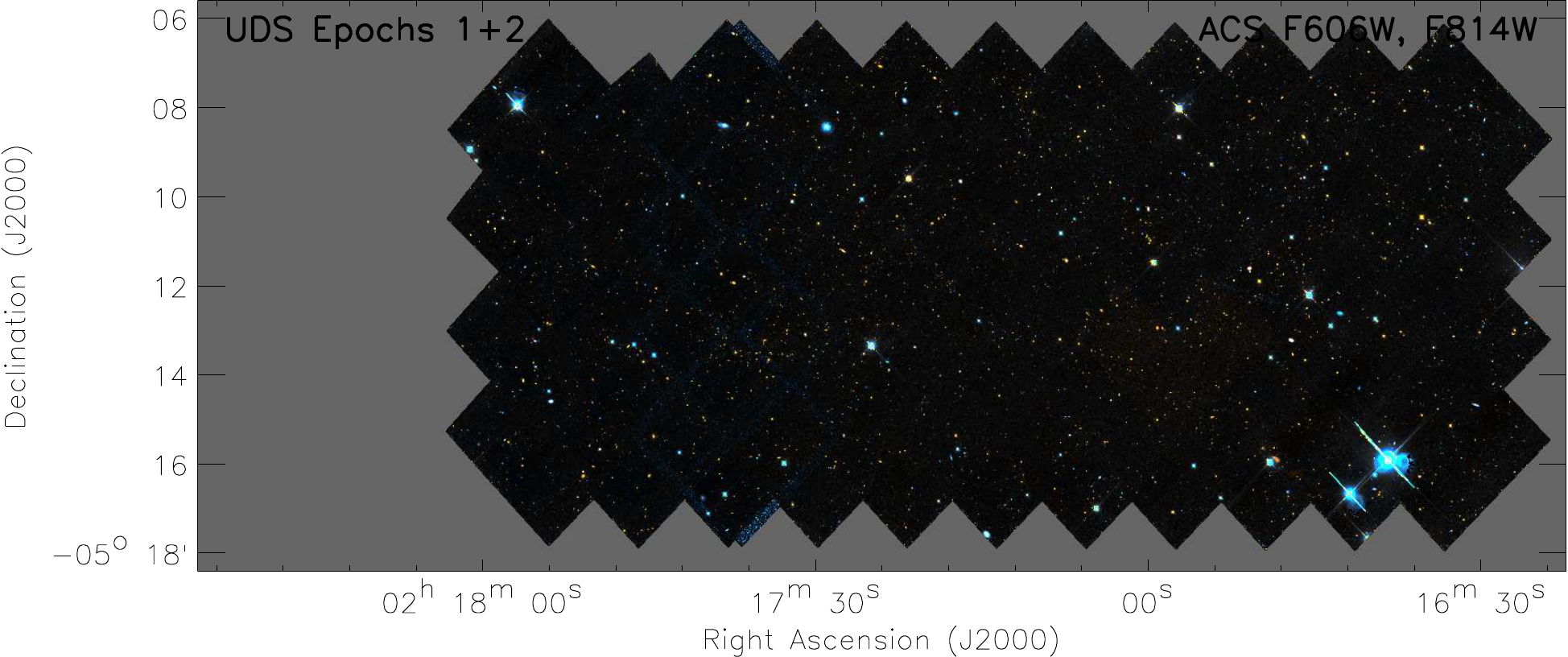}\vspace{0.5in}
  \includegraphics[height=3in]{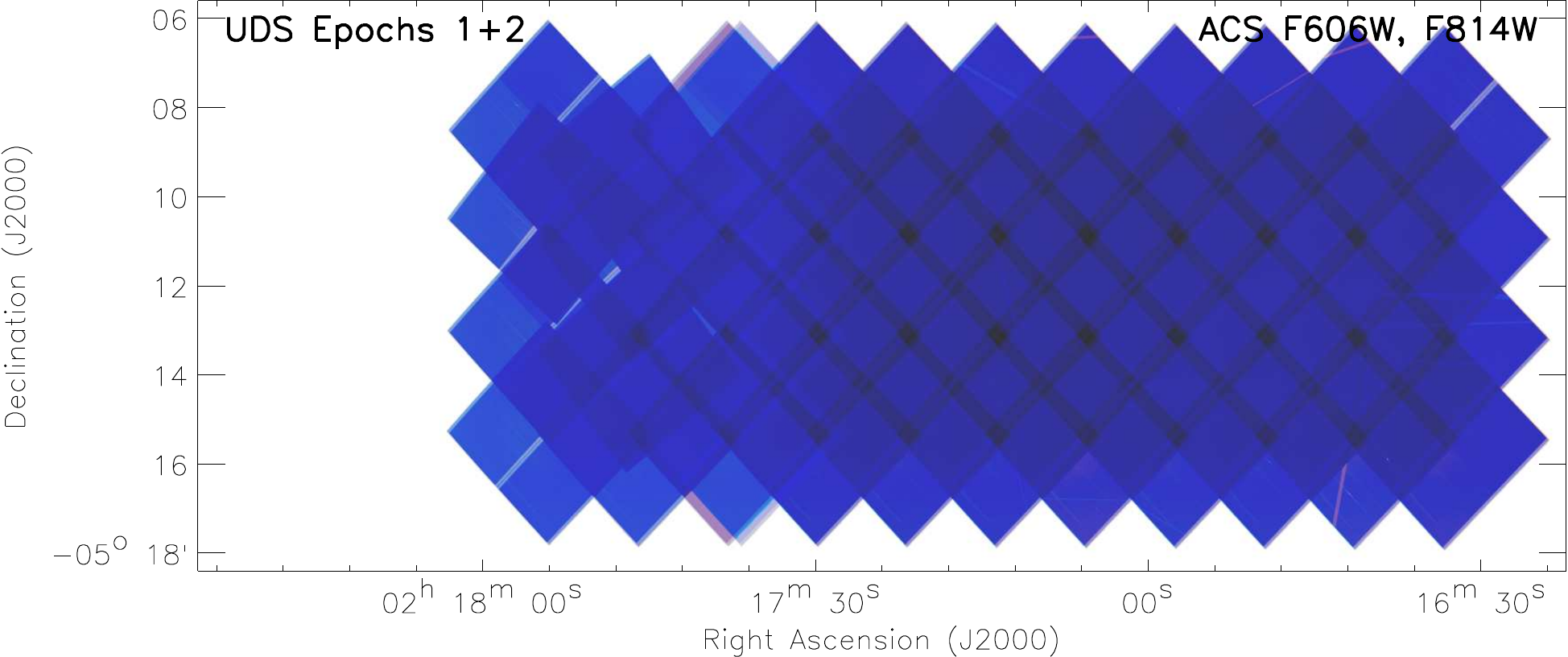}\vspace{0.5in}
\fi
\figcaption{\label{fig:uds_full_acs}%
As for Figure~\ref{fig:gsd01_acs}, but showing the full, final accumulated CANDELS dataset on the UDS field (including epochs 1 and 2).}
\end{center}
\end{figure*}

\end{document}